\documentclass[tighten, nolinenumbers, notrackchanges, twocolumn]{aastex631}

\usepackage{amsmath}
\usepackage{amssymb}
\usepackage{amssymb}
\usepackage{mathtools}
\usepackage{mathrsfs}
\usepackage{comment}

\usepackage{morefloats}
\usepackage{longtable}
\usepackage{afterpage}

\hypersetup{linkcolor=red,citecolor=blue,urlcolor=magenta}

\usepackage{soul}

\usepackage[makeroom]{cancel}

\graphicspath{{figs/}}

\usepackage{xspace}
\newcommand{\project}[1]{\textsl{#1}\xspace}
\newcommand*{\XPSI}{\project{X-PSI}}
\newcommand*{\NICER}{\project{NICER}}
\newcommand*{\XMM}{\project{XMM-Newton}}
\newcommand*{\MultiNest}{\textsc{MultiNest}\xspace}
\newcommand*{\TEMPO}{\textsc{tempo}2\xspace}
\newcommand{\msol}{M$_\odot$\xspace}
\newcommand{\jdbl}{PSR~J0030$+$0451\xspace}
\newcommand{\joh}{PSR~J0740$+$6620\xspace}

\newcommand{\TT}[1]{\texttt{#1}}

\received{}
\revised{}
\accepted{}
\published{}

\submitjournal{The Astrophysical Journal Letters}

\shorttitle{A \NICER VIEW OF THE MASSIVE PULSAR PSR~J0740$+$6620}
\shortauthors{Riley~et~al.}

\begin{document}

\title{A \NICER VIEW OF THE MASSIVE PULSAR PSR~J0740$+$6620 INFORMED BY RADIO TIMING AND \textit{XMM-NEWTON} SPECTROSCOPY}

\correspondingauthor{T.~E.~Riley}
\email{t.e.riley@uva.nl}

\author[0000-0001-9313-0493]{Thomas~E.~Riley}
\affil{Anton Pannekoek Institute for Astronomy, University of Amsterdam, Science Park 904, 1090GE Amsterdam, the Netherlands}

\author[0000-0002-1009-2354]{Anna~L.~Watts}
\affil{Anton Pannekoek Institute for Astronomy, University of Amsterdam, Science Park 904, 1090GE Amsterdam, the Netherlands}

\author[0000-0002-5297-5278]{Paul~S.~Ray}
\affil{Space Science Division, U.S. Naval Research Laboratory, Washington, DC 20375, USA}

\author[0000-0002-9870-2742]{Slavko~Bogdanov} 
\affil{Columbia Astrophysics Laboratory, Columbia University, 550 West 120th Street, New York, NY 10027, USA}

\author[0000-0002-6449-106X]{Sebastien~Guillot}
\affil{IRAP, CNRS, 9 avenue du Colonel Roche, BP 44346, F-31028 Toulouse Cedex 4, France}
\affil{Universit\'{e} de Toulouse, CNES, UPS-OMP, F-31028 Toulouse, France.}

\author[0000-0003-4357-0575]{Sharon~M.~Morsink}
\affil{Department of Physics, University of Alberta, 4-183 CCIS, Edmonton, AB, T6G 2E1, Canada}

\author[0000-0002-7177-6987]{Anna~V.~Bilous}
\affil{ASTRON, the Netherlands Institute for Radio Astronomy, Postbus
2, 7990 AA Dwingeloo, The Netherlands}

\author{Zaven~Arzoumanian}
\affil{X-Ray Astrophysics Laboratory, NASA Goddard Space Flight Center, Code 662, Greenbelt, MD 20771, USA}

\author[0000-0002-2651-5286]{Devarshi~Choudhury}
\affil{Anton Pannekoek Institute for Astronomy, University of Amsterdam, Science Park 904, 1090GE Amsterdam, the Netherlands}

\author{Julia~S.~Deneva}
\affil{George Mason University, Resident at the U.S. Naval Research Laboratory, Washington, DC 20375, USA}

\author{Keith~C.~Gendreau}
\affil{X-Ray Astrophysics Laboratory, NASA Goddard Space Flight Center, Code 662, Greenbelt, MD 20771, USA}

\author{Alice~K. Harding}
\affil{Theoretical Division, Los Alamos National Laboratory, Los Alamos, NM 87545, USA}

\author[0000-0002-6089-6836]{Wynn~C.~G.~Ho}
\affil{Department of Physics and Astronomy, Haverford College, 370 Lancaster Avenue, Haverford, PA 19041, USA}

\author{James M. Lattimer}
\affil{Department of Physics and Astronomy, Stony Brook University, Stony Brook, NY 11794-3800, USA}

\author{Michael Loewenstein}
\affil{Department of Astronomy, University of Maryland, College Park, MD 207423}
\affil{X-Ray Astrophysics Laboratory, NASA Goddard Space Flight Center, Code 662, Greenbelt, MD 20771, USA}

\author{Renee~M.~Ludlam}
\affil{Cahill Center for Astronomy and Astrophysics, California Institute of Technology, Pasadena, CA 91125, USA}
\affil{NASA Hubble Fellowship Program Einstein Postdoctoral Fellow}

\author{Craig B. Markwardt}
\affil{X-Ray Astrophysics Laboratory, NASA Goddard Space Flight Center, Code 662, Greenbelt, MD 20771, USA}

\author{Takashi Okajima}
\affil{X-Ray Astrophysics Laboratory, NASA Goddard Space Flight Center, Code 662, Greenbelt, MD 20771, USA}

\author[0000-0002-6742-4532]{Chanda Prescod-Weinstein}
\affil{Department of Physics and Astronomy, University of New Hampshire, Durham, New Hampshire 03824, USA}

\author[0000-0003-4815-0481]{Ronald A. Remillard}
\affil{MIT Kavli Institute for Astrophysics \& Space Research, MIT, 70 Vassar St., Cambridge, MA 02139, USA}

\author[0000-0002-4013-5650]{Michael~T.~Wolff}
\affil{Space Science Division, U.S. Naval Research Laboratory, Washington, DC 20375, USA}

\author[0000-0001-8384-5049]{Emmanuel Fonseca}
\affil{Department of Physics, McGill University, 3600 rue University, Montr\'eal, QC H3A 2T8, Canada}
\affil{McGill Space Institute, McGill University, 3550 rue University, Montr\'eal, QC H3A 2A7, Canada}
\affil{Department of Physics and Astronomy, West Virginia University, P.O. Box 6315, Morgantown, WV 26506, USA}
\affil{Center for Gravitational Waves and Cosmology, West Virginia University, Chestnut Ridge Research Building, Morgantown, WV 26505, USA}

\author[0000-0002-6039-692X]{H. Thankful Cromartie}
\affil{Cornell Center for Astrophysics and Planetary Science and Department of Astronomy, Cornell University, Ithaca, NY 14853, USA}
\affil{NASA Hubble Fellowship Program Einstein Postdoctoral Fellow}

\author{Matthew Kerr}
\affil{Space Science Division, U.S. Naval Research Laboratory, Washington, DC 20375, USA}

\author[0000-0001-5465-2889]{Timothy T. Pennucci}
\affil{National Radio Astronomy Observatory, 520 Edgemont Road, Charlottesville, VA 22903, USA}
\affil{Institute of Physics, E\"otv\"os Lor\'and University, P\'azm\'any P.s. 1/A, 1117 Budapest, Hungary}

\author[0000-0002-4140-5616]{Aditya Parthasarathy}
\affil{Max Planck Institute for Radio Astronomy, Auf dem H\"{u}gel 69, D-53121 Bonn, Germany}

\author[0000-0001-5799-9714]{Scott Ransom}
\affil{National Radio Astronomy Observatory, 520 Edgemont Road, Charlottesville, VA 22903, USA}

\author{Ingrid Stairs}
\affil{Dept. of Physics and Astronomy, University of British Columbia, 6224 Agricultural Road, Vancouver, BC V6T 1Z1 Canada}

\author{Lucas Guillemot}
\author{Ismael Cognard}
\affil{Laboratoire de Physique et Chimie de l'Environnement et de l'Espace - Université d'Orléans/CNRS, 45071, Orléans Cedex 02, France}
\affil{Station de radioastronomie de Nançay, Observatoire de Paris, CNRS/INSU, 18330, Nançay, France}

\begin{abstract}
We report on Bayesian estimation of the radius, mass, and hot surface regions of the massive millisecond pulsar PSR~J0740$+$6620, conditional on pulse-profile modeling of \project{Neutron Star Interior Composition Explorer} X-ray Timing Instrument (\NICER XTI) event data. We condition on informative pulsar mass, distance, and orbital inclination priors derived from the joint NANOGrav and CHIME/Pulsar wideband radio timing measurements of \citet{Fonseca20}. We use \XMM European Photon Imaging Camera spectroscopic event data to inform our X-ray likelihood function. The prior support of the pulsar radius is truncated at 16~km to ensure coverage of current dense matter models. We assume conservative priors on instrument calibration uncertainty. We constrain the equatorial radius and mass of PSR~J0740$+$6620 to be $12.39_{-0.98}^{+1.30}$~km and $2.072_{-0.066}^{+0.067}$~M$_{\odot}$ respectively, each reported as the posterior credible interval bounded by the 16\% and 84\% quantiles, conditional on surface hot regions that are non-overlapping spherical caps of fully-ionized hydrogen atmosphere with uniform effective temperature; \textit{a posteriori}, the temperature is $\log_{10}(T\textrm{~[K]})=5.99_{-0.06}^{+0.05}$ for each hot region. All software for the X-ray modeling framework is open-source and all data, model, and sample information is publicly available, including analysis notebooks and model modules in the Python language. Our marginal likelihood function of mass and equatorial radius is proportional to the marginal joint posterior density of those parameters (within the prior support) and can thus be computed from the posterior samples.

\end{abstract}


\section{Introduction}\label{sec:intro}

The nature of supranuclear density matter, as found in neutron star cores, is highly uncertain. Possibilities include both neutron-rich nucleonic matter and stable states of strange matter in the form of hyperons or deconfined quarks \citep[for recent reviews see][]{Oertel17,Baym18,Tolos20,YangJ20,Hebeler21}. One way to determine the dense matter Equation of State (the EOS, a function of both composition and inter-particle interactions) is to measure neutron star masses and radii \citep{Lattimer16,Ozel16b}. There are several possible methods, but in this Letter we focus on pulse-profile modeling \citep[see][and references therein]{Watts16,Watts19b}. This requires precise phase-resolved spectroscopy, a technique that motivated the design and development of NASA's \project{Neutron Star Interior Composition Explorer} (\NICER).

The \NICER X-ray Timing Instrument (XTI) is a payload installed on the \project{International Space Station}. The primary observations carried out by \NICER are order megasecond exposures of rotation-powered X-ray millisecond pulsars (MSPs) that may be either isolated or in a binary system \citep[][]{bogdanov19a}. Surface X-ray emission from the heated magnetic poles propagates to the \NICER XTI through the curved spacetime of the pulsar, and the compactness affects the signal registered by the instrument. However, these pulsars also spin at relativistic rates. So with a precisely measured spin frequency derived from radio timing and high-quality spectral-timing event data, we are also sensitive to rotational effects on surface X-ray emission, and therefore to the radius of the pulsar independent of the compactness \citep[see][and references therein]{bogdanov19b}.

The first joint mass and radius inferences conditional\footnote{For an introduction to the concept of conditional probabilities within Bayesian inference see \citet{Sivia06,Trotta08,Hogg2012,Gelman2013,Clyde19,Hogg2020}.} on pulse-profile modeling of \NICER observations of a MSP were reported by \citet{miller19} and \citet{riley19a}.

 The target was PSR~J0030$+$0451, an isolated\footnote{No binary companion has ever been detected despite 20 years of intensive radio timing \citep{Lommen00,Arzoumanian18}.} source spinning at approximately $205$~Hz. Being isolated, the radio timing model for this MSP has no dependence on its mass, in contrast to the radio timing model for an MSP in a binary. This meant that a wide prior on the mass had to be assumed in the pulse-profile modeling, which nevertheless - due to the high quality of the data set in terms of the number of pulsed counts - delivered credible intervals on the mass and radius posteriors at the $\sim 10\%$ level. These posterior distributions have been used to infer properties of the dense matter EOS (in combination with constraints from radio timing, gravitational wave observations, and nuclear physics experiments). To give a few examples, there have been follow-on studies constraining both parameterized EOS models \citep{miller19,raaijmakers19,raaijmakers20,Dietrich20,Jiang20,AlMamun21} and non-parameterized EOS models \citep{Essick20,Landry20}, some focusing particularly on the neutron star maximum mass \citep{LimY20,Tews20b}. Others have focused on specific nuclear physics questions: hybrid stars and phase transitions to quark matter \citep{Tang20,LiA20,Christian20,XieWJ20c,Blaschke20,Alvarez20b}; the three nucleon potential \citep{Maselli20}; relativistic mean-field models \citep{Traversi20}; muon fraction content \citep{ZhangNB20}; and the nuclear symmetry energy \citep{Zimmerman20,Biswas20}. This is by no means an exhaustive review of the citing literature, but serves to give a flavor of how the previous \NICER result has been used.

Pulse-profile modeling also yields posterior distributions for the properties of the hot X-ray emitting regions on the star's surface, which are assumed to be related to the star's magnetic field structure \citep{Pavlov97}. The analysis of PSR~J0030$+$0451 implied a complex non-dipolar field \citep{Bilous19} and the posteriors have been used in follow-on studies of pulsar magnetospheres and radiation mechanisms \citep{ChenAY20,Suvorov20,Kalapotharakos20}.  

The subject of this Letter is the rotation-powered millisecond pulsar PSR~J0740$+$6620, spinning at approximately 346~Hz as it orbits with a binary companion \citep{Cromartie19}. Being in a binary at a favorable inclination for measurement of the Shapiro delay allows the mass of this source to be measured independently via radio timing \citep[e.g.,][]{HbkPulsAstro}. \citet{Cromartie19} reported a mass of $2.14^{+0.10}_{-0.09}$ \msol, making this the highest (well-constrained) mass neutron star. High-mass neutron stars (with the highest central densities) are particularly powerful in terms of their potential to constrain the dense matter EOS. The mass alone can cut down parameter space, but a measurement of radius adds far more \citep[see for example][]{HanS20,XieWJ20}.

The \project{North American Nanohertz Observatory for Gravitational Waves} (NANOGrav) and \project{Canadian Hydrogen Intensity Mapping Experiment} (CHIME) Pulsar collaborations recently joined forces to perform wideband radio timing of PSR~J0740$+$6620 \citep{Fonseca20}.  They derived an updated measurement of the pulsar mass ($2.08\pm 0.07$ \msol), its distance from Earth, and the orbital inclination. From these informative measurements and \NICER observations \citep{Wolff20} comes the potential for synergistic constraints on X-ray pulse-profile parameters that do not appear in the wideband radio timing solution; in particular, the radius of PSR~J0740$+$6620 and the properties of the hot surface X-ray emitting regions conditional on a model. We report such inferences in this Letter.

We organize this Letter as follows. In Section~\ref{sec:modeling procedure} we summarize the pulse-profile model components; we provide additional detail about novel model components to augment the information in \citet{riley19a} and \citet{bogdanov19b}. Section~\ref{sec:modeling procedure} also covers the X-ray likelihood function---which is the probability of the \NICER XTI event data and the spectroscopic event data acquired by the \XMM European Photon Imaging Camera (EPIC)---and details about the prior probability density functions of model parameters that are fundamentally shared by all models or multiple models. In Section~\ref{sec:inferences} we report model inferences and details about the prior probability density functions that are specific to a given surface hot region model. In Section~\ref{sec:discussion} we discuss these inferences in detail, covering their physical implications and potential systematic errors. In Section~\ref{sec:conclusion} we conclude by reporting the mass and radius constraints and commenting on the outlook for future pulse-profile modeling efforts.  EOS inference using our derived mass-radius posterior is carried out in a companion paper \citep{Raaijmakers21}.  

\section{Modeling procedure}\label{sec:modeling procedure}

The methodology in this Letter is largely shared with that of \citet{riley19a}, \citet{bogdanov19b}, and \cite{bogdanov19c}. In this section, we summarize that methodology and give a more detailed report of the new model components.

We formulate the general form of the likelihood function shared by all models and also the prior probability density functions (PDFs) of parameters that are shared by all models. We reserve definitions of prior PDFs of phenomenological surface hot region parameters for Section \ref{sec:inferences} where posterior inferences are reported. All posterior PDFs are computed using the \textit{X-ray Pulse Simulation and Inference} (\XPSI) \texttt{v0.7} framework \citep{xpsi}, an updated version of the package used by \citet{riley19a}.\footnote{\url{https://github.com/ThomasEdwardRiley/xpsi}} The analysis files may be found in the persistent repository of \citet{zenodo}: the data products; the numeric model files including the telescope calibration products; model modules in the Python language using the \XPSI framework; posterior sample files; and Jupyter analysis notebooks. We begin by introducing the data sets used in the analysis, including the most important aspects of the data selection and preparation.

\subsection{X-ray event data}
In this section we summarize the event data sets reported by \citet{Wolff20}, including any pre-processing tailored to this present Letter.

\subsubsection{\NICER XTI}

The PSR J0740$+$6620 analysis is based on a sequence of exposures with \NICER XTI (hereafter \NICER) acquired in the period 2018 September 21 -- 2020 April 17. The event data was obtained using similar filtering criteria to the previously analyzed \NICER data set of PSR J0030+0451. We only used good time intervals when all 52 active detectors were collecting data but rejected all events from DetID 34, as it often shows enhanced count rates relative to the other 51 detectors. We excluded time intervals when PSR J0740$+$6620 was situated at an angle $\le80^{\circ}$ from the Sun to reduce the increase in background in the lowest channels due to optical loading. We further excised events collected at low cut-off rigidity (COR\_SAX values $<5$) to minimize particle background contamination. The resulting cleaned event list has an on-source exposure time of 1602683.761\,s. 

We model events registered in the PI channel subset $[30,150)$, corresponding to the nominal photon energy range $[0.3,1.5]$~keV.\footnote{A photon that deposits all of its energy, $E\in[0.3,1.5)$~keV, in a detector with perfect energy resolution is registered in channel subset $[30,150)$ after mapping according to the gain-scale calibration product.
} The quoted nominal photon energy for a channel is the energy mid-point for that channel.

Below nominal energy $0.3$~keV, the instrument calibration products have greater \textit{a priori} uncertainty due to the sharp lower-energy threshold cutoff, as well as the presence of unrejected instrumental noise that can extend above 0.25~keV.  We therefore neglect the information below channel $30$ in order to reduce the risk of inferential bias. \citet{Wolff20} detect pulsations with the highest significance when considering event data up to nominal energy $\sim\!1.2$~keV; in this Letter we include the additional information at nominal energies in the range $[1.2,1.5]$~keV. The number of counts generated by the PSR~J0740$+$6620 hot regions in channels $150$ and above, however, is small and diminishes relative to the counts generated by background processes (including a non-thermal component from the environment in the near-vicinity of PSR~J0740$+$6620); we therefore neglect this higher-energy information, and focus energy resolution at nominal energies below $1.5$~keV.

The \NICER event data are phase-folded according to the NANOGrav radio timing solution presented in \citet{Fonseca20}. The phase-binned count numbers are displayed in Figure~\ref{fig:J0740 count data}.

{
    \begin{figure*}[t!]
    \centering
    \includegraphics[clip, trim=0cm 0cm 0cm 0cm, width=0.75\textwidth]{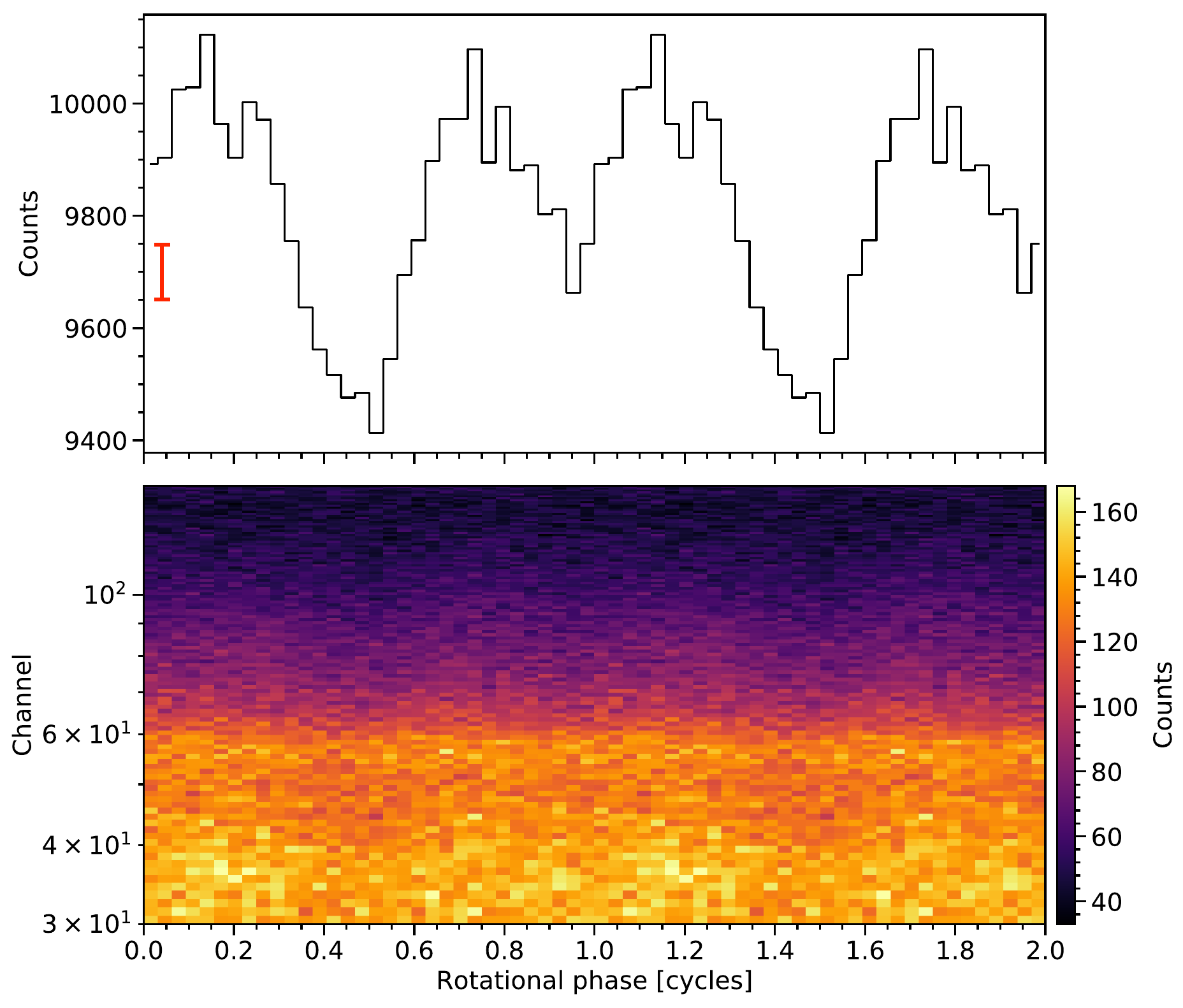}
    \caption{\small{Phase-folded PSR~J0740$+$6620 event data for two rotational cycles (for clarity): we use $32$ phase intervals (bins) per cycle and the count numbers in bins separated by one cycle---in a given channel---are identical. The total number of counts is given by the sum over all phase-channel pairs. The \textit{top} panel displays the pulse-profile summed over the contiguous subset of channels $[30,150)$. We use the \textit{red} bar to indicate the typical standard deviation that an adequately performing Poisson count model will exhibit. The \textit{bottom} panel displays the phase-channel resolved count numbers for channel subset $[30,150)$. For likelihood function evaluation we group all event data registered in a given channel into phase intervals spanning a single rotational cycle. The description in this caption is based on that given by \citet{riley19a} about the corresponding count-number figure for PSR~J0030$+$0451.}}
    \label{fig:J0740 count data}
    \end{figure*}
}

\subsubsection{\XMM EPIC}

The \XMM (hereafter \textit{XMM}) telescope observed PSR~J0740$+$6620 as part of a Director's Discretionary Time program in three visits: 2019 October 26 (ObsID 0851181601), 2019 October 28 (ObsID 0851181401), and 2019 November 1 (ObsID 0851181501). The European Photon Imaging Camera (EPIC) instruments (pn, MOS1, and MOS2) were employed in `Full Frame' imaging mode with the `Thin' optical blocking filters in place. Due to the insufficiently fast read-out times (73.4~ms for EPIC-pn and 2.6~s for EPIC-MOS1/2), the data do not provide useful pulse timing information, so only phase-averaged spectral information is available. The \textit{XMM} data were reduced using the Science Analysis Software (SAS\footnote{The \XMM SAS is developed and maintained by the Science Operations Centre at the European Space Astronomy Centre and the Survey Science Centre at the University of Leicester.
}) using the standard set of analysis threads. 

The event data were first screened for periods of strong soft proton background flaring. The resulting event lists were then further cleaned by applying the recommended PATTERN ($\le 12$ for MOS1/2 and $\le 4$ for pn) and FLAG ($0$) filters. The final source event lists were obtained by extracting events from a circular region of radius 25$''$ centered on the radio timing position of PSR~J0740$+$6620. This resulted in effective exposure times of 6.81, 17.96, and 18.7\,ks for the pn, MOS1, and MOS2 instruments, respectively.

\subsection{Radiation propagation from surface to telescope}\label{sec:radiation propagation}

\subsubsection{Design of the equatorial radius prior}\label{sec:radius prior design}

One of the ultimate aims of this Letter is to report a likelihood function of mass and (equatorial) radius that can be used for EOS posterior computation. As highlighted by \citet[][]{Riley18}, it is desirable to define a prior PDF which is jointly flat with respect to two parameters within the prior support; these parameters can simply be mass and radius, or some deterministically related variables, such as mass and compactness. The marginal joint posterior PDF of these parameters is then proportional to the marginal likelihood function of these parameters, meaning that the marginal likelihood function can be estimated from posterior samples, for use in subsequent inferential analyses.

In this Letter we follow \citet{riley19a} by defining a joint prior PDF of mass and radius that is flat within the prior support, which is maximally inclusive in regards to theoretical EOS predictions. The prior support is zero for $R>16$~km because we are not aware of any EOS models predicting a radius higher than this limit that would be compatible with current constraints from nuclear physics, or with the constraints posed by the gravitational wave measurement of tidal deformability for the binary neutron star merger GW170817 \citep[see, e.g.,][]{Reed21}. A difference to \citet{riley19a} is that we define the prior support using a higher compactness limit as discussed in Section~\ref{sec:ray tracing} (see Table~\ref{table: ST-U} for the prior PDF and support). The prior support is also subject to the condition that the effective gravity lies within a bounded range at every point of the rotating oblate surface, in order to conform to bounds on the atmosphere models we condition on (see Section~\ref{sec:atmosphere setup} and Table~\ref{table: ST-U}).

In Section~\ref{sec:radio} we introduce information about the mass in the form of a marginal PDF whose shape approximates the marginal likelihood function of mass derived in a recent radio timing analysis; we multiply this likelihood function with the jointly flat prior PDF of mass and radius described above, thereby defining an updated prior PDF for the X-ray pulse-profile modeling in this Letter. Our prior PDF for pulse-profile modeling is therefore not jointly flat with respect to mass and radius, but all shape information is likelihood-based, satisfying the requirements of \citet{Riley18}.

\subsubsection{Mass, inclination, and distance priors}\label{sec:radio}

Radiation is transported from the oblate X-ray emitting surface to distant static telescopes by relativistic ray-tracing, as described by \citet[][]{bogdanov19b} and references therein. The gravitational mass of \joh, the distance of \joh from Earth, and the inclination of the Earth's line-of-sight to the \joh spin axis are all parameters of the source-receiver system that need to be specified in order to simulate an X-ray signal incident on a telescope, and can be inferred via pulsar radio timing. Note that these static model X-ray telescopes are fictitious constructs: the real telescopes are in motion relative to the pulsar \citep[see, e.g.,][]{riley_thesis}. The \NICER event data pre-processing operations include the phase-folding of on-board arrival times according to a NANOGrav radio timing solution; the phase-folded events are then the events that would be registered by a telescope that is static relative to the pulsar.

The NANOGrav and CHIME/Pulsar collaborations jointly performed wideband radio timing of PSR~J0740$+$6620 \citep{Fonseca20,FonsecaZen21}. A product of this radio timing work was a joint posterior PDF of the pulsar gravitational mass, Earth distance, and the inclination Earth subtends to the orbital direction\footnote{The pulsar spin angular momentum is assumed to be parallel to the orbital angular momentum, but we also test sensitivity to an isotropic spin-direction prior.} that we can condition on as an informative prior PDF for joint \NICER and \textit{XMM} X-ray pulse-profile modeling. The synergy---due to degeneracy breaking---between radio timing and X-ray modeling yields a higher sensitivity to the pulsar radius and the parameters of a surface X-ray emission model. We used two sets of joint posterior PDFs provided by \citet{Fonseca20} as prior information over the course of our analysis: one set that was numerically estimated using a weighted least-squares solver, used in this Letter for an exploratory analysis (Section~\ref{sec:exploratory}); and a second that instead used a generalized least-squares (GLS) algorithm for determining timing model parameters from wideband timing data, used in this Letter for a production analysis (Section~\ref{sec:production analysis}). \citet{Fonseca20} report on results obtained with GLS fitting of wideband timing data, though they publicly provide both sets of PDFs as they were used in this work.

Systematic error in the parameter estimates by NANOGrav and CHIME/Pulsar is dominated by sensitivity to the choice of the dispersion-measure variability (DMX) model. Posterior PDFs were computed for three DMX variants, and the systematic error implied by the posterior variation was substantially smaller than the formal posterior spread conditional on any one DMX model. We average (marginalize) the posterior PDFs over the three DMX models with a uniform weighting.\footnote{Equivalent to choosing a prior mass function of the DMX models that yields posterior probability ratios of unity between those models.}

The prior PDFs we implement for every model that conditions on joint NANOGrav and CHIME/Pulsar wideband radio timing are as follows. The distance is separable from the pulsar mass and the Earth inclination to the orbital axis, but the mass and (cosine of the) inclination are correlated. For the distance, we implement the NANOGrav~$\times$~CHIME/Pulsar measurement by first marginalizing the PDFs over the DMX models and then reweighting from the flat distance prior conditioned on by NANOGrav and CHIME/Pulsar to a physical distance prior in the direction of PSR~J0740$+$6620 following the exemplar treatment of distance information by \citet{Igoshev16}.\footnote{The ecliptic coordinates of PSR~J0740$+$6620 reported by \citet{Arzoumanian18} were transformed to galactic coordinates using Astropy \citep[\url{http://www.astropy.org};][]{astropy2013,astropy2018}.} The physical distance prior, however, remains relatively flat in the context of the likelihood function---meaning the likelihood function is dominant in the measurement---and the modification to the original DMX-averaged PDF is entirely minor. Additional distance likelihood information derives from the Shklovskii effect which effectively truncates the PDF, putting an upper limit on the distance \citep{Shklovskii70}; such truncation however occurs well into the tail of the NANOGrav~$\times$~CHIME/Pulsar distance distribution, so it is also an unimportant detail. Finally, we approximate the NANOGrav~$\times$~CHIME/Pulsar marginal PDF of distance as a skewed and truncated Gaussian distribution (please see Table~\ref{table: ST-U} for details). We display the marginal distance PDF variants referenced above in Figure~\ref{fig:marginal distance PDFs}.

For the mass and the (cosine of the) orbital inclination we implement a multi-variate Gaussian distribution with the covariance matrix of the DMX-averaged PDF. Implicit in this PDF is a prior PDF defined by the radio timing collaboration. The marginal prior PDF of the pulsar mass is \textit{not} flat: the prior PDF of the cosine of the orbital inclination and of the mass of the white dwarf companion of PSR~J0740$+$6620 is jointly flat, and these variables map deterministically to the pulsar mass through the binary mass function and (precise) radio timing of the system. We do not modify the joint prior PDF of the pulsar mass and the inclination despite the likelihood function of the pulsar mass (marginalized over all other radio timing parameters) being formally desirable for EOS parameter estimation (see Section~\ref{sec:radius prior design} and \citealt{Riley18}). In this case the posteriors are sufficiently dominated by the likelihood function to ignore the small structural modifications that would result from tweaking relatively diffuse priors. Moreover, changing the pulsar mass prior PDF would change the prior PDF of the companion, which may be undesirable.

Ultimately, handling the detailed structure\footnote{Please see Figure~\ref{fig:marginal distance PDFs} for reference. The form of the DMX-marginalized joint mass and inclination prior PDF is the dominating factor in the posterior PDF rendered in Figure~\ref{fig:ST-U spacetime corner}; for supplementary plots of the variation of the joint prior PDF of mass and inclination with DMX model, please refer to the analysis notebooks released with this Letter, in which the model components are constructed.} of these prior PDFs---i.e., beyond the location and marginal spread of the parameters---is not important because the \NICER likelihood function is not sufficiently sensitive to some combination of these radio-timing parameters and its native parameters (such as equatorial radius) for posterior inferences to change to any discernable degree. That is, the changes will be difficult to resolve from sampling noise and implementation error for instance \citep{higson2018sampling,higson2019diagnostic}, and relative to model-to-model systematic variations that, when marginalized over, further broaden the posteriors of shared parameters of interest. Also note that the distance only enters in the likelihood function in combination with the effective-area scaling parameter defined in Section~\ref{sec:response models} that operates on all X-ray instrument response models because of global calibration error; the distance to PSR~J0740$+$6620 could therefore be combined with this absolute scaling parameter, further diminishing the importance of treating fine details in the prior PDF of the distance.

{
    \begin{figure}[t!]
    \centering
    \includegraphics[clip, trim=0cm 0cm 0cm 0cm, width=\columnwidth]{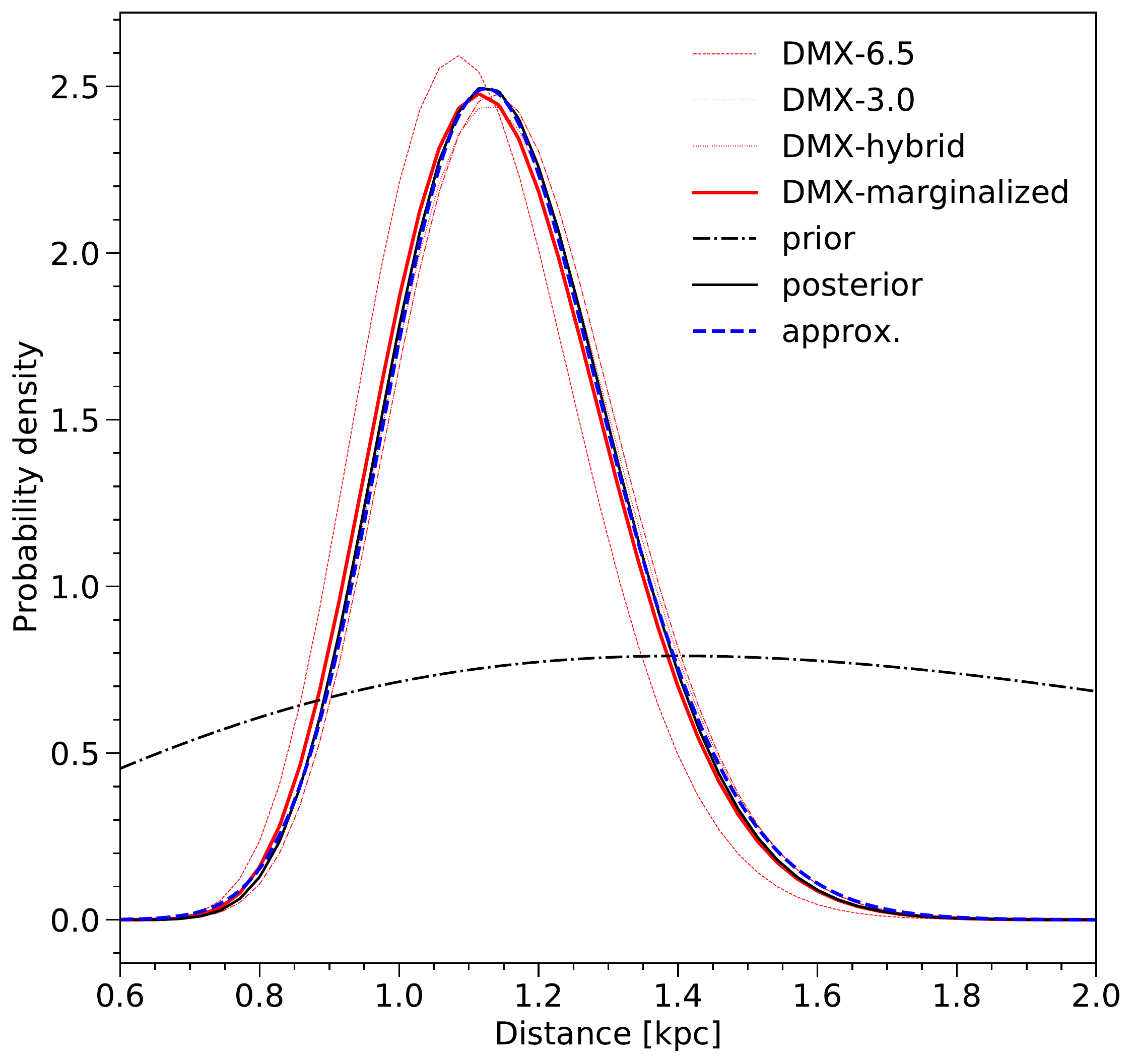}
    \caption{\small{Implementation of the joint NANOGrav and CHIME/Pulsar posterior PDF of distance \citep{Fonseca20} as a prior PDF in the context of joint \NICER and \textit{XMM} X-ray pulse-profile modeling. The \textit{red} distributions are marginal posterior PDFs of the distance derived by NANOGrav and CHIME/Pulsar, conditional on a flat prior PDF; the \textit{solid red} distribution is the posterior PDF marginalized over the three DMX models, and each of the lightweight \textit{red} distributions is a posterior PDF conditional on a particular DMX model. These posterior PDFs of distance are proportional to the marginal likelihood function of distance. The \textit{black dash-dot} PDF is the prior distribution of galactic pulsars in the direction of PSR~J0740$+$6620, adopted from \citet{Igoshev16} and renormalized to the interval $D\in[0.6,2.0]$~kpc on which the NANOGrav~$\times$~CHIME/Pulsar PDF is supported. The \textit{black solid} distribution is the posterior PDF of distance conditional on the \textit{black dash-dot} prior PDF. The \textit{blue dashed} distribution is an approximating PDF that we condition on as a prior PDF for the pulse-profile modeling in this Letter, after renormalizing to be supported on the interval $D\in[0.0,1.7]$~kpc; please refer to Table~\ref{table: ST-U} for details needed to reproduce this approximating PDF.}}
    \label{fig:marginal distance PDFs}
    \end{figure}
}

\subsubsection{Relativistic ray-tracing}\label{sec:ray tracing}

The original version of the \XPSI package used to model \jdbl \citep{riley19a} via relativistic ray-tracing assumed that no more than one ray connected the telescope to each point on the stellar surface. However, photons emitted from a star with compactness greater than $GM/Rc^2 = 0.284$ can have a deflection angle larger than $\pi$, which makes multiple images of small regions at the ``back" of the star possible. This was not an issue for the analysis of \jdbl, because the 95\% credible region only allowed compactness values in the range of $GM/Rc^2 \le 0.171$. Additionally, the independent analysis by \citet{miller19} allowed for the possibility of multiply-imaged regions on the star but still led to a credible range on compactness that is too small to allow for multiple images.

The radio observations of Shapiro delay in the \joh binary system by NANOGrav and CHIME/Pulsar \citep{Fonseca20} lead to the marginal 68\% credible interval of $M=2.08\pm0.07$~\msol. For reference, a small selection of EOS that cover a realistic range of stiffness are shown in Figure~\ref{fig:compactness-vs-mass}. The EOS shown include a set of three EOS (HLPS soft, intermediate, and stiff) constructed by \citet{Hebeler13} that span a range of stiffness allowed by nuclear experiments, as well as the A18$+\delta(v)$+UIX* EOS \citep{Akmal98} (abbreviated to APR). Each of the four curves shows the values of the equatorial compactness and mass for stars rotating at the rate of 346~Hz, computed using the code {\tt rns} \citep{Stergioulas95}. Figure~{\ref{fig:compactness-vs-mass}} shows that for this set of EOS, the large mass values ($>2.0$\msol) lead to a significant range of models that have large enough compactness (i.e., are above the horizontal line at 0.284) that multiple images of some part of the star are possible. Because the stars are oblate, the compactness at the spin poles is larger than the equatorial compactness, so the range of multiply-imaged stars extends to slightly lower values of compactness, however the change is not perceptible on this figure. 

Due to the expectation that \joh could be very compact, the \XPSI code was extended to allow multiple rays from any point to reach the telescope. This improvement to \XPSI as well as details of code validation are discussed in \citet{bogdanov19c}. \XPSI sums over the primary and the visible higher-order images of any regions on a star that lie in a multiply-imaged region. The \XPSI package can typically detect up to the quaternary, quinary, or senary order depending on resolution settings. In practice only secondary images, and potentially the tertiary images for some configurations, might be important; but omission of secondary images can lead to large errors of $\mathcal{O}(10\%)$ in the light-curve calculation (see, e.g., the \XPSI documentation\footnote{\url{https://thomasedwardriley.github.io/xpsi/multiple_imaging.html}}).

{
    \begin{figure}[t!]
    \centering
    \includegraphics[clip, trim=0cm 0cm 0cm 0cm, width=\columnwidth]{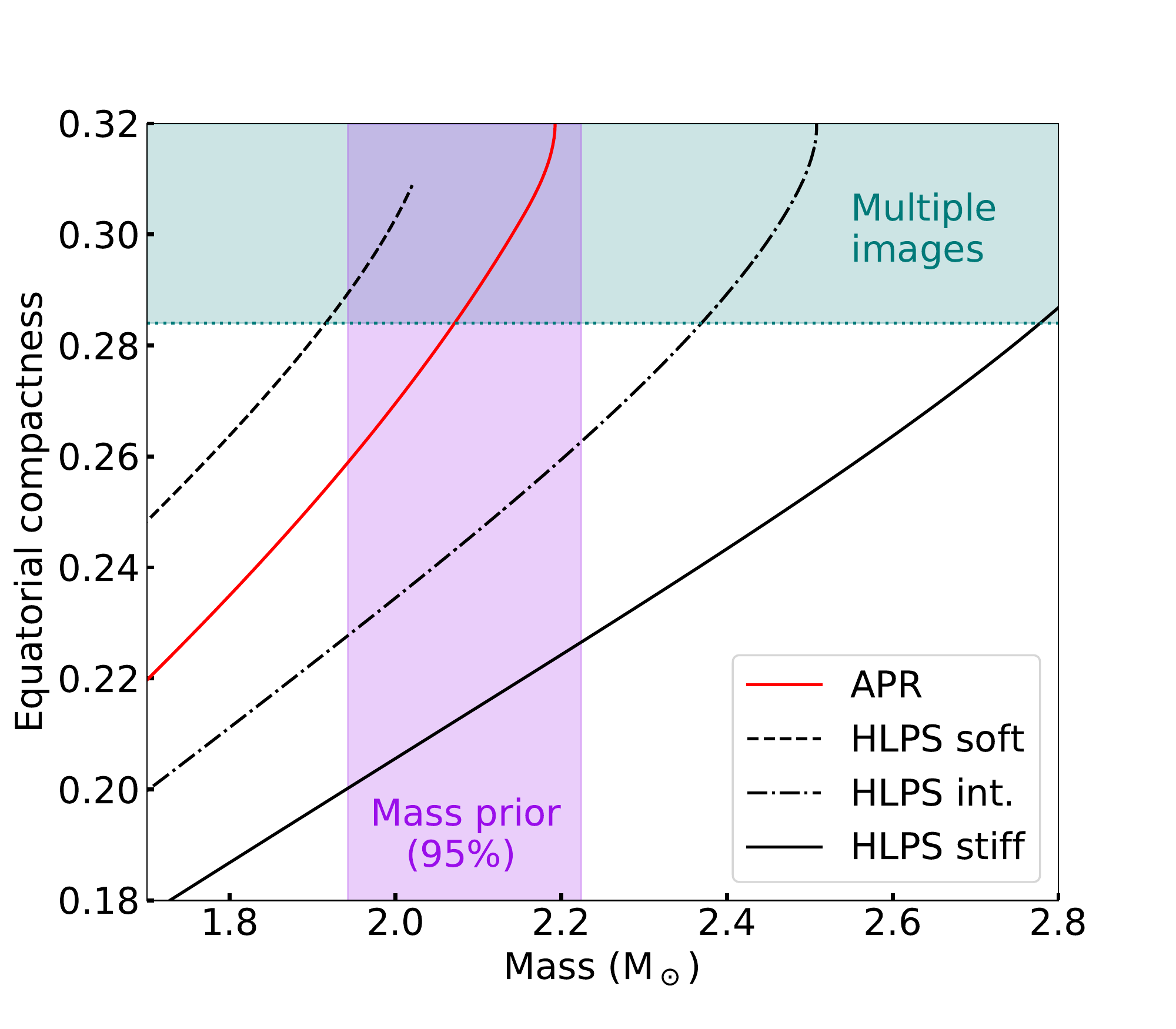}
    \caption{\small{Equatorial compactness ($GM/Rc^2$) versus mass for stars spinning at a rate of 346~Hz. The relations for four example EOS models (see text for description) are shown. The 95\% interval for the mass prior is shaded, as is the region above compactness 0.284 where multiple images of the equator can appear. Note that for a range of compactness below 0.284, non-equatorial surface regions are multiply imaged due to oblateness. For the high masses of interest, multiple imaging is clearly relevant.}}
    \label{fig:compactness-vs-mass}
    \end{figure}
}

\subsubsection{Interstellar attenuation of X-rays}

The X-ray signal is attenuated to some degree by the intervening interstellar medium along the line of sight to the pulsar. In all models, the attenuation physics is parameterized solely by the neutral hydrogen column density $N_{\rm H}$, relative to which the abundances of all other attenuating gaseous elements, dust, and grains are fixed by the state-of-the-art \texttt{TBabs} model \citep[updated in 2016]{Wilms2000}. We implement this attenuation model as a one-dimensional lookup table with respect to energy at a fiducial column density, and then raise the attenuation factor to the power of the ratio of the column density to the fiducial value.

The column density along the line of sight to PSR~J0740$+$6620 can be estimated using several techniques. The HEASARC neutral hydrogen map tool\footnote{\url{https://heasarc.gsfc.nasa.gov/cgi-bin/Tools/w3nh/w3nh.pl}} yields $N_{\rm H}\approx3.5\times10^{20}$~cm$^{-2}$ given the most modern map \citep{HI4PIcollab2016}.\footnote{Neutral hydrogen maps provide an integrated estimate along the line of sight through the Milky Way, and can therefore be interpreted as an approximate upper limit.}.

Using the relation between dispersion measure (DM) and $N_{\rm H}$ from \citet{He2013} together with the DM measurement reported by \citet{Cromartie19}, $N_{\rm H}\approx4.5\times10^{20}$~cm$^{-2}$. Finally, estimates based on 3D $E\left(B-V\right)$ extinction maps \citep{Lallement2018}, together with the relation between extinction and $N_{\rm H}$ from \citet{Foight2016}, yield values of $N_{\rm H}\approx4.5\times10^{20}$~cm$^{-2}$.

Given that $N_{\rm H}$ could be as large as $4.5\times 10^{20}$~cm$^{-2}$ based on these estimates, and given the large uncertainties in the relations between $N_{\rm H}$ and other quantities such as the DM, a conservative prior of $N_{\rm H}\sim U(0,10^{21})$~cm$^{-2}$ is warranted.

\subsection{Surface hot regions}

\subsubsection{Temperature field}\label{sec:temp field}

The surface effective temperature field of a rotation-powered MSP is one of---if not the---primary source of uncertainty \textit{a priori} regarding the physical processes that generate the X-ray event data. The image of a neutron star cannot be resolved with any current X-ray telescope, so all our knowledge about the surface temperature field comes from models. These models heavily rely on the assumptions about magnetic field configuration, which is a crucial part in calculating heating by space-like or return current in some sub-region of the open field line footprints \citep{Kalapotharakos20} or anisotropic thermal conductivity of the outer star layers \citep[e.g.,][]{Kondratyev20, deGrandis20}. There is, therefore, an essentially unknown degree of complexity in the structure of the temperature field. 

A given likelihood function is insensitive to complexity beyond some degree. We focus on a simple model of the surface temperature field: two disjoint hot regions that are simply-connected spherical caps (when projected onto the unit sphere) wherein the effective temperature of the atmosphere is uniform. This model may be identified as \texttt{ST-U} in \citet{riley19a}. The phase-folded \NICER pulse-profile is suggestive of two phase-separated hot regions which may or may not be disjoint.

We could begin with an even simpler model that restricts the hot regions to be related via antipodal reflection symmetry \citep[\texttt{ST-S} in][]{riley19a}. Computing a posterior conditional on such a simple surface hot region model is justifiable, and it also delivers a lower-dimensional target distribution to check for egregious model implementation error \citep[as reasoned by][]{riley19a}. However, for the analysis reported in this Letter, we immediately explored the \texttt{ST-U} model because antipodal reflection symmetry is physically unrealistic and the \texttt{ST-U} model is sufficiently simple and inexpensive to condition on; nevertheless, if one is interested in the question of whether we are sensitive to deviations from this symmetry, we open-source the entire analysis package for this Letter, and the online \XPSI documentation offers guidance on how to condition on the \texttt{ST-S} model.

Nodes can be added to the model space by incrementing the complexity of the surface hot regions, following \citet{riley19a}, wherein symmetries between hot regions are broken\footnote{That is, the breaking of antipodal reflection symmetry and of shape symmetry.} and temperature components are added.\footnote{All models in \citet{riley19a}, two of which we condition on in this Letter, can be labeled by the number of disjoint hot regions they define. Note that in practice the prior PDFs of temperature and solid angle subtended at the stellar center are diffuse and permit the contribution of a hot region to be negligible \textit{a posteriori} in the \NICER and \textit{XMM} wavebands.
} In this Letter we introduce a new binary flag for the atmosphere composition (hydrogen or helium as discussed in Section~\ref{sec:atmosphere setup}) within the (single-temperature) hot regions. However, we consider only one higher-complexity model than \texttt{ST-U}. We opt to do this because subject to limited (computational) resources, \texttt{ST-U} performs well and is ultimately determined to be sufficiently complex.

We now define prior PDFs that are shared by all models. Every hot region has one effective temperature component. The prior PDF of that effective temperature is flat in its logarithm, diffuse,\footnote{For example, with prior support bounds of $10^{5.1}$~K and $10^{6.8}$~K for the NSX fully-ionized hydrogen atmosphere---please see Section~\ref{sec:atmosphere setup}.} and is separable from the joint prior PDF of all other model parameters. Every hot region also has a colatitude coordinate and an azimuth coordinate (i.e., phase shift) at the center of a constituent spherical cap with a finite effective temperature component \citep{riley19a}. The prior PDF of these coordinates is non-trivial because it is not separable from the prior PDF of other parameters when there are two surface hot regions in the model, as we proceed to explain.

In order to eliminate a trivial form of hot region exchange-degeneracy,\footnote{Where precisely the same hot region configuration exists twice in parameter space, only one of which the prior support should include.} we define the prior support by imposing that one hot region object in the \XPSI model has a center colatitude that is always less than or equal to the center colatitude of the companion hot region. We also impose that the hot regions are disjoint, meaning that they cannot overlap in the prior support, because otherwise the model cannot always be characterized by the number of disjoint hot regions; the non-overlapping condition is a function of the center colatitudes, the center azimuths, and the hot region angular radii, meaning that the joint prior is not separable, and the marginal prior PDFs are modulated relative to the PDFs that were used to construct the form of the joint PDF in six dimensions. For instance, the joint prior PDF of the azimuthal coordinates exhibits rarefaction where the azimuths are close in value.

We now construct the prior PDF specifically for the \texttt{ST-U} model (Section~\ref{sec:production analysis}) that is the focus of this Letter. First, as a construction tool, define a flat PDF of the cosine of each hot region center colatitude, with support being the interval $\cos(\Theta)\in[-1,1]$; such a PDF is founded on the argument of isotropy, \textit{a priori}, of the direction to Earth (see below). Second, define the PDF of each hot region center azimuth\footnote{A periodic parameter, also referred to as cyclic or wrapped.} to be flat on the interval $2\pi\phi\in[0,2\pi]$~radians. Third, define a flat PDF of each angular radius on the interval $[0,\pi/2]$. To calculate the marginal prior PDF of one of these parameters, marginalize over the other five parameters subject to the prior support condition of non-overlapping hot regions. For instance, the prior PDF of a hot region angular radius is given by marginalizing over the angular radius of the companion hot region, and the hot region center colatitudes and azimuths. Marginalization yields a marginal prior PDF that deviates in form from the initially flat PDF of the parameter---the support of the PDF remains unchanged however.
 
Our hot region models are phenomenological and to a degree, agnostic, despite being influenced by temperature fields implied by pulsar magnetospheric simulations. Our choice here is contrary to \citet{riley19a}, wherein the prior PDF of the colatitude at the center of a hot region in isolation---meaning one hot region on the surface---was flat. A flat PDF of colatitude, combined with a flat prior in azimuth yields an anisotropic probability density field on the unit sphere, weighted towards polar regions. One reason this might be a sub-optimal assumption is because of X-ray selection effects: pulsed emission from two hot regions will exhibit a larger amplitude if the hot regions are closer to the equatorial zone than a polar zone, for equatorial observers as is the case assumed for PSR~J0740$+$6620 based on the NANOGrav~$\times$~CHIME/Pulsar wideband radio timing measurements (Section~\ref{sec:radio}). However, such arguments are tenuous, taking no heed of radio selection effects and magnetospheric physics, for instance.

\subsubsection{Atmosphere}\label{sec:atmosphere setup}

The specific intensity is determined from lookup tables generated using the \texttt{NSX} atmosphere code assuming either fully ionized hydrogen or helium \citep{Ho01} or partially ionized hydrogen \citep{Ho09}.\footnote{Note that only the composition within the hot regions is explicitly defined for the purpose of signal generation.} In this work, we consider only fully ionized models since the limited parameter ranges of existing opacity tables \citep{OPAL,OP,OPLIB} reduce the accuracy of atmosphere models employing these tables (see \citealt{bogdanov19c} and \citealt{IMJ0740} for further discussion and comparisons).

\subsubsection{Exterior of the hot regions}\label{sec:exterior of the hot regions}

The surface exterior to the hot regions does not explicitly radiate in any model we condition on. The signal that would be generated by a cooler exterior surface can be robustly subsumed in the phase-invariant count rate terms described in Section~\ref{sec:XTI backgrounds}, provided that the angular scale of the hot regions (whose images are explicitly integrated over) is small and that exterior emission is soft and thus dominated by other \NICER backgrounds in the channels with low nominal photon energies. Explicitly including radiation from the exterior surface would add a very weakly informative mode of dependence on the mass, radius, and other parameters of interest, and thus the likelihood function is assumed insensitive to its exact treatment. Explicit treatment would also require handling of the ionization state of the cooler atmosphere.

\subsection{Instrument response models}\label{sec:response models}

For \NICER and each of three \textit{XMM} EPIC cameras we use a tailored ancillary response file (ARF) and redistribution matrix file (RMF) to compose an on-axis response matrix. For each of the \textit{XMM} cameras, the ARF and RMF are those specific to the PSR~J0740$+$6620 extraction regions. 

For \NICER, we use the most recent calibration products made available by the instrument team, namely the nixtiref20170601v002 RMF and the nixtiaveonaxis20170601v004 ARF. For the latter, the effective areas per energy channel were rescaled by a factor of $51/52$ to account for the removal of all events from DetID 34. The instrument response products (RMF and ARF files) for the \textit{XMM} pn, MOS1, and MOS2 cameras tailored to the PSR~J0740$+$6620 observations were produced with the \texttt{rmfgen} and \texttt{arfgen} commands in SAS and products from the Current Calibration Files repository.

Calibration of the performance of \NICER is conducted mainly with observations of the Crab pulsar and nebula. The energy-dependent residuals in the fits to the Crab spectrum are generally $\lesssim2\%$\footnote{See \url{https://heasarc.gsfc.nasa.gov/docs/heasarc/caldb/nicer/docs/xti/NICER-xti20200722-Release-Notesb.pdf} for further details.}. The calibration accuracy for the \textit{XMM} MOS and pn instruments is reported to be less than 3\% and 2\% (at 1$\sigma$), respectively \footnote{See in particular Table 1 in \url{https://xmmweb.esac.esa.int/docs/documents/CAL-TN-0018.pdf}.}
However, in the absence of a suitable absolute calibration source, the absolute energy-independent effective area of \NICER and the three \textit{XMM} detectors is uncertain at an estimated level of $\pm10\%$.

We define a free parameter that operates as an energy-independent scaling factor shared by all instruments due to the lack of a perfect astrophysical calibration source. Further, for each of \NICER and \textit{XMM}, we define a free parameter that also operates as an energy-independent effective area scaling. The \textit{overall} scaling factor for each telescope is a coefficient of the response matrix, defined as the product of the shared scaling factor and the \textit{a priori} statistically independent telescope-specific scaling factor. The overall scaling factors of different instruments are therefore correlated \textit{a priori} to simulate---in an albeit simple way---the absolute uncertainty of X-ray flux calibration and the instrument-to-instrument calibration product errors. However, the correlated prior PDF is such that the effective area for \NICER is permitted to decrease from the nominal effective area whilst the effective area for \textit{XMM} can increase from its respective nominal effective area, and vice versa. We choose the statistically independent telescope-specific scaling factors to have equal spread \textit{a priori} to the scaling factor shared by all instruments of $\pm10\%$, therefore yielding a more conservative joint prior PDF of the overall scaling factors that we approximate as a bivariate Gaussian (see Table~\ref{table: ST-U}). We discuss posterior sensitivity to effective area prior information in Section~\ref{sec:XMM constraining power}.

The response models are implicitly assumed to accurately represent the time-averaged operation of the instruments during the (composite) exposures to PSR~J0740$+$6620.

\subsection{Likelihood function and backgrounds}

In this section we formulate the likelihood function as the conditional probability of the \NICER and \textit{XMM} event data sets.\footnote{The domain of the likelihood function is a subset of the model parameter space (which is in general discrete-continuous mixed). The domain of the more general parameterized sampling distribution is a subset of the Cartesian product of the model parameter space and the data space; the likelihood function is the function that this sampling distribution reduces to when the data-space variables become fixed by observation.} The relative constraining power offered by the \textit{XMM} likelihood function is weak, but we provide detail about the methodology with the overarching aim of supporting the use of imaging observations in ongoing and future \NICER pulse-profile modeling efforts.

The expected photon specific flux signal generated by the surface hot regions is calculated as a function of rotational phase (over a single rotational cycle) and energy, by integrating over the photon specific intensity image on the sky of a distant static instrument \citep{bogdanov19b}. We then assume that an adequately performing model of both the \NICER and \textit{XMM} event data sets can be constructed by operating on the same incident signal. The count number statistics for both \NICER and all \textit{XMM} cameras is Poissonian: for every detector channel of the four instruments---and then for \NICER every phase interval associated with a channel---the sampling distribution from which the registered count-number variate is drawn is assumed to be a Poisson distribution with an expectation that is a function of parameters of the incident signal and parameters of the model instrument response. 

The \NICER event time-tagging resolution is state-of-the-art, and PSR~J0740$+$6620 is sufficiently faint that in the absence of backgrounds, the event arrival process statistics would deviate in an entirely minor way from the incident photon Poisson point process; in reality we contend with background radiation, and deadtime corrections to the integrated exposure time are calculated during event data pre-processing. Pile-up, the registering of multiple photons as a single event during a detector readout interval, is not an issue for \textit{XMM} because the source is so faint.

As we expound upon in Sections~\ref{sec:XTI backgrounds}~and~\ref{sec:XMM background}, the background event data for all four instruments---after implementation of filters during data pre-processing \citep{Wolff20}---is modeled with a set of count rate variables, one per detector channel. As a corollary of the assumptions stated above, the background processes contributing to these events are also assumed to be Poissonian event arrival processes. The expectations of the sampling distributions of the registered count-number variates are simply sums of the expected count numbers from the PSR~J0740$+$6620 surface hot regions and the expected background count numbers. Refer to \citet{riley19a} for details about the \NICER count-number sampling distribution---the form of the \textit{XMM} camera count-number sampling distribution follows by a phase-averaging operation.

\subsubsection{NICER}\label{sec:XTI backgrounds}

Let $d_{\rm N}$ be the \NICER count matrix over phase interval--channel pairs. We now define variables that the \NICER likelihood is a function of: let $s$ be a vector of parameters, collected from Section~\ref{sec:radiation propagation} above, of which the incident signal generated by the pulsar is a function; let \textsc{nicer} denote a (parameterized) model for the response of the instrument in response to incident radiation; and let $\{\mathbb{E}[b_{\rm N}]\}$ be a set of statistically independent nuisance variables, one per detector channel that contains event data to be modeled, defined as the phase-invariant expected count rate. The \NICER event data is phase-resolved, so there are multiple random variates---distributed in phase---per background random variable. The background model with variables  $\{\mathbb{E}[b_{\rm N}]\}$ is free-form as discussed by \citet{riley19a}. The set of these variables is designed to handle complex channel-to-channel variations in the expected phase-invariant count rate. A physical background model should need (far) fewer random variables for the underlying background-generating process to capture these complexities. The likelihood function is denoted by $p(d_{\rm N} \,|\, s, \{\mathbb{E}[b_{\rm N}]\}, \textsc{nicer})$.

We numerically marginalize over the variables $\{\mathbb{E}[b_{\rm N}]\}$ to yield a marginal likelihood function that is combined with a joint prior PDF to define a target distribution to draw samples from. The count rate variables have separable flat prior PDFs that are strictly improper because we do not explicitly define upper-bounds on the prior support of each variable \citep{riley19a}; the posterior is considered integrable however, so these ill-defined prior PDFs do not result in posterior pathologies. The separable prior PDF of the count rate variables is overly-diffuse, with extremely high prior-predictive complexity, such that inferences will be insensitive to minor changes to the function: $\mathbb{E}[b_{\rm N}] \sim U(0,\mathcal{U})$,
where the upper-bound $\mathcal{U}$ of the prior support is left unspecified.\footnote{The posterior should be integrable---without proof here---even if the prior is improper, but if an upper-bound were to be required, it could for instance be based on \NICER count rate limitations.}

The marginalization operation is separable over channels, yielding a product of one-dimensional integrals. We need to perform fast numerical marginalization over the $\{\mathbb{E}[b_{\rm N}]\}$ in order to compress the dimensionality of the sampling space. The simpler the form of the integrands---the functions of the $\{\mathbb{E}[b_{\rm N}]\}$---the more straightforward fast numerical marginalization is. If the prior PDF $p(\mathbb{E}[b_{\rm N}])$ is flat, the integrand has a single global maximum, and would be highly Gaussian if the peak of the conditional likelihood function $p(d_{\rm N} \,|\, s, \mathbb{E}[b_{\rm N}], \textsc{nicer})$ lies within the support of $p(\mathbb{E}[b_{\rm N}])$.

As discussed by \citet{riley19a}, a major open question regards the total expected spectral signal that is attributable to the surface hot regions in reality.\footnote{Assuming that the pulsed emission is dominated by rotationally modulated emission from the surface.} In lieu of a physical background model---which as discussed above is difficult to formulate for \textit{NICER}---independent information conditional on data acquired with an imaging X-ray telescope such as \textit{XMM} can be fundamentally valuable for deriving robust inferences if the exposure time is sufficiently long---it will however be substantially shorter than the order megasecond exposures of the near-dedicated \NICER telescope.

Note that in order to infer the contribution from contaminating sources to the \NICER event data,\footnote{That is, contamination that cannot be robustly filtered out during event data pre-processing.} we would need to separate out the $\{\mathbb{E}[b_{\rm N}]\}$ into an environmental background model---with some informative prior PDF including space weather contributions---and some model for the contribution from contaminating (point) sources in the field that cannot be resolved from the point-spread function (PSF) of PSR~J0740$+$6620. However, for the principal purpose of constraining the physical properties of the pulsar, we are uninterested in the distinction between \NICER backgrounds. Moreover, we do not have the statistical power to distinguish the contributions if the priors for the components are all rather diffuse. In other words, if some of the priors are informative, we do not have the statistical power to gain much information \textit{a posteriori}.

\subsubsection{\XMM}\label{sec:XMM background}

Information about the total background contribution to the \NICER event data---including contaminating (point) sources in the field \citep[see][for a breakdown of components]{bogdanov19a}---is encoded in the \textit{XMM} spectroscopic imaging event data. Due to the relatively brief exposure times of the observations, very few background counts are available for a reliable background estimate. Thus, we obtained representative background estimates with higher photon statistics from the blank-sky event files provided by the \XMM Science Operations Centre\footnote{See \url{https://www.cosmos.esa.int/web/xmm-newton/blank-sky}.}. The blank-sky images were filtered in the same manner as the PSR~J0740$+$6620 field images and the background was extracted from the same location on the detector as the pulsar. The resulting background spectrum was then rescaled so that the exposure times and BACKSCAL factors match those of the PSR~J0740$+$6620 exposures.

Leveraging spatial resolving power, we can derive posterior inferences about the signal from the pulsar in isolation with little confusion about the expected contributions from the pulsar and background sources.\footnote{There remains confusion, however, about what contribution from the pulsar and its near vicinity is generated by surface hot regions.} When this information about the pulsar signal is injected into modeling of \NICER observations---either explicitly as a prior or as a likelihood factor---the contribution from the pulsar to the \NICER event data is informed; here we work towards a likelihood function factor for the \textit{XMM} telescope.

Let $d_{\rm X}$ be the \textit{XMM} time-integrated count vector (over detector channels) from a region $\mathcal{S}$ of a CCD of a camera, that is some sufficiently large subset of the support of the PSF of the PSR~J0740$+$6620 whilst optimizing signal-to-noise.\footnote{Imaging observations are configured such that the target point source is confined to a single CCD, for instance to avoid masking part of the source PSF with gaps between CCDs.} We now define variables that the \textit{XMM} likelihood is a function of. The vector of parameters of the incident signal generated by the pulsar remains as $s$, shared with the \NICER likelihood function. Let \textsc{xmm} denote a (parameterized) model for the response of an \textit{XMM} camera in response to incident radiation. Once more, in lieu of a physical model for the underlying background-generating process, let us define one statistically independent random nuisance variable per detector channel: the expected count rate.\footnote{Where there is now one variate per channel because the count numbers are phase-averaged.} These variables are collectively denoted $\{\mathbb{E}[b_{\rm X}]\}$. The likelihood function for each \textit{XMM} camera is denoted by $p(d_{\rm X} \,|\, s, \{\mathbb{E}[b_{\rm X}]\}, \textsc{xmm})$. We numerically marginalize over these model variables in the same vein as for the \NICER background-marginalized likelihood function; however, to constrain the signal from PSR~J0740$+$6620 we are in need of a more informative prior PDF of $\{\mathbb{E}[b_{\rm X}]\}$.

To form the prior PDF of $\{\mathbb{E}[b_{\rm X}]\}$ for an \textit{XMM} camera, we consider a set of independent Poisson-random  variates $\{\mathscr{B}_{\rm X}\}$ defined as time-integrated astrophysical (sky) background count numbers in detector channels. The events constituting $\{\mathscr{B}_{\rm X}\}$ are extracted from a blank-sky\footnote{That is, a region of sky devoid of any (bright) non-diffuse X-ray sources.} region $\mathcal{B}$ of the camera CCD array that is disjoint from region $\mathcal{S}$ associated with PSR~J0740$+$6620. The \textit{XMM} cameras are composed of multiple side-by-side CCD detectors. For each camera, it is preferable to choose the region $\mathcal{B}$ on the same detector as $\mathcal{S}$ because the different CCD detectors have slightly different responses to incident radiation and therefore exhibit slightly different astrophysical (sky) backgrounds. For blank-sky exposures, the conditional sampling distribution in the space of the data is $p(\{\mathscr{B}_{\rm X}\} \,|\, \{\mathbb{E}[\mathscr{B}_{\rm X}]\})$, where the variables $\{\mathbb{E}[\mathscr{B}_{\rm X}]\}$ are per-channel expected count \textit{numbers}.

To constrain the background in the \textit{XMM} images of PSR~J0740$+$6620, we use blank-sky estimates generated using the \XMM SAS tools. The blank-sky exposures are much longer in duration than the exposures to PSR~J0740$+$6620 by almost two orders of magnitude, allowing us to constrain the astrophysical (sky) background count rate in the on-source exposures more tightly (in the absence of systematic error) than possible from blank-sky extraction regions from the same CCD during the shorter on-source exposure. For each XMM camera the respective region $\mathcal{B}$ is not only localized to the same CCD as the PSR~J0740$+$6620 PSF, but \textit{is} the same region of the CCD. However, the blank-sky estimates are based on an ensemble of pointings over the sky. Our cross-check of the sky background in these reference exposures against the sky background in the vicinity of PSR~J0740$+$6620 did not yield evidence of systematic difference.

We now formally derive the constraints on the expected number of background counts per detector channel registered within the source region $\mathcal{S}$ of an XMM camera, $\{\mathscr{B}_{\rm X}\}$, conditional on the counts registered in region $\mathcal{B}$ over the ensemble of blank-sky exposures. A prior PDF of the variables $\{\mathbb{E}[\mathscr{B}_{\rm X}]\}$ is needed. In the same vein as for the \NICER background variables, we define a separable prior PDF
\begin{equation}
    \mathbb{E}[\mathscr{B}_{\rm X}] \sim U(0,\mathscr{V}),
    \label{eqn:blank-sky prior}
\end{equation}
where the upper-bound $\mathscr{V}$ of the prior support is left unspecified. This trivial model over-fits the background data, with posterior PDF
\begin{equation}
p(\{\mathbb{E}[\mathscr{B}_{\rm X}]\} \,|\, \{\mathscr{B}_{\rm X}\})
\propto
p(\{\mathscr{B}_{\rm X}\} \,|\, \{\mathbb{E}[\mathscr{B}_{\rm X}]\});
\end{equation}
the vector of random variates $\{\mathscr{B}_{\rm X}\}$ is coincident in value with the maximum \textit{a posteriori} vector in the space of 
$\{\mathbb{E}[\mathscr{B}_{\rm X}]\}$.

{
    \begin{figure}[t!]
    \centering
    \includegraphics[clip, trim=0cm 0cm 0cm 0cm, width=\columnwidth]{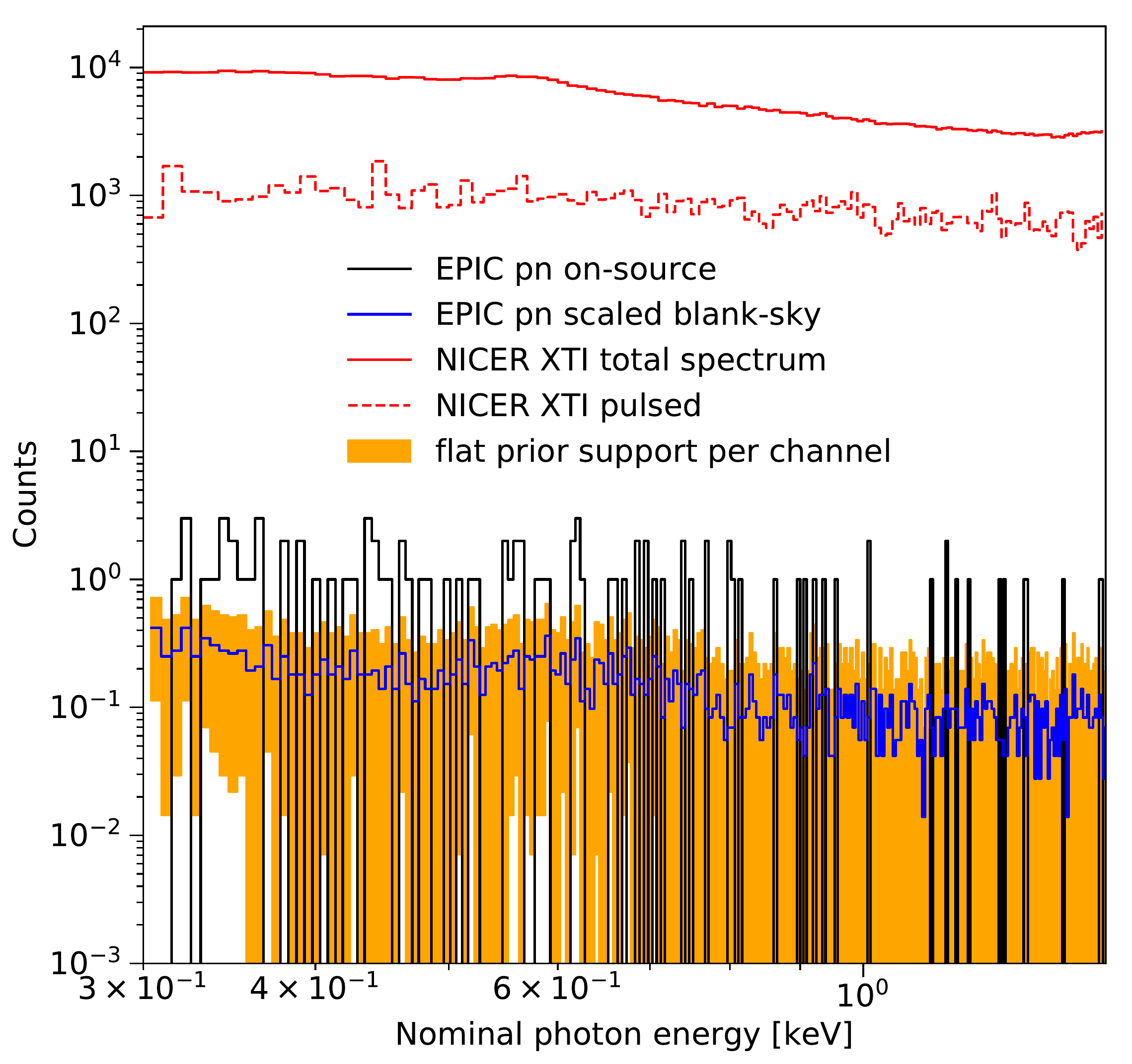}
    \caption{\small{The \textit{black} step function is the \textit{XMM} pn camera \joh count number spectrum as a function of detector channel nominal photon energy. The \textit{XMM} events contain source events and events from diffuse background that can be estimated from blank-sky information. The \NICER count number spectrum is the \textit{red} solid step function, including \joh events and all backgrounds; the \textit{red} dashed step function is the empirical pulsed count number in each channel. The \textit{XMM} pn events are clearly sparse in comparison to the \NICER event data, but note that the number of pulsed \NICER counts is lower than indicated here, as shown in Figure~\ref{fig:J0740 count data}. The \textit{blue} step function is the count number spectrum derived from blank-sky exposure, scaled down to the exposure time on PSR~J0740$+$6620 and also scaled for the CCD extraction region area ratio $A_{\mathcal{S}}/A_{\mathcal{B}}$ as described in the main text---please see Equation~(\ref{eqn:expected count-rate variable transformation}). The \textit{orange} shaded region is the support of the joint prior PDF of the expected count rate variables $\{\mathbb{E}[b_{\rm X}]\}$, derived from the blank-sky exposures. The figure elements rendered here are representative of the corresponding information from the \textit{XMM} MOS1 and MOS2 cameras, as can be seen in the online figure set associated with Figure~\ref{fig:XMM background}.}}
    \label{fig:pn count spectrum}
    \end{figure}
}

As is the case for the \NICER background marginalization operation described in Section~\ref{sec:XTI backgrounds}, if the PDF $p(\{\mathbb{E}[\mathscr{B}_{\rm X}]\} \,|\, \{\mathscr{B}_{\rm X}\})$ is flat, the marginalization is more straightforward to execute. For example, we could form a flat prior PDF spanning the highest-density $x\%$ posterior credible interval in $\mathbb{E}[b_{\rm N}]$, given $\{\mathscr{B}_{\rm X}\}$. We form a flat prior PDF in a simpler way. The PDF $p(\{\mathbb{E}[\mathscr{B}_{\rm X}]\} \,|\, \{\mathscr{B}_{\rm X}\})$ has the approximate structure of a truncated Gaussian with deviation dependent on the number of counts $\mathscr{B}_{\rm X}$. We let the lower- and upper-bounds of the support respectively be $\mathscr{L}\coloneqq\textrm{max}\left(0,\mathscr{B}_{\rm X}-n\sqrt{\mathscr{B}_{\rm X}}\right)$ and $\mathscr{U}\coloneqq\mathscr{B}_{\rm X}+n\sqrt{\mathscr{B}_{\rm X}}$, where $n$ is a setting that controls the degree of conservatism. For some channels the number of counts is low and the PDF $p(\{\mathbb{E}[\mathscr{B}_{\rm X}]\} \,|\, \{\mathscr{B}_{\rm X}\})$ deviates substantially from being a truncated Gaussian, but we nevertheless adopt the same procedure---the lower-bound is pushed to zero or far into the lower-tail of the distribution, but the upper-bound remains sensible. For some channels, however, $\mathscr{B}_{\rm X}=0$. In these cases, we simply set $\mathscr{L}\coloneqq0$ and then iterate upwards in channel number to locate the next finite value of $\mathscr{B}_{\rm X}+n\sqrt{\mathscr{B}_{\rm X}}$.\footnote{This procedure is sufficient for the channel cuts we make because there is always a higher channel with a finite number of counts $\mathscr{B}_{\rm X}$.} We choose $n=4$. With the flat PDF of $\mathbb{E}[\mathscr{B}_{\rm X}]$ defined, we transform variables to derive the prior PDF $p(\mathbb{E}[b_{\rm X}] \,|\, \{\mathscr{B}_{\rm X}\})$ according to
\begin{equation}
\mathbb{E}[b_{\rm X}]
=
\frac{A_{\mathcal{S}}}{A_{\mathcal{B}}}
\frac{\mathbb{E}[\mathscr{B}_{\rm X}]}{T_{\mathscr{B}}},
\label{eqn:expected count-rate variable transformation}
\end{equation}
where $T_{\mathscr{B}}$ is the blank-sky exposure time needed to transform to a count rate, and $A_{\mathcal{S}}/A_{\mathcal{B}}$ is the ratio of the areas of the extraction regions $\mathcal{S}$ (encompassing the PSR~J0740$+$6620 PSF) and region $\mathcal{B}$.

In many respects this flat PDF of $\mathbb{E}[\mathscr{B}_{\rm X}]$ is a remarkably conservative choice for the prior constraint on $p(\mathbb{E}[b_{\rm X}] \,|\, \{\mathscr{B}_{\rm X}\})$. A reason it is not conservative is the assumption of zero prior mass above an upper count-rate limit $\mathscr{U}$ somewhere in the upper-tail of the true probability PDF $p(\{\mathbb{E}[\mathscr{B}_{\rm X}]\} \,|\, \{\mathscr{B}_{\rm X}\})$, especially because the count-number data $d_{\rm X}$ for each \textit{XMM} camera are moderately consistent with being generated by background processes. The upper-limit $\mathscr{U}$ can be decreased so that $p(\mathbb{E}[b_{\rm X}] \,|\, \{\mathscr{B}_{\rm X}\})$ is more informative at the risk of bias. On the other hand, $\mathscr{U}$ can be increased, which naturally weakens the constraining power but can be justified as a safety precaution to capture a contribution such as a power-law component originating from the magnetosphere of PSR~J0740$+$6620 that is not explicitly modeled.\footnote{Note that the binary companion of PSR~J0740$+$6620 has been inferred to be an ultra-cool white dwarf \citep{Beronya19} whose thermal surface emission would not contribute X-ray events, unless there is some interaction with higher-energy winds from PSR~J0740$+$6620. } Alternatively, we could justify a higher upper-limit in terms of systematic error due to blank-sky estimates derived from an ensemble of exposures over the sky and variation in time of the \textit{XMM} camera response matrices between the PSR~J0740$+$6620 exposure and blank-sky exposure ensemble. The setting of $n=4$ seems like a reasonable---albeit arbitrary---choice to balance information loss versus bias; we probe posterior sensitivity to this hyperparameter and conclude our inferences are insensitive to its value (see Section~\ref{sec:inferences}). We display the \textit{XMM} pn camera count number spectrum in Figure~\ref{fig:pn count spectrum} together with the blank-sky derived prior PDF of the expected count rate variables $\{\mathbb{E}[\mathscr{B}_{\rm X}]\}$; we also display the \NICER count number spectrum for data quality comparison.

\subsubsection{The likelihood function}

The likelihood function given the \NICER and \textit{XMM} data sets may be written as
\begin{widetext}
\begin{equation}
    p(d_{\rm N}, d_{\rm X}, \{\mathscr{B}_{\rm X}\} \,|\, s, \{\mathbb{E}[b_{\rm N}]\},
    \{\mathbb{E}[b_{\rm X}]\})
    =
    p(d_{\rm N} \,|\, s, \{\mathbb{E}[b_{\rm N}]\}, \textsc{nicer})
    p(d_{\rm X} \,|\, s, \{\mathbb{E}[b_{\rm X}]\}, \textsc{xmm})
    p(\{\mathscr{B}_{\rm X}\} \,|\, \{\mathbb{E}[b_{\rm X}]\}),
\end{equation}
\end{widetext}
where we combine the expected background count rate variables over the \textit{XMM} cameras into $\{\mathbb{E}[b_{\rm X}]\}$, we combine the count numbers in the PSR~J0740$+$6620 exposures into $d_{\rm X}$, and we combine the blank-sky count numbers into $\{\mathscr{B}_{\rm X}\}$. Introducing the flat prior densities from Equation~(\ref{eqn:blank-sky prior}), approximating the posterior PDF $p(\{\mathbb{E}[\mathscr{B}_{\rm X}]\} \,|\, \{\mathscr{B}_{\rm X}\})$ as flat and bounded as described in Section~\ref{sec:XMM background}, and marginalizing over all expected background count rate variables yields the background-marginalized likelihood function
\begin{widetext}
\begin{equation}
    p(d_{\rm N}, d_{\rm X}, \{\mathscr{B}_{\rm X}\} \,|\, s)
    \propto
    \mathop{\int}_{\{0\}}^{\{\mathcal{U}\}}
    p(d_{\rm N} \,|\, s, \{\mathbb{E}[b_{\rm N}]\}, \textsc{nicer})
    d\{\mathbb{E}[b_{\rm N}]\}
    \mathop{\int}_{\{\mathscr{L}\}}^{\{\mathscr{U}\}}
    p(d_{\rm X} \,|\, s, \{\mathbb{E}[b_{\rm X}]\}, \textsc{xmm})d\{\mathbb{E}[\mathscr{B}_{\rm X}]\}.
    \label{eqn:background-marginalized likelihood}
\end{equation}
\end{widetext}
The background-marginalized likelihood function is fed as a callback to a sampling process, together with a joint prior PDF callback for the pulsar signal parameters $s$ and parameters associated with the \textsc{nicer} and \textsc{xmm} instrument response models.

\subsection{Model space summary}

All nodes in the model space share some underlying physics. Namely, the machinery for relativistic ray-tracing: an oblate surface is embedded in an ambient Schwarzschild spacetime, and the X-ray emission emergent from the atmosphere is attenuated by the interstellar medium as it is transported to a distant static telescope. The nodes in the model space differ first and foremost in their surface hot region parameterization complexities and atmosphere composition flag values, but also in terms of the prior PDF and the likelihood function factors. The prior PDFs for the parameters controlling the shared processes (i.e., mass, equatorial radius, viewing angle to the spin axis, distance, column density) are either informative---such as the joint NANOGrav and CHIME/Pulsar measurement of mass, distance, and viewing angle---or diffuse if there is limited prior knowledge or we aim to probe the consistency of likelihood function factors across telescopes.

\subsection{Posterior computation}\label{sec:posterior computation}
We implement nested sampling to compute the posterior distribution conditional on each model. Namely, we use \XPSI to construct the likelihood function and the prior PDFs and then couple them to \MultiNest \citep{MultiNest_2008,MultiNest_2009, PyMultiNest}. For the headline model we report in this Letter, including \NICER and \textit{XMM} likelihood factors, the dimensionality of the sampling space is $15$. Details about our nested sampling protocol are given in \citet[][see the appendix matter in particular]{riley19a} and also in \citet[][see chapter 3 and the associated appendix]{riley_thesis}. In summary, our minimum resolution settings are as follows: $10^{3}$ live points; a bounding hypervolume expansion factor of $0.1^{-1}$; and an estimated remaining log-evidence of $10^{-1}$. Regarding live points, most posteriors for sensitivity analyses are computed with $2\times10^{3}$ or $4\times10^{3}$ live points, and production calculations used $4\times10^{4}$ live points. The number of live points is the most fundamental parameter that should be changed to probe sampling resolution sensitivity---the expansion factor can be fixed at some value similar to those recommended in the literature \citep{MultiNest_2009}. Likelihood function evaluation time is the dominant sink, being several seconds per core for the processor speeds typical on a cluster or supercomputer. We do \textit{not} use the constant efficiency algorithm variant for any sampling process, and we do \textit{not} use the mode-separation algorithm variant unless stated otherwise.\footnote{For additional details about these variants, refer to \citet{riley19a} and references therein. We also do \textit{not} use the importance sampling algorithm variant \citep{MultiNest_2013}.} Regarding the mode-separation variant, if there are multiple modes of commensurate posterior mass, sampling resolution gets distributed between those modes. It follows that the bounding approximation to the constant likelihood hypersurfaces in each mode is lower than if the global resolution settings were consumed solely by that mode. We eliminate hot region exchange degeneracy from the prior support as discussed in Section~\ref{sec:temp field}, which eliminates mirrored modes and thus improves the resolution of that mode in terms of bounding approximation error.

For most posteriors reported in this work, the nested samples are considered high-resolution in the context of literature recommendations and, for the production analysis, were costly to generate given resource limitations. It is important to remark that our posterior computation procedure has not been validated in any meaningful way via simulation-based calibration because at present it is basically intractable for any one group to calibrate credible region coverage on a model-by-model basis (see the discussion in chapter 3 of \citealt{riley_thesis} and in \citealt{riley19a}). For discussion on the level of calibration we have attained by cross-checking against independent calculations performed by another group \citep{miller19}, we refer to \citet{bogdanov19c}. However, we open-source the entire analysis package for this Letter, so another group with resources is free to modify, cross-check, and improve upon the posterior computation.

\section{Inferences}\label{sec:inferences}

In this section we report our inferences. We first summarize the measures used to assess model performance. We then discuss an exploratory analysis that examined sensitivity to resolution settings and selected model assumptions, in particular the effects of different assumed atmospheric composition and hot region configuration. We then report high-resolution posterior inferences for the superior model.

\subsection{Performance measures}\label{sec:performance measures}

For pulse-profile modeling with \XPSI a set of performance measures should be considered for each model in the model space, largely following the protocol of \citet{riley19a}. The first measure, given a set of posterior samples, is graphical and the most practical: basic posterior-checking by inspecting for inconsistency between the statistically independent \NICER count number variates and the separable sampling distribution from which those variates are assumed to be drawn \textit{a posteriori}. We estimate the expectation with respect to the posterior of the expected count numbers and form Poisson sampling distributions from these expected count numbers. If there is clear structural difference---namely residual correlations in the joint space of energy and phase---then the model cannot generate data with the structure of the real count numbers. Supposing there are no discernable correlations, then because the sampling distribution for each variate is Poissonian with a sufficiently large expectation for the distribution to be well-approximated as Gaussian, we can inspect the distribution of the standardized residuals to identify any clear deviation from a normal distribution---e.g., too much or too little weight in the tails---that would be indicative of noise-model inaccuracies. If this check also passes, then the model has sufficient complexity to generate synthetic data with the structure of the \NICER PSR~J0740$+$6620 event data and is adequate for simulation purposes---e.g., for statistical forecasts of the constraining power achievable with future X-ray space telescope concepts such as the \textit{enhanced X-ray Timing and Polarimetry mission} \citep[\textit{eXTP};][]{Zhang19,Watts19} and the \textit{Spectroscopic Time-Resolving Observatory for Broadband Energy X-rays} \citep[\textit{STROBE-X};][]{Ray19}.

The second measure we inspect is the maximum likelihood estimate reported by a nested sampling process. Note that nested sampling does not target maximum likelihood estimation, but the drawing of samples from the typical set of a target distribution---the maximum likelihood estimate is therefore also subject to the concentration of prior mass in parameter space. More specifically, we are working with the estimated maximum of a background-marginalized likelihood function given by Equation~(\ref{eqn:background-marginalized likelihood}) or one of the likelihood factors (i.e., the \NICER or \textit{XMM} likelihood function). We can use these point estimates to compare, in a simple way, models of the same count number variates. We only graduate to comparison of models using maximum likelihood estimates if the graphical posterior checking described above does not reveal failures. The model that reports the highest maximum likelihood estimate amongst those that model the same count number variates has a sampling distribution\footnote{Within the continuous set of such distributions associated with the model.} that captures the most structure in the set of count number variates. It is plausible, therefore, that a data-generating process defined by that model is the closest approximation of physical reality attained by the models considered. For models that can \textit{a posteriori} generate data with the structure of the real count number variates, the maximum likelihood estimate can be used to resolve small differences that human inspection fails to uncover.

The third measure that we examine when comparing models of the same set of count number variates is the evidence (the prior predictive probability distribution evaluated using the real count number variates). Whilst maximum likelihood estimates are mere point estimates, the evidence is the expectation of the likelihood function with respect to the joint prior PDF. In one respect, this is powerful because unwarranted prior predictive complexity is penalized: if too much complexity is added to model $\mathcal{M}$ to construct model $\mathcal{M}^{+}$, then for $\mathcal{M}^{+}$ the likelihood function over a large swathe of prior mass is smaller than the expected likelihood (with respect to the prior) of model $\mathcal{M}$, which can entirely negate any localized increases in the likelihood function. In other words, much of the additional complexity is unhelpful because the data generated does not have a similar structure to the real data. On the other hand, penalizing complexity in this way is arguably misleading if the model has phenomenological components: in this Letter the surface hot region models \textit{are} phenomenological.

The evidence, together with a prior mass function of nodes of the model space, may be a biased model selector in our context. Unfortunately, it is necessary in a formal Bayesian framework to use the evidence to marginalize over nodes of the model space in order to compute a posterior PDF of parameters of interest that are shared between nodes. Marginalizing over nodes that differ solely by the atmosphere composition is not problematic. If we do not formally marginalize in such a manner over all models, however, then we can only report the posterior distribution (marginalized over atmosphere composition) for each hot region model that satisfies the graphical posterior checking criterion, together with the maximum likelihood estimate. The reader is then free to interpret the model-to-model posterior variation as a systematic error estimate by, for instance, weighting the posteriors equally which would roughly lead to credible regions with near-maximum hypervolume (width in one dimension, area in two dimensions, and so on); formally, this is equivalent to defining an implicit prior mass distribution over model space nodes that happens to nullify evidence differences, leading to a uniform posterior mass function of models.\footnote{And more formally still, this would mean the prior mass function is dependent on the data, which is a fallacy.} Alternatively, any other weighting can be interpreted as the reader choosing their own prior mass function of model space nodes.

Lastly, when comparing models of the same set of count number variates, we also consider the tractability of the model. If two models pass graphical posterior predictive checks, and supposing one model is less complex, that model is almost by definition more straightforward to implement and to compute the posterior for. The adequately performing model that requires fewest resources to reproduce or prove erroneous---thereby increasing the robustness and potentially the computation accuracy---can be reasoned to be the most useful in practice.

\subsection{Exploratory analysis}\label{sec:exploratory}

In this section we report on posterior sensitivity to various features of the analysis pipeline. Although one can attempt to probe sensitivity using importance sampling, we opted for nested sampling for every variant of interest. In our sensitivity analyses, our nested sampling resolution settings are lower than for the production analysis because we were ultimately resource-limited. We varied the number of nested sampling live-points and the bounding hypervolume expansion factor; we explored \textit{XMM} background prior hyperparameter variation; we switched the atmosphere composition from hydrogen to helium; and we varied likelihood function resolution settings. We have not probed sensitivity to event data set selection (namely, \NICER detector channel cuts) nor sensitivity to approximation of the atmosphere ionization state as fully-ionized.

The posteriors we report in this section were computed using at least $2\times10^{3}$ live points and (except where explicitly noted) condition on the \texttt{ST-U} model and either the \NICER likelihood function or the \NICER and \textit{XMM} likelihood function. For a full description of this model, and a schematic diagram, see \citet{riley19a}. Briefly, however, \texttt{ST-U} assumes each hot region is a single-temperature spherical cap. The two regions can have completely independent properties (temperature and size) and are free to take any location on the star's surface provided that they do not overlap (see also the discussion in Section \ref{sec:temp field}). Our exploratory analysis indicated that this model provided an adequate description of the \joh data set using the performance measures outlined in Section~\ref{sec:performance measures}. Finally, note that posteriors reported in this section are conditional on a \NICER exposure time that was erroneously high by $\sim2\%$. We corrected this number for a subset of posteriors reported in this section, and for the production analysis (Section~\ref{sec:production analysis}); our posteriors are however insensitive to this level of error in exposure time.

\subsubsection{Impact of radio timing prior information}\label{sec:radio impact}

The informative joint NANOGrav and CHIME/Pulsar prior is critical for deriving a useful constraint on the radius of \joh. For comparison, we compute a posterior conditional on a fully-ionized hydrogen atmosphere and a diffuse, separable prior PDF of the mass, the distance, and the cosine of inclination angle. The mass prior is such that the joint prior PDF of mass and radius is flat within the prior support (see Table~\ref{table: ST-U}). The distance prior PDF is adopted from \citet{Igoshev16} and displayed in Figure~\ref{fig:marginal distance PDFs}, with support $D\in[0.1,10.0]$~kpc. The prior PDF of the cosine of the inclination angle is isotropic, meaning flat with support $\cos i\in[0,1]$.

\figsetstart
\figsetnum{5}
\figsettitle{Exploratory analysis.}

\figsetgrpstart
\figsetgrpnum{5.1}
\figsetgrptitle{Impact of radio timing prior information}
\figsetplot{f5_1.png}
\figsetgrpnote{Posterior sensitivity to radio timing prior information. See the figure in the main text.}
\figsetgrpend

\figsetgrpstart
\figsetgrpnum{5.2}
\figsetgrptitle{Effect of atmospheric composition}
\figsetplot{f5_2.png}
\figsetgrpnote{Posterior sensitivity to the hot region atmospheric composition being fully-ionized hydrogen versus fully-ionized helium. Note that there is a finite degree of Monte Carlo noise introducing minor differences in the posterior PDFs because the RNG seeds are different.}
\figsetgrpend

\figsetgrpstart
\figsetgrpnum{5.3}
\figsetgrptitle{Hot region complexity}
\figsetplot{f5_3.png}
\figsetgrpnote{Posterior sensitivity to hot region complexity, probed by conditioning on the \texttt{ST-U} model versus the \texttt{ST+PST} model, both assuming a fully-ionized hydrogen atmosphere. Note that there is a finite degree of Monte Carlo noise introducing minor differences in the posterior PDFs because the RNG seeds are different.}
\figsetgrpend

\figsetgrpstart
\figsetgrpnum{5.4}
\figsetgrptitle{\textit{XMM-Newton} background prior sensitivity}
\figsetplot{f5_4.png}
\figsetgrpnote{Posterior sensitivity to the hyperparameter $n\in[0.01,8]$ controlling the \textit{XMM} background prior. Note that there is a finite degree of Monte Carlo noise introducing minor differences in the posterior PDFs because the RNG seeds are different.}
\figsetgrpend

\figsetgrpstart
\figsetgrpnum{5.5}
\figsetgrptitle{Likelihood function resolution sensitivity}
\figsetplot{f5_5.png}
\figsetgrpnote{Posterior sensitivity to \textsl{X-PSI} discretization degrees when integrating the photon specific flux signal incident on the telescopes as a function of time (rotational phase) and photon energy. Note that there is a finite degree of Monte Carlo noise introducing minor differences in the posterior PDFs because the RNG seeds are different.}
\figsetgrpend

\figsetgrpstart
\figsetgrpnum{5.6}
\figsetgrptitle{Nested sampling resolution sensitivity}
\figsetplot{f5_6.png}
\figsetgrpnote{Posterior sensitivity to nested sampling live-point number and hypervolume expansion factor. The \NICER exposure time was $\sim2\%$ in error for three of the posteriors, and corrected for the remaining two; the effect of the exposure time correction is shown by the $4\times10^{3}$ live-point posteriors. The number of live points is a more fundamental sampling process setting than the hypervolume expansion factor: the two settings do not have orthogonal effects on resolution, with the expansion factor is applied to a union of hyperellipsoids that bounds the set of live points. If the number of live points is increased, the bounding approximation to constant-likelihood hypersurfaces improves and yields finer sampling of the posterior. Increasing the expansion factor improves the approximation to the constant-likelihood hypersurfaces but results in more rejected points---and therefore more unused information---after an equal number of core hours. Note that there is a finite degree of Monte Carlo noise introducing minor differences in the posterior PDFs because the RNG seeds are different, but this noise level decays with increasing live-point number.}
\figsetgrpend

\figsetgrpstart
\figsetgrpnum{5.7}
\figsetgrptitle{Nested sampling resolution sensitivity}
\figsetplot{f5_7.png}
\figsetgrpnote{Posterior sensitivity to nested sampling live-point number. Note that there is a finite degree of Monte Carlo noise introducing minor differences in the posterior PDFs because the RNG seeds are different, but this noise level decays with increasing live-point number.}
\figsetgrpend

\figsetend

{
    \begin{figure*}[t!]
    \centering
\includegraphics[clip, trim=0cm 0cm 0cm 0cm, width=0.95\textwidth]{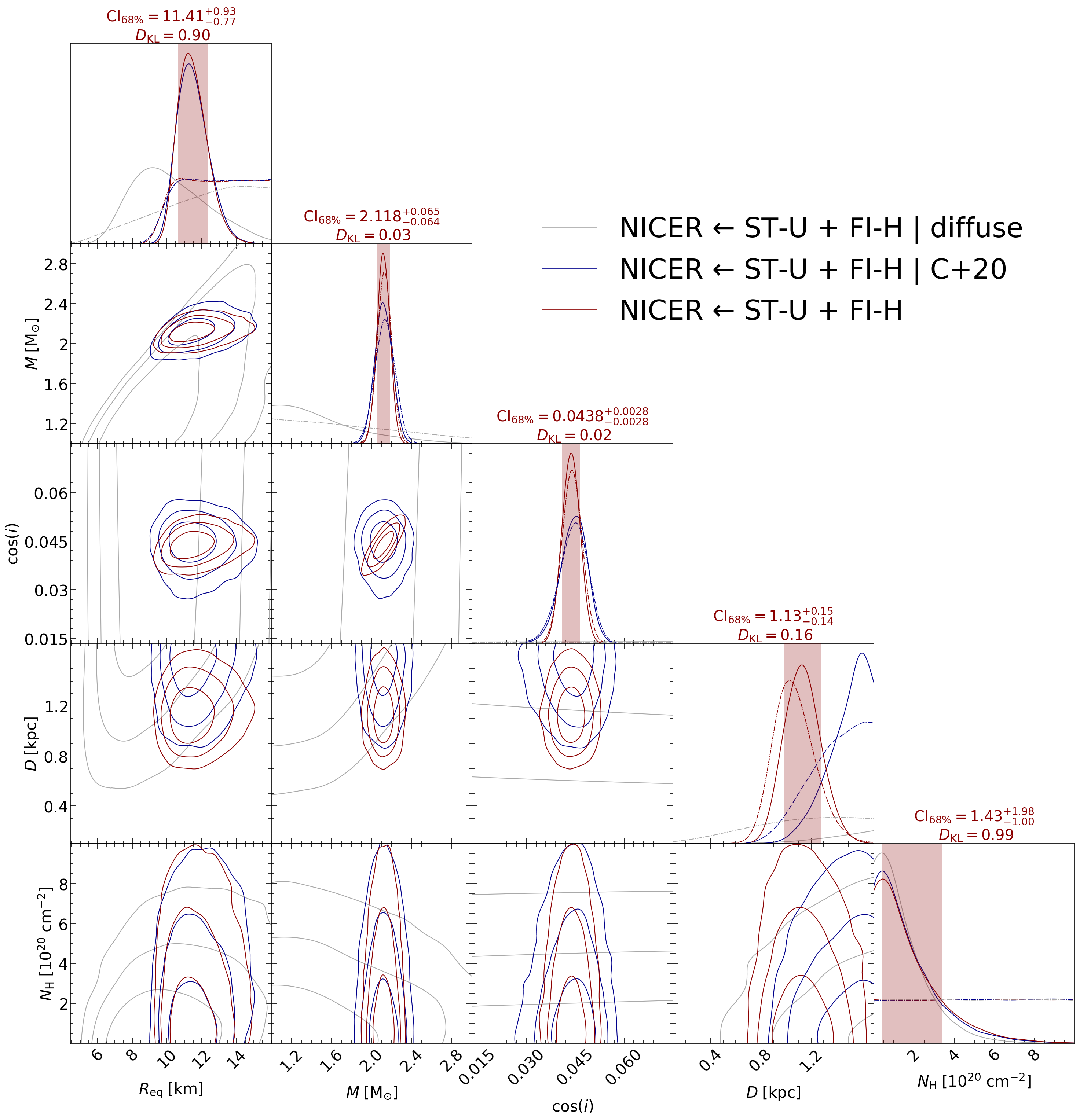}
    \caption{\small{
    One- and two-dimensional marginal PDFs conditional on the \TT{ST-U} model, the \NICER likelihood function alone, and one of three prior PDFs to probe the impact of radio timing information. From leftmost to rightmost in each panel, the parameters are the equatorial radius, the gravitational mass, the cosine of viewing angle subtended to pulsar spin axis, the distance, and the column density. We display the marginal prior PDFs for each parameter as the \textit{dash-dot} functions; the informative priors encode the information from NANOGrav~$\times$~CHIME/Pulsar and \citet[][denoted by the conditional argument C+$20$]{Cromartie19}, and the diffuse prior is described in Section~\ref{sec:radio impact}. We report estimators for the \NICER posterior conditional on the joint NANOGrav and CHIME/Pulsar prior. We report the KL-divergence, $D_{\mathrm{KL}}$, from prior to posterior in \textit{bits} for each parameter. The shaded credible intervals $\mathrm{CI}_{68\%}$ for each parameter are symmetric in marginal posterior mass about the median, containing $68.3\%$ of the mass. The credible regions in the off-diagonal panels, on the other hand, are uniquely the \textit{highest-density}---and thus the smallest possible---credible regions, containing $68.3\%$, $95.4\%$, and $99.7\%$ of the posterior mass. In the appendix of \citet{riley19a} we provide additional information regarding posterior kernel density estimation (KDE), error analysis, and the estimators displayed here; note that here we use an automated Gaussian KDE bandwith optimized by \project{GetDist}~\citep{Lewis19}. The complete figure set for the exploratory analysis ($7$ images) is available in the online journal.}}
    \label{fig:radio prior reversion}
    \end{figure*}
}

The radius posterior conditional on the diffuse prior is much broader than when we condition on the joint NANOGrav and CHIME/Pulsar prior, and there is no independent indication from the pulse-profile modeling for a high mass (see Figure~\ref{fig:radio prior reversion}). Once the models are conditioned on the joint NANOGrav and CHIME/Pulsar prior PDF, we gain very little additional information \textit{a posteriori} from the X-ray likelihood function about the pulsar mass, distance, and inclination angle with respect to the spin axis. We can verify this by examining the marginal posterior distributions in comparison to the respective prior distributions, which is summarized for each parameter by the Kullback-Leibler divergence estimate \citep{kullback1951}.

\subsubsection{Impact of X-ray telescopes}

The \NICER likelihood function is sensitive to the basic configuration of the surface hot regions and their temperatures, despite the relatively small number of pulsed counts (those above the phase-invariant background). The \textit{XMM} likelihood function is less informative both due to the lack of phase information, and because the events are sparse for all three EPIC cameras and moderately consistent with the expected background signal derived from blank-sky exposures. However, the \textit{XMM} likelihood function acts to reduce the \NICER posterior mode volume substantially, affecting the inferred radius and geometry. The reason for this is because the \textit{XMM} likelihood is sensitive to the combined phase-averaged signal from the hot regions being too bright. Therefore, given the \NICER likelihood function, we constrain the contribution to the unpulsed portion of the pulse-profile that must be generated by the hot regions rather than the backgrounds. Posterior figures demonstrating this are reserved for Section~\ref{sec:production analysis}.

\subsubsection{Effect of atmospheric composition}

The atmospheric composition of \joh is not known \textit{a priori}. We therefore compared \texttt{ST-U} posteriors for hydrogen and helium atmospheres assuming full ionization---see the discussion in Section~\ref{sec:atmosphere setup}---in the second online figure of the set associated with Figure~\ref{fig:radio prior reversion}. The marginal radius posteriors were indistinguishable, although there were some small changes in the properties of the hot regions. However, given the apparent lack of sensitivity to atmospheric composition, inferences reported hereafter are conditioned on a fully-ionized hydrogen atmosphere---we do not need to marginalize over the binary atmosphere parameter. Both hot regions are inferred to have effective temperatures $T\approx 10^6$~K, at which partial ionization effects should be small. 

\subsubsection{Hot region complexity}

The \citet{riley19a} analysis of \jdbl reported a number of hot region models that provided an adequate description of the data according to their performance measures (largely adopted here). These included \texttt{ST-U} and variants in which one of the hot regions was permitted increasingly complex forms, including rings and crescents. The inferred radius changed as model complexity increased, but evidence calculations showed a substantial improvement in model performance. As a result of this, we deemed the \texttt{ST+PST} model---in which one hot region is a single temperature spherical cap and the other is, \textit{a posteriori}, a crescent---to be superior for \jdbl.

For \joh the \texttt{ST-U} model also provides an adequate description of the data. In order to assess whether additional complexity is useful we also condition on the \texttt{ST+PST} model. The posterior configuration and properties of the hot regions conditional on this more complex model (which includes the possibility of hot regions that are simply spherical caps) did not differ in an important way from the configuration inferred from \texttt{ST-U}: the hot region for which more complexity was permissible exhibited degeneracy \textit{a posteriori}---we were not sensitive to the existence of additional emission structure beyond that of a simple spherical cap, and the evidence estimates are consistent. There was therefore no extended crescent as inferred for \jdbl; the likelihood function degeneracy included some subset of possible crescent structures---those on smaller angular scales \citep[see][for discussion about hot region structure degeneracy]{riley19a}---which may be of interest to pulsar modelers. The inferred radius changed very little (see the third online figure of the set associated with Figure~\ref{fig:radio prior reversion}), and there was no increase in model performance. For this reason we hereafter report inferences exclusively for the \texttt{ST-U} model. 

\subsubsection{\XMM background prior sensitivity}

As described in Section~\ref{sec:XMM background}, the \textit{XMM} background is free-form, but each variable (one per channel) has a prior with compact support. For each variable, a flat prior PDF is defined whose width is controlled by a hyperparameter $n$. For the headline inferences reported in this Letter we used $n=4$, having tested sensitivity to varying $n$ in the range $n\in[0.01,8]$. In the limit that $n$ tends to zero, the background information would be treated as a point estimate of the \textit{XMM} background. The posterior distribution of the radius is insensitive to $n$ being varied through the range $n\in[0.01,4]$; see the fourth online figure of the set associated with Figure~\ref{fig:radio prior reversion}. It broadens slightly for $n=8$ because fainter combined signals from the hot regions have greater background-marginalized likelihoods, yielding additional posterior weight for higher-radius configurations that reduce the unpulsed component whilst conserving the pulsed component to satisfy the \NICER event data. However, the value $n=8$ is arguably too conservative even when considering potential systematic error.

\subsubsection{Likelihood function resolution sensitivity}

The \XPSI likelihood function has a number of resolution settings, most notably settings that control the discretization of the computational domain for computation of signals (pulse-profiles) incident on telescopes. The photon specific flux signal we require is an integral over a distant observer's sky of the photon specific intensity from the hot regions, yielding a two-dimensional function of time (rotational phase) and photon energy. The level of discretization with respect to four variables in the domain of the incident photon specific intensity field generally controls the computational expense of likelihood evaluation. These variables are the number of rotational phases and energies the photon specific flux signal is computed at; the number of hot region surface elements; and the number of rays calculated. Please see the \XPSI documentation\footnote{\url{https://thomasedwardriley.github.io/xpsi/}} for additional information.

We tested posterior sensitivity to increasing the discretization degrees for these variables by recomputing a posterior PDF with a new nested sampling process. We found that doubling these discretization degrees does not yield a change in the posterior PDF that is clearly distinguishable from Monte Carlo sampling noise;\footnote{The nested sampling seed was set based on the system clock for each sampling process and therefore not held constant as would be ideal.} see the fifth online figure of the set associated with Figure~\ref{fig:radio prior reversion}.

\subsubsection{Nested sampling resolution sensitivity}

For a fixed bounding hypervolume expansion factor of $0.1^{-1}$, the posterior PDFs were insensitive to doubling live-point number from $10^{3}$ up to $4\times10^{3}$. Following sampler comparisons within the \NICER collaboration, we then increased resolution to $4\times10^{4}$ live points, leading to broadening of the radius posterior; see the sixth and seventh online figures in the set associated with Figure~\ref{fig:radio prior reversion}. Increasing the sampling resolution by using $8\times10^{4}$ live points led to some further broadening, but doubled an already large computational cost. Given the computational resources required for posterior computation with such a large number of live points, we were not able to rigorously prove convergence with live-point number. We decided to adopt $4\times10^{4}$ live points for the production analysis---all information necessary to reproduce and improve upon our posterior computation is available in open-source repositories.

Such posterior mode broadening with increased nested sampling resolution is naturally expected because nested sampling algorithms approximate hypersurfaces in parameter space of constant likelihood; these approximations improve with sampling resolution but their sufficiency is difficult to prove for non-trivial likelihood functions encountered in real problems and when subject to resource limitations. It is desirable to transform away non-linear modal degeneracies so that an approximation conforms more efficiently to structure in the sampling space; however, this can in practice be intractable task for a given problem. Moreover, when sampling from a target distribution with two or more modes of commensurate posterior mass, the live-point resolution is roughly split between the modes, reducing the resolution of a given mode due to the approximations alluded to above. For \joh, posterior bimodality arises due to the near-equatorial inclination of the source, leading to two competitive geometric configurations of the hot regions (see Section~\ref{sec:production analysis}, where we discuss this further).

\subsection{Production analysis}\label{sec:production analysis}

Our exploratory analysis indicates that model \texttt{ST-U} provides an adequate description of the data and that the posteriors are largely insensitive to either atmospheric composition or increased hot region complexity. In this section, we present high-resolution posteriors---using $4\times10^{4}$ live points---conditional on the \texttt{ST-U} model, and fully-ionized hydrogen atmosphere. For each posterior we use either the \NICER likelihood function alone, the \NICER and \textit{XMM} likelihood function, or the \textit{XMM} likelihood function alone. The posterior PDF conditional on the \NICER likelihood function alone is derived by importance-sampling another posterior PDF, thereby updating a deprecated radio timing prior PDF (see Section~\ref{sec:radio} and \citealt{Fonseca20}). The weighted and equally-weighted samples from the marginal joint posterior distribution of mass and radius may be found in the persistent repository of \citet{zenodo}, together with credible region contour point-sequences and marginal posterior PDFs of the radius (to facilitate plotting).

Figure \ref{fig:ST-U residuals} provides a simple graphical posterior predictive check on the model performance, demonstrating that the \texttt{ST-U} model can generate synthetic event data that is commensurate with the \NICER data. No unexpected structure---such as large deviations or correlations---is emergent in the residuals.   

Marginal posterior distributions of the spacetime parameters---in particular the radius---are shown in Figure \ref{fig:ST-U spacetime corner}. The figure displays posteriors conditional on the \NICER data alone, conditional on the \textit{XMM} data alone, and conditional on both \NICER and \textit{XMM} data. As expected, the phase-resolved \NICER likelihood function is far more constraining, in isolation, than the phase-averaged \textit{XMM} likelihood function. However, the likelihood function product over telescopes excludes regions of the \NICER posterior modes where the contribution to the unpulsed component of the pulse-profile from the hot regions is too bright. The unpulsed component is brighter for models in the \NICER posterior modes where the star is more compact. Restricting to lower compactness increases the inferred radius for the combined data set by $\sim 1$~km.

The inferred hot region parameters, again comparing the likelihood functions in isolation to the likelihood function, are shown in Figure \ref{fig:ST-U geometry corner}. The effect of including the \textit{XMM} data can be seen in the joint posterior PDF of the stellar radius and the angular radii of the two hot regions ($\zeta_p$ and $\zeta_s$). The \NICER-only posteriors (at the 99.7\% level) include models with a smaller stellar radius ($<10.5$ km) and hot regions with a larger angular radius ($\zeta \sim 1.2$ rad). The large hot regions on very compact stars lead to a bright unpulsed component of the combined signal from those regions. The inclusion of the \textit{XMM} data means that more of the unpulsed signal is associated with the background instead of the hot regions. As a result, these smaller stars with large hot regions are excluded when the \textit{XMM} data is included.

{
    \begin{figure*}[th!]
    \centering
    \includegraphics[clip, trim=0cm 0cm 0cm 0cm, width=0.75\textwidth]{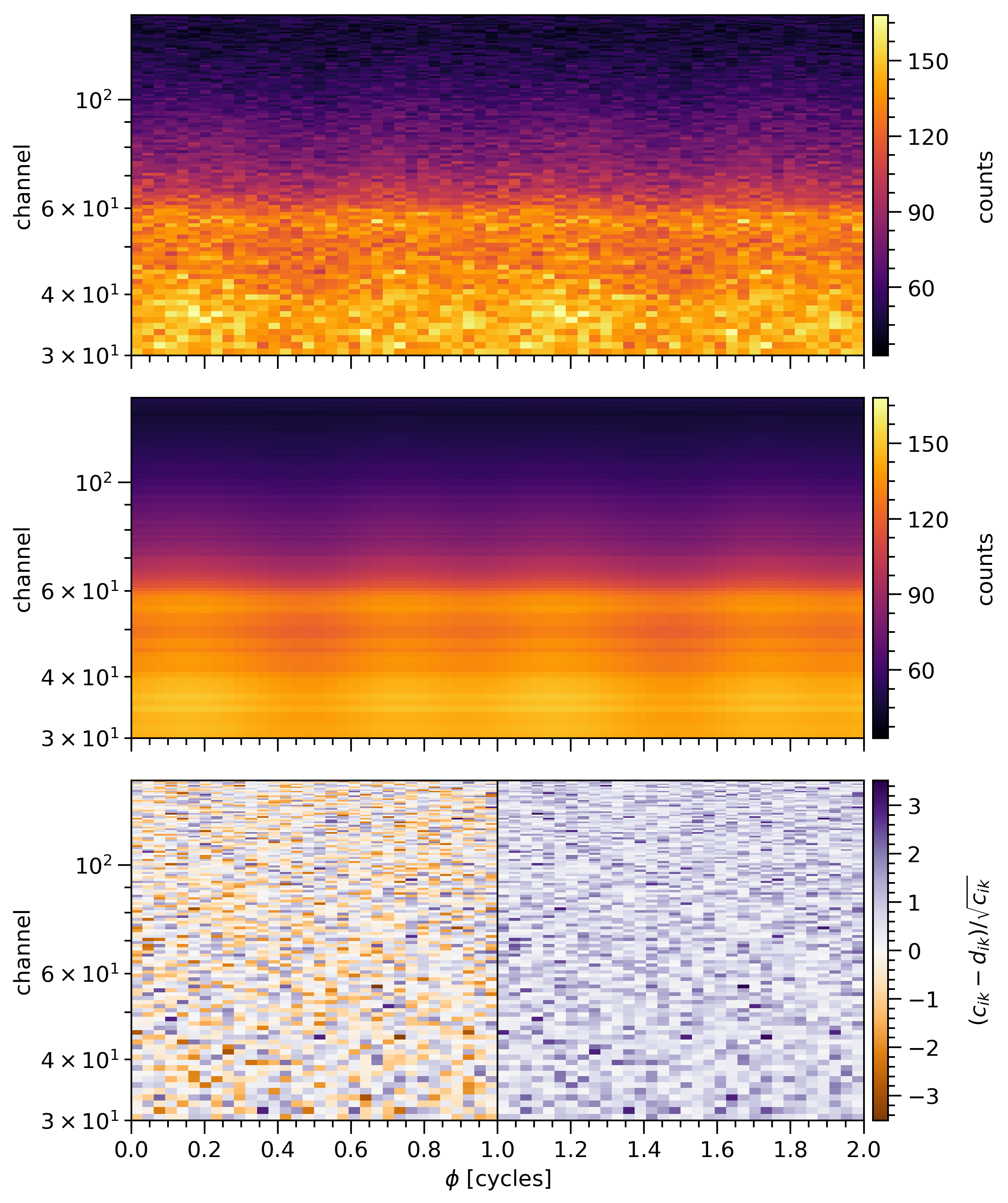}
    \caption{\small{\NICER count data $\{d_{ik}\}$, posterior-expected count numbers $\{c_{ik}\}$, and (Poisson) residuals for \TT{ST-U}. Note that we split the count numbers in the upper two panels over two rotational cycles, such that the information on phase interval $\phi\in[0,1]$ is identical to the information on $\phi\in(0,2]$; our data sampling distribution, however, is defined as the (conditional) joint probability of all event data grouped into phase intervals on $\phi\in[0,1]$. We display the standardized (Poisson) residuals in the \textit{bottom} panel: the residuals for the rotational cycle $\phi\in[0,1]$ were calculated in terms of all event data on that interval (as for likelihood definition), and simply cloned onto the interval $\phi\in(1,2]$. In Section~\ref{sec:performance measures} and in the appendix of \citet{riley19a} we elaborate on the information displayed here. 
    }}
    \label{fig:ST-U residuals}
    \end{figure*}
}

{
    \begin{figure*}[t!]
    \centering
\includegraphics[clip, trim=0cm 0cm 0cm 0cm, width=\textwidth]{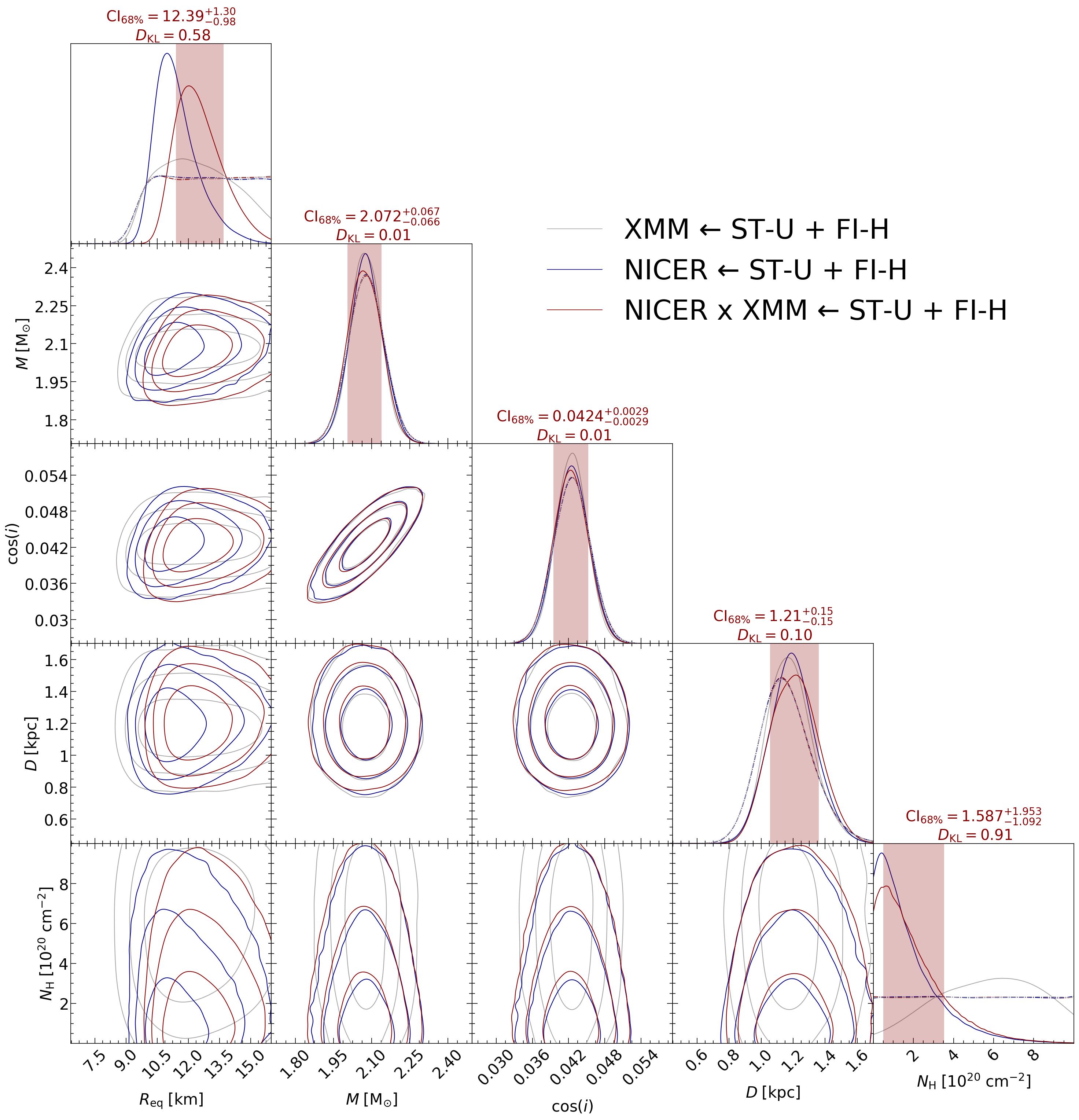}
    \caption{\small{
    One- and two-dimensional marginal PDFs for fundamental parameters, conditional on the \TT{ST-U} model and either the \textit{XMM} likelihood function alone, the \NICER likelihood function alone, or the \NICER and \textit{XMM} likelihood function. From leftmost to rightmost in each panel, the parameters are the equatorial radius, the gravitational mass, the cosine of viewing angle subtended to pulsar spin axis, the distance, and the column density. We display the marginal prior PDFs for each parameter as the \textit{dash-dot} functions; the informative priors encode the NANOGrav~$\times$~CHIME/Pulsar information. We report estimators for the \NICER~$\times$~\textit{XMM} posterior. We use an automated Gaussian KDE bandwith optimized by \project{GetDist}~\citep{Lewis19}. See the caption of Figure~\ref{fig:radio prior reversion} for additional details about the figure elements.
    }}
    \label{fig:ST-U spacetime corner}
    \end{figure*}
}

{
    \begin{figure*}[t!]
    \centering
    \includegraphics[clip, trim=0cm 0cm 0cm 0cm, width=\textwidth]{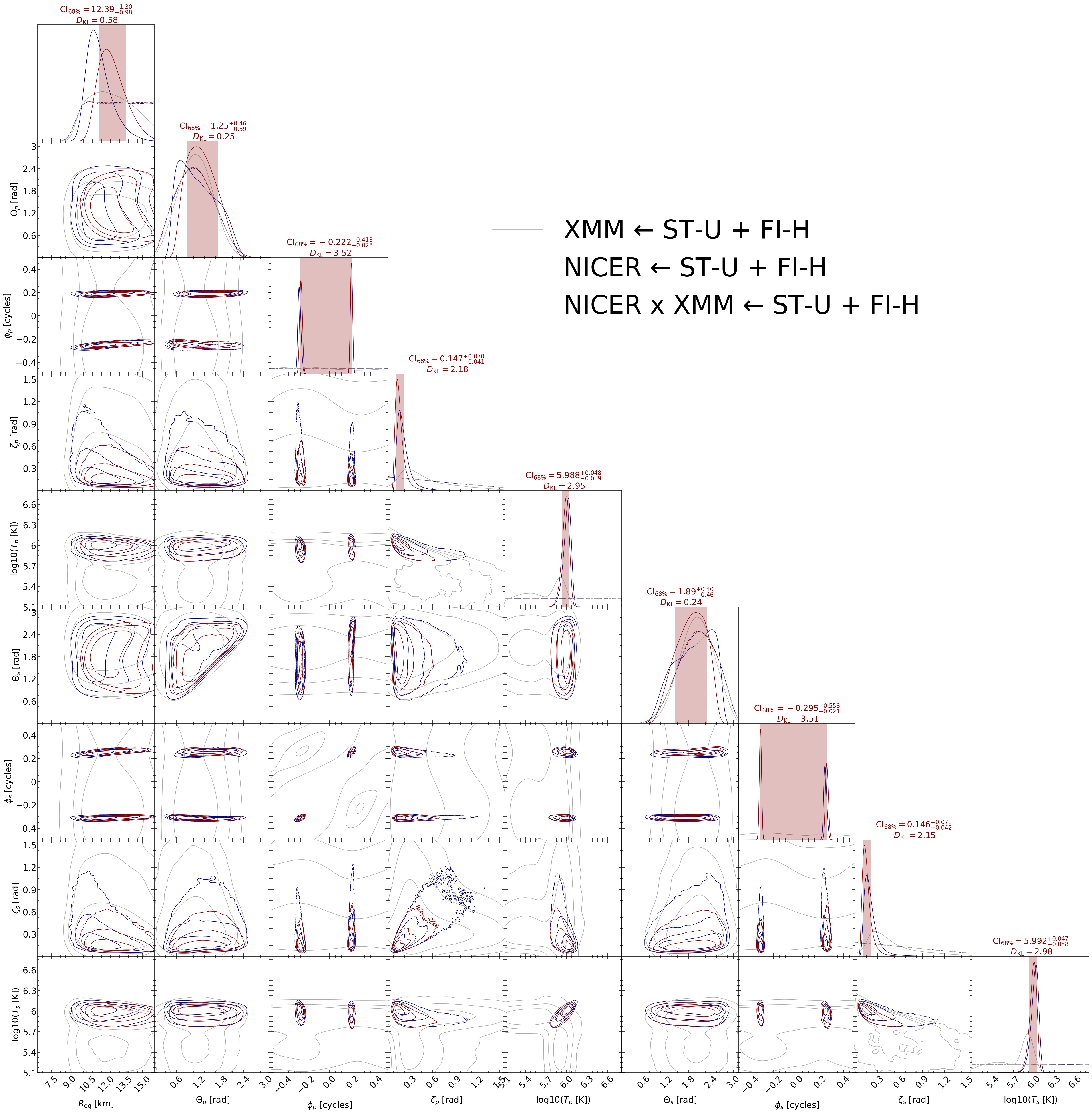}
    \caption{\small{One- and two-dimensional marginal posterior distributions of hot region parameters conditional on the \texttt{ST-U} model. Three types of posterior distribution are rendered: one conditional only on the \NICER likelihood function; one conditional only on the \textit{XMM} likelihood function; and one conditional on the \NICER and \textit{XMM} likelihood function. 
    }}
    \label{fig:ST-U geometry corner}
    \end{figure*}
}

{
    \begin{figure*}[t!]
    \centering
\includegraphics[clip, trim=0cm 0cm 0cm 0cm, width=0.495\textwidth]{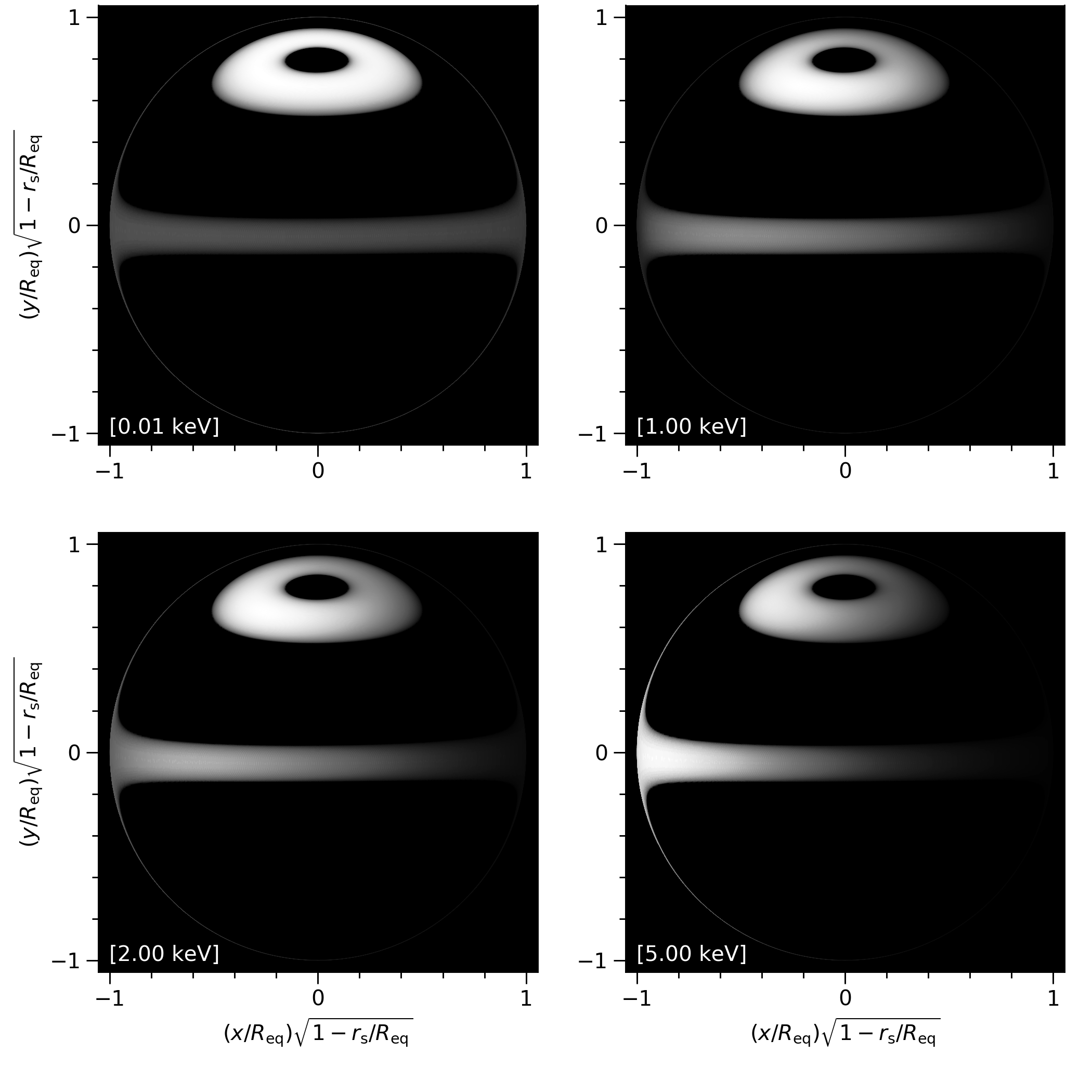}
\includegraphics[clip, trim=0cm 0cm 0cm 0cm, width=0.495\textwidth]{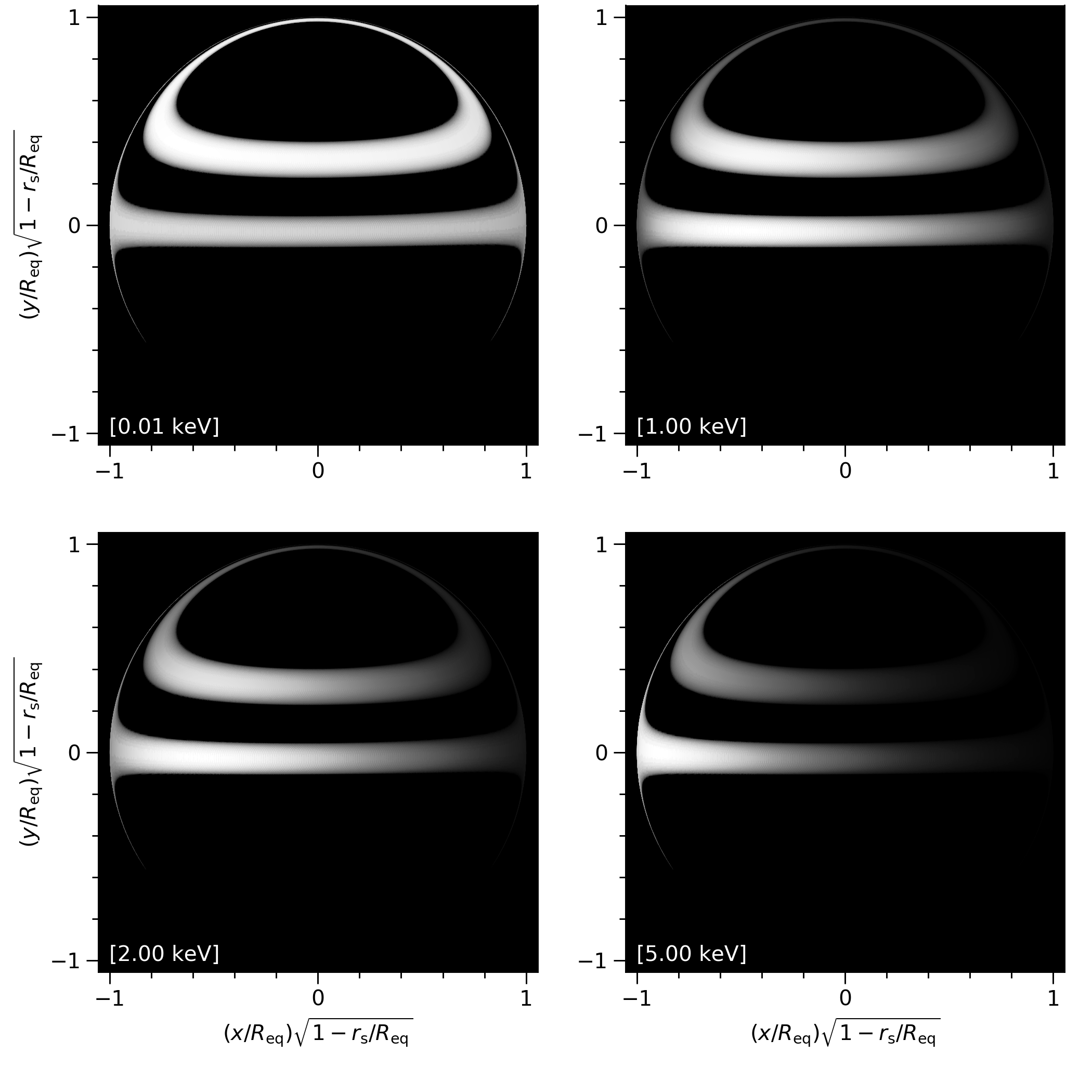}
\includegraphics[clip, trim=0cm 0cm 0cm 0cm, width=0.495\textwidth]{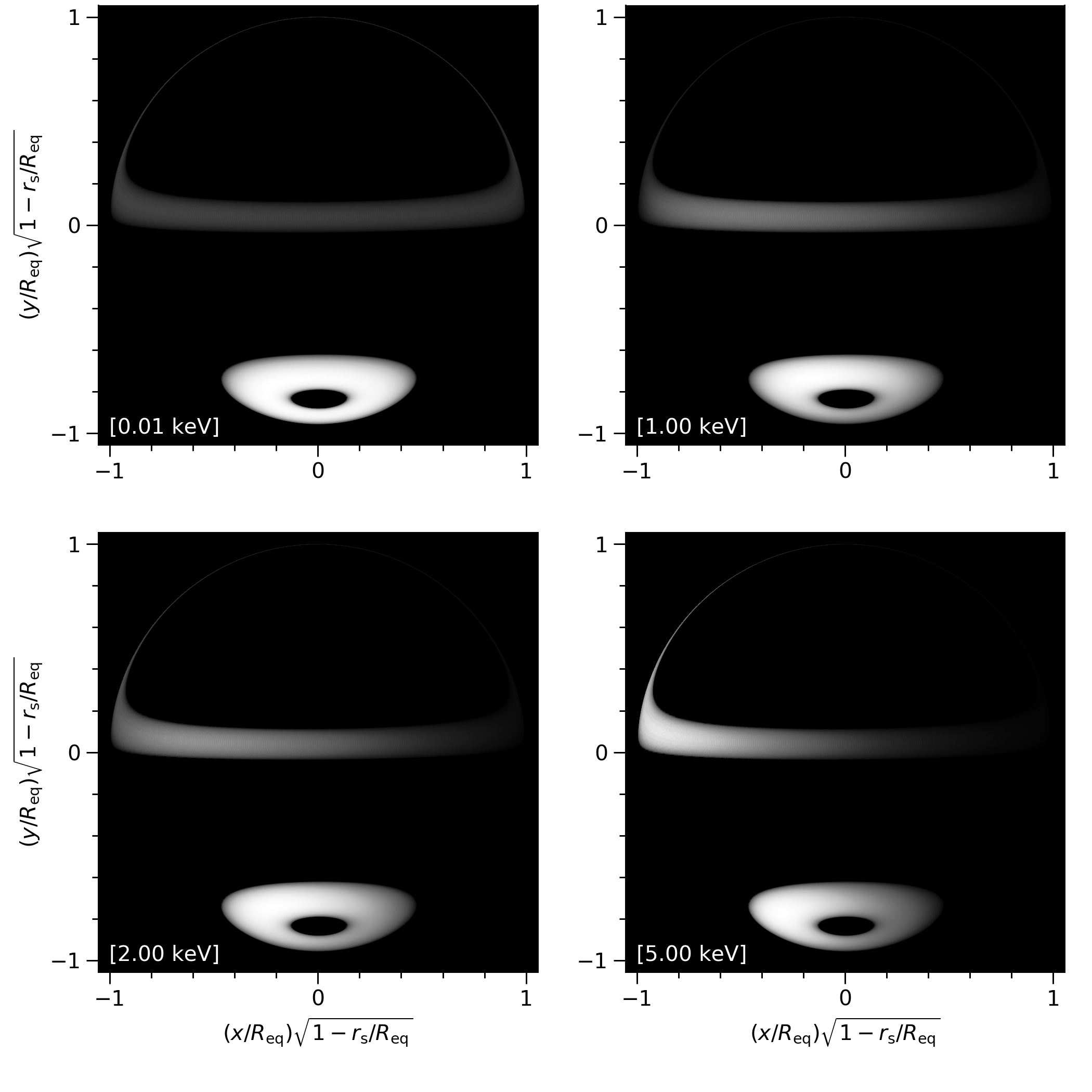}
\includegraphics[clip, trim=0cm 0cm 0cm 0cm, width=0.495\textwidth]{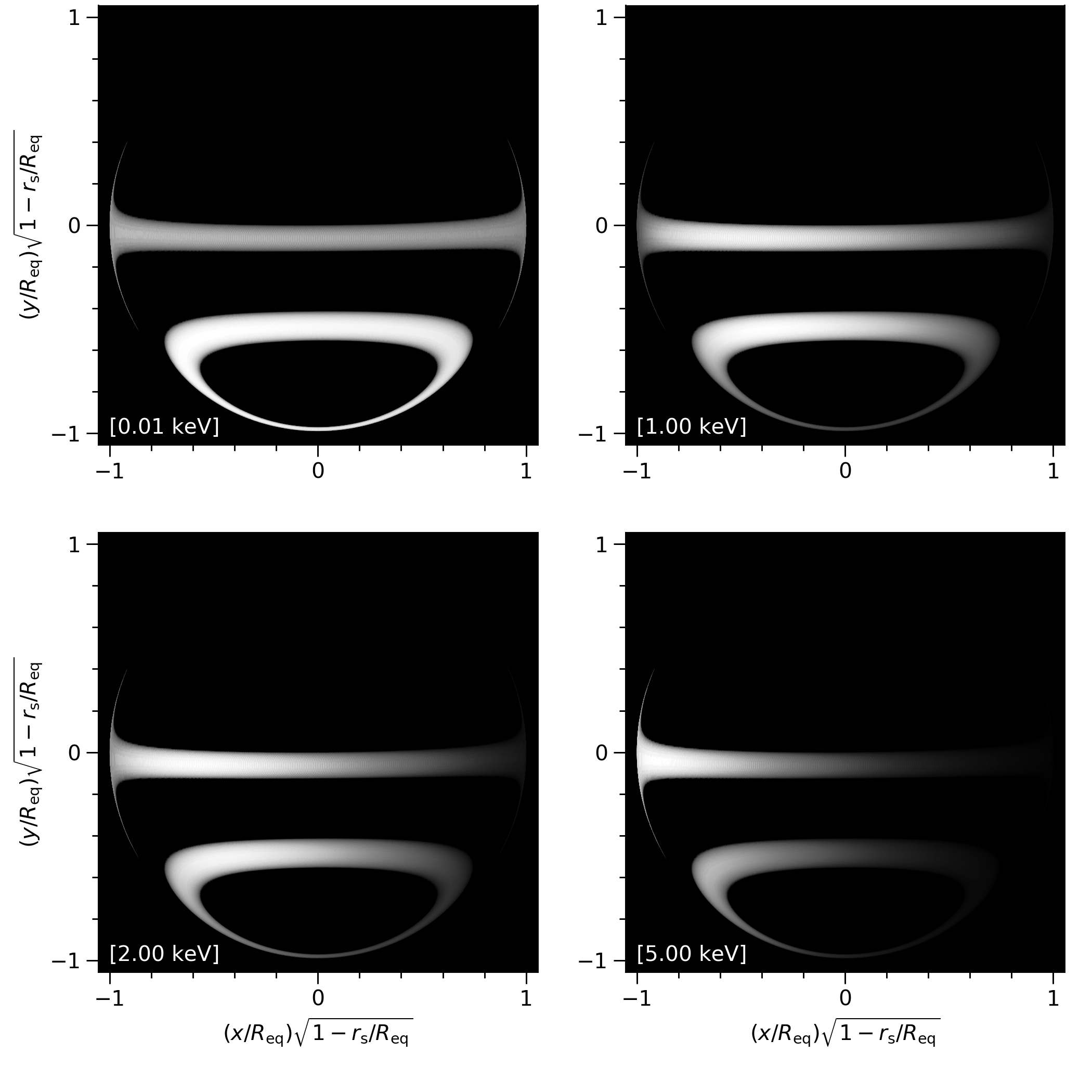}
    \caption{\small{A novel type of figure rendering the phase-averaged expected (photon) specific intensity as a function of sky direction for the source-receiver configuration estimated to maximize the (background-marginalized) \texttt{ST-U} likelihood function given by Equation~(\ref{eqn:background-marginalized likelihood}), showing the two different posterior modes and the effect of \textit{XMM} likelihood function inclusion. \textit{Top-left set of four panels}: \NICER likelihood function---mode one. \textit{Bottom-left set of four panels}: \NICER likelihood function---mode two. \textit{Top-right set of four panels}: product of \NICER and XMM likelihood functions---mode one. \textit{Bottom-right set of four panels}: product of \NICER and \textit{XMM} likelihood functions---mode two. The expected photon specific flux spectrum registered by \NICER if we phase-average, and (when included) the \textit{XMM} cameras, is implicitly formed from a fine set of these images. These representative images at four photon energies include all relativistic effects in the likelihood function; note that we extend slightly beyond the \textit{XMM} waveband used for likelihood evaluation in order to render the (relativistic) rotational effects more vividly. Each panel is normalized to the maximum phase-average specific intensity over sky direction at that energy. The background sky has the same intensity---zero---as neighborhoods of the image that a hot region never traverses because the surface exterior of the hot regions is not explicitly radiating in the models (please see Section~\ref{sec:exterior of the hot regions}). For animated (photon) specific intensity sky maps corresponding to these four variants (two posterior mode variants for each of two likelihood function variants), together with pulse-profile traces and spectral evolution, please refer to the online journal. 
    }}
    \label{fig:ST-U skymap}
    \end{figure*}
}

{
\begin{figure*}[t!]
\centering
\begin{interactive}{animation}{NICER_x_XMM__STU__FIH__ML__skymap_animated.mp4}
\includegraphics[clip, trim=0cm 0cm 0cm 0cm, width=\textwidth]{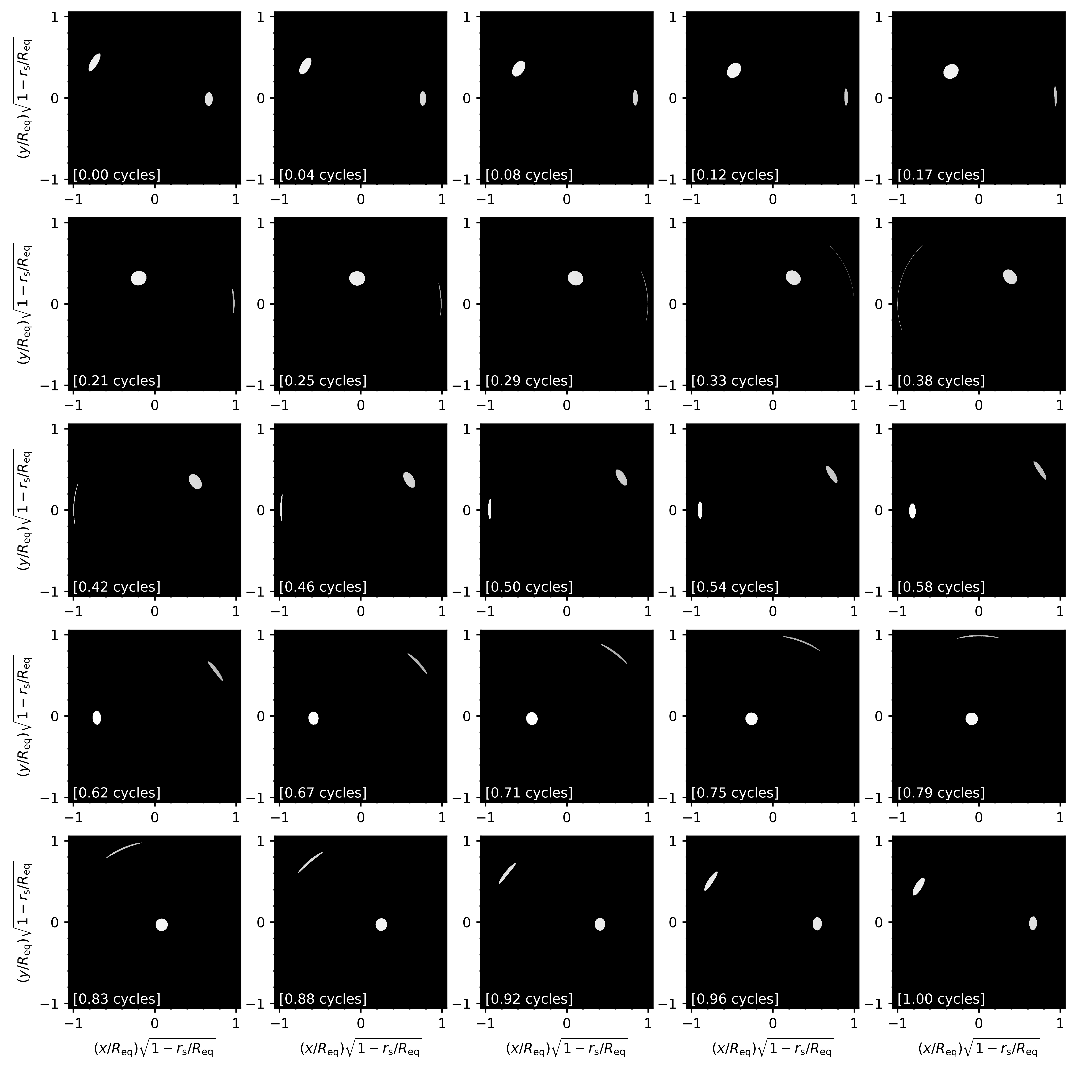}
\end{interactive}
\caption{A summary of the animated figure available in the online journal. In the animated figure, the \textit{top} three panels show the (photon) specific intensity as a function of sky direction at three different photon energies as the star rotates. The \textit{bottom-left} panel displays the (photon) specific flux pulse-profiles traced out by the skymaps, each normalized to its respective maximum. The \textit{bottom-right} panel displays the (photon) specific flux spectrum, where the energy bounds each correspond to a skymap energy, as does the vertical line; the trace of the vertical line intersecting the spectrum is one of the pulse profiles. The star rotates 16 times during the 48 second animation. In this summary figure we aim to display the gravitationally lensed geometric configuration of the surface hot regions from our Earthly viewing perspective, over the course of one rotational cycle. We display the (photon) specific intensity as a function of sky direction at the lowest energy, as the star rotates through the panels from left to right and from top to bottom.
}
\label{fig:animated}
\end{figure*}

}

{
    \begin{figure*}[t!]
    \centering
\includegraphics[clip, trim=0cm 0cm 0cm 0cm, width=0.495\textwidth]{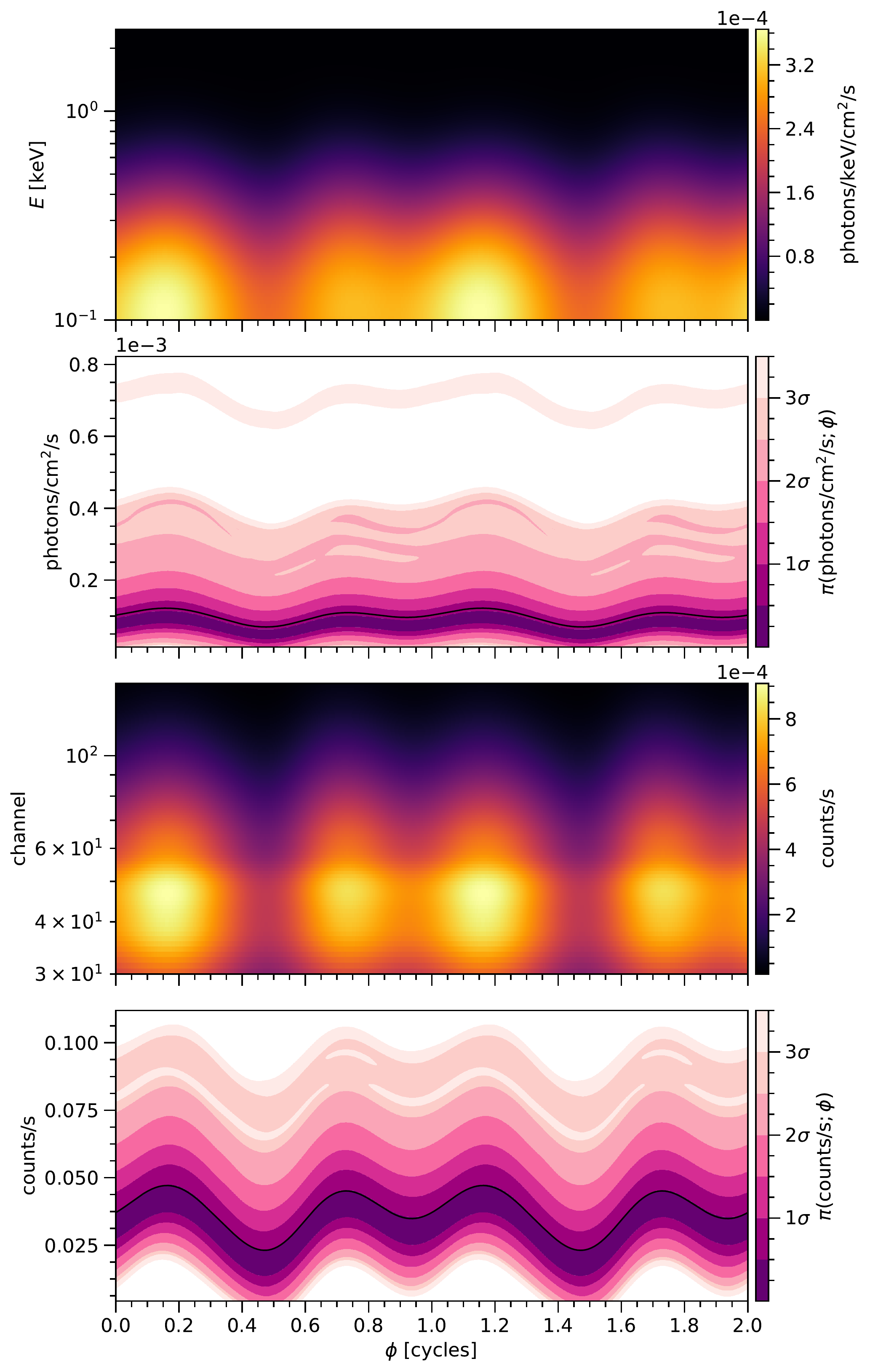}
\includegraphics[clip, trim=0cm 0cm 0cm 0cm, width=0.495\textwidth]{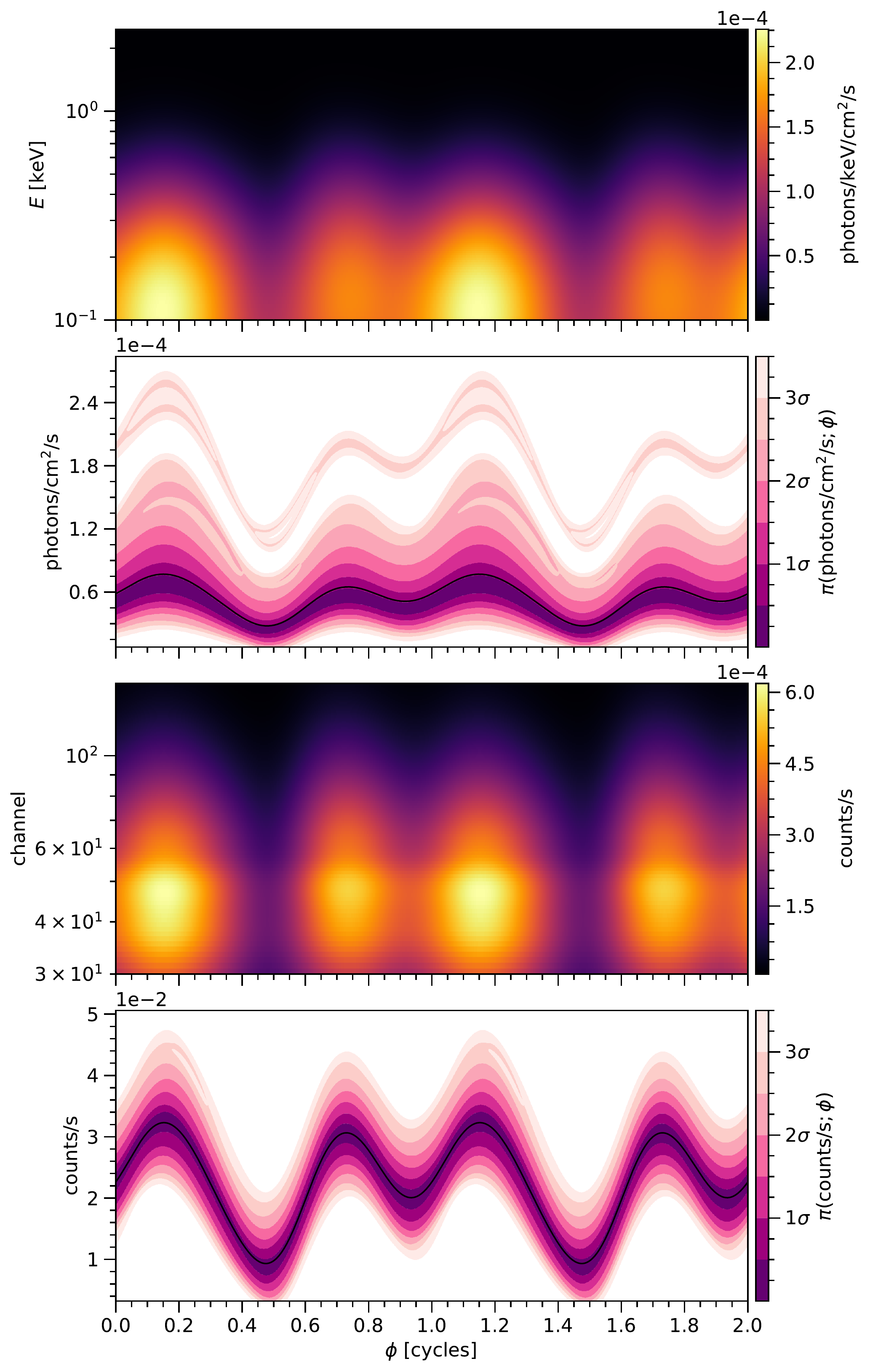}
    \caption{\small{Posterior-expected pulse-profiles conditional on the \TT{ST-U} hot region model and either the \NICER likelihood function (\textit{left} panels) or the \NICER and \textit{XMM} likelihood function (\textit{right} panels). We show the signal incident on the telescopes (\textit{top} and \textit{top-center} panels) and as registered by \NICER (\textit{bottom-center} and \textit{bottom} panels). The signal in the \textit{top} panels has been integrated over the linearly-spaced instrument energy intervals, and is effectively proportional to the photon specific flux. The \textit{black} rate curves are the posterior-expected signals generated by the hot regions in combination. We also represent the conditional posterior distribution of the incident photon flux (\textit{top-center} panels) and the \NICER count rate (\textit{bottom} panels) at each phase as a set of one-dimensional highest-density credible intervals, and connect these intervals over phase via the contours; these distributions are denoted by $\pi(\mathrm{photons/cm^{2}/s};\phi)$ and $\pi(\mathrm{counts/s};\phi)$. Note that the fractional width of the credible interval at each phase is usually higher for $\pi(\mathrm{photons/cm^{2}/s};\phi)$ than for $\pi(\mathrm{counts/s};\phi)$ because of the variation permitted for the instrument model; in combination, the signal registered by the instrument is more tightly constrained. To generate the conditional posterior bands we apply the \XPSI package, which in turn wraps the \TT{fgivenx}~\citep{fgivenx} package. 
    }}
    \label{fig:ST-U pulse}
    \end{figure*}
}

{
    \begin{figure*}[t!]
    \centering
    \includegraphics[clip, trim=0cm 0cm 0cm 0cm, width=0.495\textwidth]{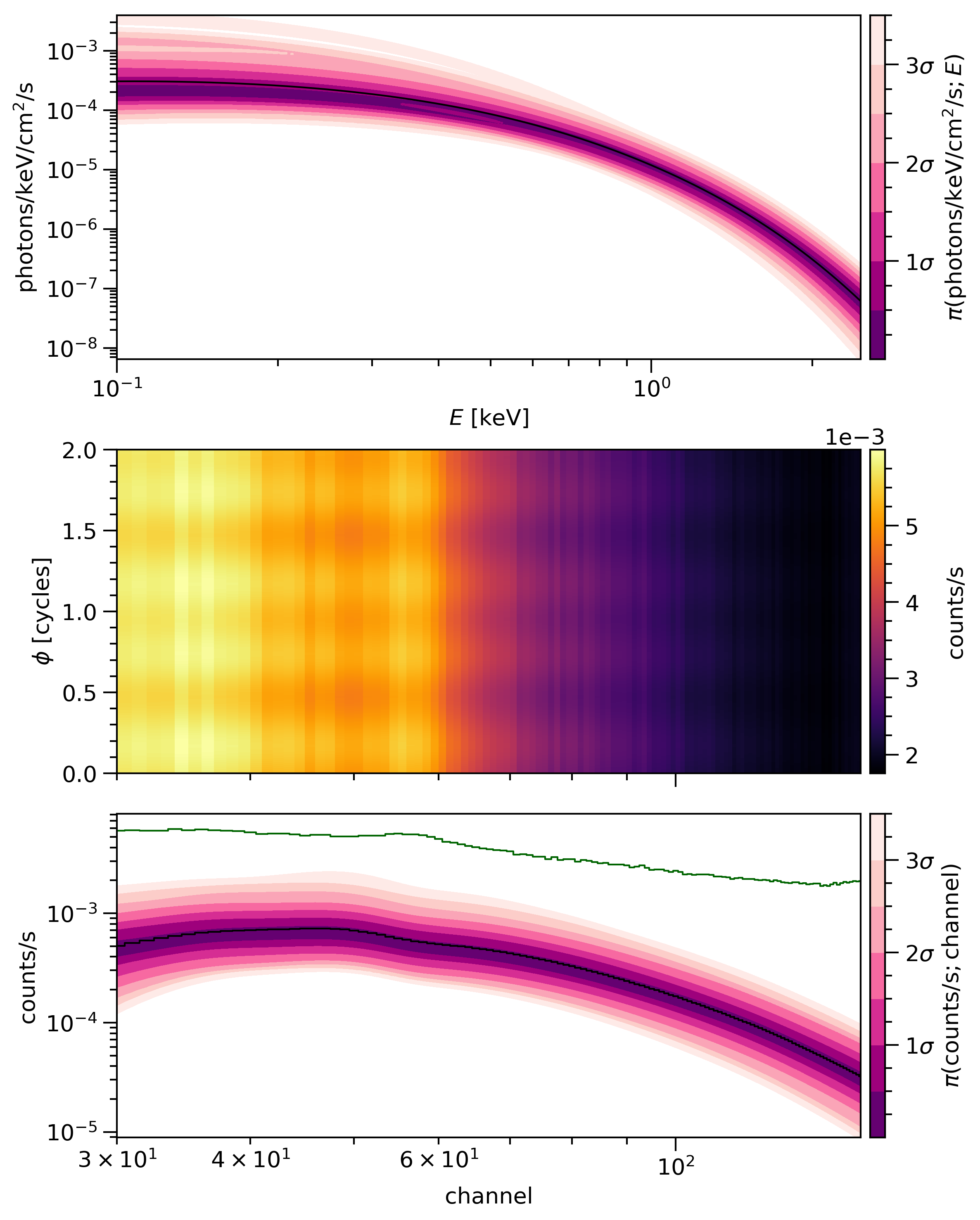}
    \includegraphics[clip, trim=0cm 0cm 0cm 0cm, width=0.495\textwidth]{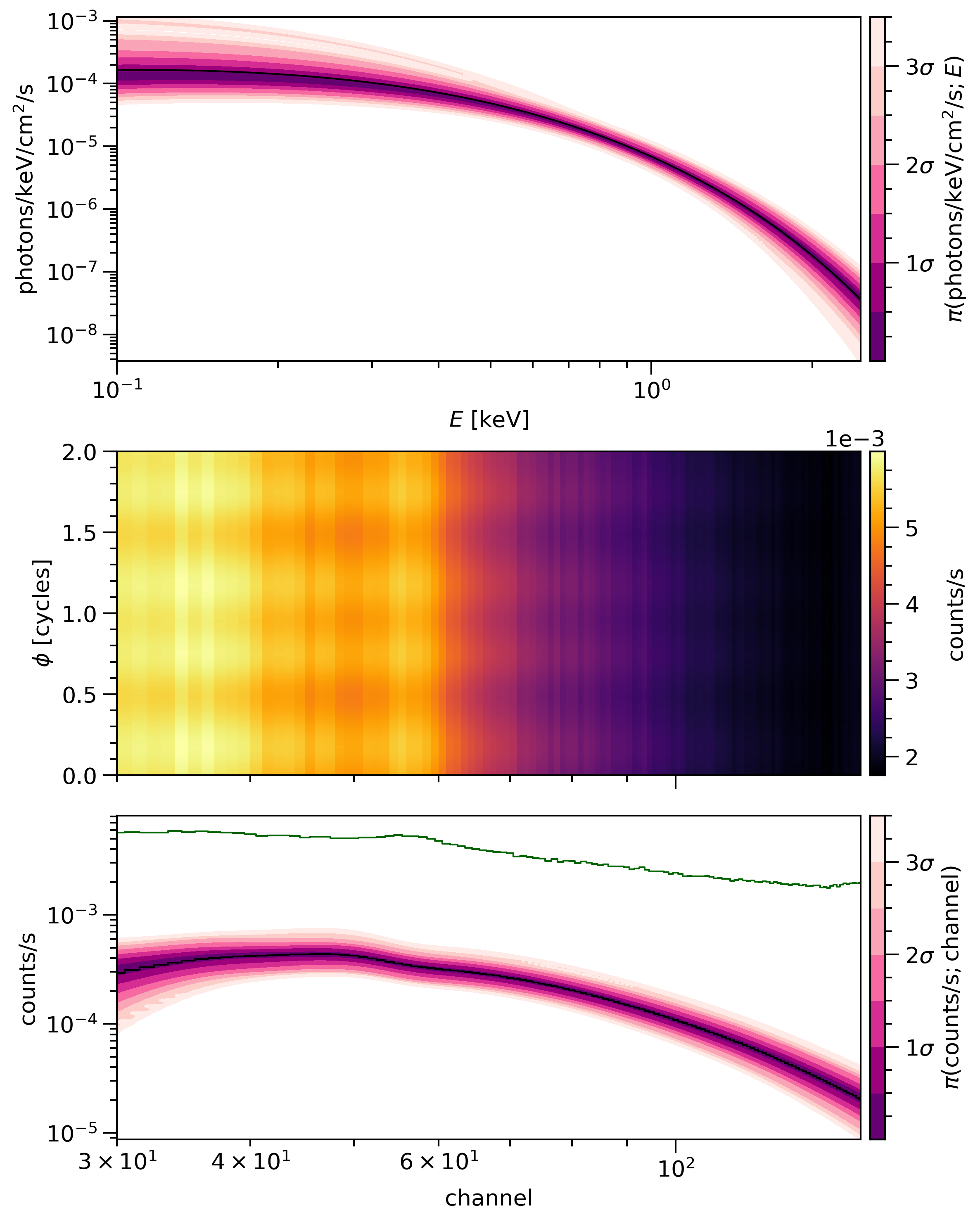}
    \caption{\small{Posterior-expected spectra conditional on the \TT{ST-U} hot region model and either the \NICER likelihood function (\textit{left} panels) or the \NICER and \textit{XMM} likelihood function (\textit{right} panels). We show the spectrum that would be incident on the telescopes if it were unattenuated by the interstellar medium (\textit{top} panels) and the signal as registered by \NICER (\textit{center} and \textit{bottom} panels). The \textit{black} rate curves are the posterior-expected spectra generated by the hot regions in combination. We represent the conditional posterior distribution $\pi(\mathrm{photons/keV/cm^{2}/s};E)$ of the unattenuated incident photon specific flux at each energy as a set of one-dimensional highest-density credible intervals, and connect these intervals over phase via the contours (\textit{top} panels); the energies displayed are those spanning the waveband of the \NICER channel subset $[30,150)$. In the \textit{center} panels we display the background-marginalized posterior-expectation of the source count-rate signals, \textit{plus} the background count-rate terms that maximize the \textit{conditional} likelihood functions; the \textit{center-right} signal is equivalent to that displayed in the center panel of Figure~\ref{fig:ST-U residuals}. In the \textit{bottom} panels we display the posterior-expected count-rate spectra generated by the hot regions in combination and individually, together with the conditional posterior \NICER count rate distribution $\pi(\mathrm{counts/s};\textrm{channel})$ for each channel. Moreover, the topmost \textit{green} step functions are the phase-average of the \textit{center} panels---each is effectively, but not exactly, the observed count-number spectrum divided by the total \NICER exposure time.
}}
    \label{fig:ST-U spectrum}
    \end{figure*}
}

Interestingly there are two different posterior modes (due to the near-equatorial inclination),\footnote{Note that these are different geometric configurations because we expressly exclude hot region exchange degeneracy from the prior support (see Section~\ref{sec:temp field}).} which can be seen in more detail in the phase-averaged skymaps in Figure \ref{fig:ST-U skymap} and the animated skymap in Figure~\ref{fig:animated}. For neither mode are the two hot regions antipodal. The effect on the inferred signal (pulse-profile and phase-averaged spectrum) of combining the two data sets is shown in Figures \ref{fig:ST-U pulse} and \ref{fig:ST-U spectrum}, respectively. The inclusion of the \textit{XMM} data reduces the contribution of the hot region emission to the unpulsed component of the pulse profile, leading to a lower count rate but an increased pulsation amplitude (see lower panels of Figure \ref{fig:ST-U pulse}) in the combined signal from the hot regions.

\section{Discussion}\label{sec:discussion}

\subsection{Radius measurement and implications for EOS}

The inferred equatorial radius of the massive pulsar \joh is $R_\mathrm{eq} = 12.39_{-0.98}^{+1.30}$~km, where the credible interval bounds are approximately the 16\% and 84\% quantiles in marginal posterior mass, given relative to the median. The inferred mass, $2.072_{-0.066}^{+0.067}$~\msol is dominated by the mass prior from the radio timing, $2.08\pm 0.07$~\msol. The 90\% credible interval for the radius is $12.39_{-1.50}^{+2.22}$~km and the 95\% credible interval is $12.39_{-1.68}^{+2.63}$~km. It is worth stressing that when we carried out pulse-profile modeling without using the mass prior from radio timing, we would not have inferred independently from the \NICER and \textit{XMM} data alone that \joh is a high-mass source (nor did we obtain any informative constraint on the radius; see Section~\ref{sec:radio impact}).

Rotation increases the equatorial radius of a neutron star. The increase in radius for fixed mass is small, as shown in Figure~\ref{fig:mass-vs-radius}, where mass-radius curves are plotted for four representative EOS that span a wide range of allowed stiffness \citep{Hebeler13}. Mass-radius curves for non-rotating stars and stars rotating at 346~Hz are shown for each EOS. For a 2.0~\msol star, the equatorial radius increases by slightly more than 0.05~km for the softest EOS, while the increase is as large as 0.2~km for the stiffest EOS. These increases in radius due to spin are smaller than our uncertainty in measuring the neutron star's radius at present. The change in radius due to spin is already incorporated into our pulse-profile models, since we assume the shape of the rotating star is deformed into an oblate shape. While the shape function that we use is an approximation, \citet{Silva20} have shown that it is sufficiently accurate for all of the rotation-powered pulsars with spin frequencies less than 400~Hz that \NICER observes.

{
    \begin{figure}[t!]
    \centering
    \includegraphics[clip, trim=0cm 0cm 0cm 0cm, width=\columnwidth]{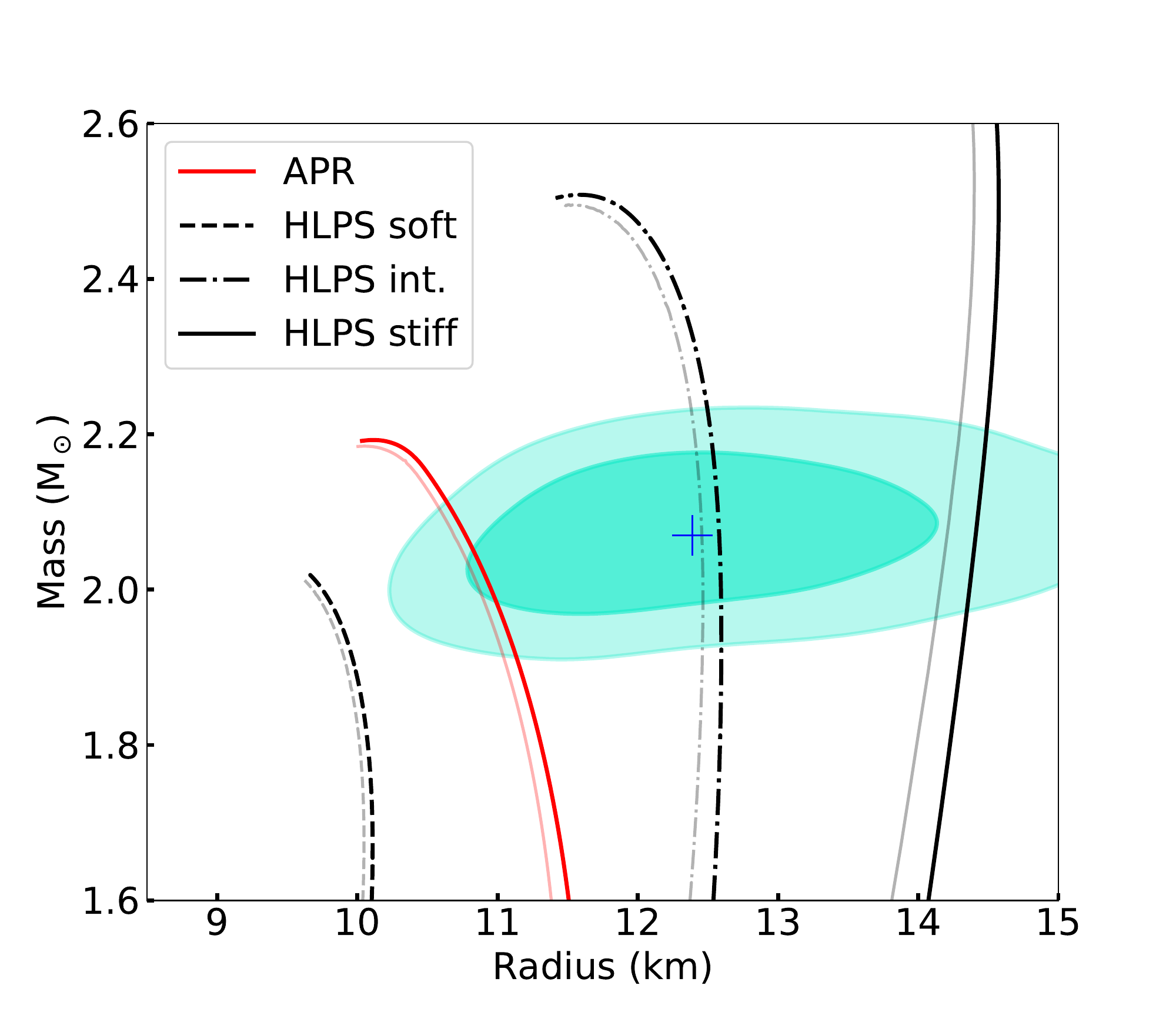}
    \caption{\small{Mass versus equatorial radius for several example EOS models from \citet{Hebeler13}, showing the difference between non-rotating stellar models and stars rotating at 346~Hz. For each EOS shown the right hand (heavier) curve is for a spin of 346~Hz, while the left-hand (lighter) curve is for zero rotation. The 68\% and 95\% credible regions for mass and radius inferred from our analysis of \joh are shown by the shaded \textit{cyan} contours. The blue crosshair shows the inferred median values.}}
    \label{fig:mass-vs-radius}
    \end{figure}
}

The radius inferred for \joh is very similar to the radius inferred from pulse-profile modeling of \NICER data for \jdbl, although for the latter the inferred mass was lower: \citet{riley19a} found $R_\mathrm{eq} = 12.71^{+1.14}_{-1.19}$~km and $M = 1.34^{+0.15}_{-0.16}$~\msol; the independent analysis of  \citet{miller19} found $R_\mathrm{eq} = 13.02^{+1.24}_{-1.06}$~km and $M = 1.44^{+0.15}_{-0.14}$~\msol. Note that the width of the credible interval on the radius for \joh is not smaller, despite the constraining mass prior: this is due to the lower number of source counts, which limits the precision of the radius measurement. The radius that we report for \joh is also consistent with the values inferred from the most recent phase-averaged spectral modeling of quiescent and bursting neutron stars \citep[][noting that for some of these analyses a neutron star mass is assumed rather than inferred or known in advance]{Nattila17,Steiner18,Baillot19,Gonzalez19}, and with indirect constraints from the inner radii of accretion disks \citep{Ludlam17}.  

The detection of gravitational waves from binary neutron star mergers provides an alternative method of constraining the dense matter EOS. A measurement of tidal deformability and neutron star masses from the late inspiral phase can be used to infer EOS parameters and hence the associated mass-radius relation. The posteriors on any mass-radius relation derived in this way depend on the EOS model and the priors on the model parameters, and this should be kept in mind when comparing them to the radii inferred directly (without reference to EOS models) from pulse profile modeling \citep{Greif19}. Nevertheless the radii derived from the tidal deformability of the binary neutron star merger GW170817 are (for a range of EOS models) lower then the value we derived for \joh \citep[see for example][]{Abbott18,Abbott19,De18,Most18,Landry20,Essick20,LiA20}. However, the credible intervals on the gravitational-wave derived radii are of similar extent, and thus the results are certainly consistent with those derived by \NICER. 

As is clear from Figure \ref{fig:mass-vs-radius}, the radius of \joh is in the center of the range considered plausible for neutron stars $\sim 2$~\msol, and appears compatible with a wide range of EOS models \citep[see e.g.][]{Hebeler13,Greif19}. However full Bayesian inference of EOS models is  required to fully quantify the constraints arising. For this we refer the reader to the companion paper by \citet{Raaijmakers21}, which carries out EOS inference using results from \NICER both individually and in combination with constraints from gravitational wave observations and their electromagnetic counterparts.  Using two different high density EOS parameterizations, and models that connect to microscopic calculations of neutron matter from chiral effective field theory interactions at nuclear densities, \citet{Raaijmakers21} show that the new \NICER results provide tight constraints, for example on the pressure of neutron star matter at around twice saturation density.  

The measurement of radius for a high mass neutron star is also interesting for the properties of potential quark cores \citep[see for example][]{Annala20}.  \citet{HanS20} consider the implications of such a measurement for our understanding of quark matter phases in neutron stars for different model types: self-bound strange quark star models; hybrid star models with different types of phase transition; and third family models where two branches with different radii are possible for the same mass \citep{Schertler00}. Our upper- and lower-bounds on the radius posterior at high mass disfavor some regions of quark matter parameter space: both the stiffness of the strange quark phase and the transition properties. The inferences are moreover quite restrictive for self-bound strange quark stars and third family stars, both of which typically have low radii at high mass. 

We can also look at the implications of the change in radius as one moves from  $M \sim 2.0 $~\msol to $M\sim 1.4$~\msol, an important distinguishing characteristic of different EOS models \citep{Greif19,HanS20,XieWJ20,Drischler20b}.  Generally, hadronic EOSs having symmetry energy parameters in the ranges predicted by nuclear mass fits and neutron matter studies and with $M_{\rm max}\lesssim 2.2$ \msol would result in $\Delta R=R_{2.0}-R_{1.4} \lesssim -1$ km.  The above studies show that matter with a phase transition around $2n_{\rm sat}$ to a relatively soft phase with sound speed squared $c_s^2 \sim 1/3$ (such as to non-interacting quark matter) would also result in $\Delta R \lesssim -1$ km. Such models also have $M_{\rm max} \lesssim 2.2$ \msol.  In contrast, larger values of $\Delta R \lesssim 0.5$ km suggest either stiffer high-density matter without a phase transition having $M_{\rm max} \sim 2.3-2.5$ \msol, or a transition to a relatively stiff phase at a transition density $\ge2.6n_{\rm sat}$ \citep{Drischler20b}.  Even larger values of $\Delta R>0.5$ km would imply a transition at a lower density $\le 2n_{\rm sat}$ to similarly stiff matter.  The companion paper by \citet{Raaijmakers21}, which utilizes parameterized EOS models constrained by theories of neutron matter, together with observations of pulsar masses, gravitational waves from mergers, and the \XPSI \NICER results for \jdbl and \joh, infers that $\Delta R\simeq-0.5^{+1.2}_{-1.5}$ km averaged over EOS models.  This is consistent with the direct observational value $R_{J0740}-R_{J0030}=-0.3^{+1.2}_{-1.5}$ km \citep[this paper, ][]{riley19a}.  Although values of $\Delta R<-1$ km and $\Delta R>+1$ km cannot be ruled out, these results suggest more moderate values of $\Delta R$ that favor relatively stiff dense matter with a large $M_{\rm max}$ or an EOS with stiffening at a density $2-3n_{\rm sat}$, which could result from a first order phase transition or a crossover transition like that due to the appearance of quarkyonic matter \citep{McLerran19}. Observational upper limits to $M_{\rm max}$, such as the value $M_{\rm max} \lesssim 2.2-2.3$ \msol suggested by GW170817 \citep{Margalit17}, could help distinguish these possibilities.

The maximum mass of neutron stars, and hence the boundary between the neutron star and black hole populations, is also a function of the EOS.  Currently feasible EOS models would permit a maximum neutron star mass in the range $2-3$~\msol; but stiffer EOS, with larger radii, are required to achieve higher masses.  Analysis of the electromagnetic counterpart of the binary neutron star merger GW170817 has suggested a maximum neutron star mass somewhere in the range $2.0-2.3$~\msol \citep{Margalit17,Rezzolla18,Ruiz18,Shibata19}, but there is a strong dependence on how the kilonova is modeled. Nevertheless this range is consistent with the relatively soft EOS inferred from the tidal deformability for GW170817 (see above). The recent detection of GW190814, a binary compact object merger involving an object with mass $\sim 2.6$~\msol \citep{Abbott20}, is however intriguing. There is considerable debate over whether this object could be a high-mass neutron star rather than a low mass black hole \citep{Fattoyev20,Sedrakian20,HuangK20,Drischler20b,TanH20,Tsokaros20,Dexheimer20,Zhang20,Tews20b,Godzieba20} and still be consistent with GW170817.  The radius that we have inferred for \joh suggests a lower maximum mass, however \citep[see][]{Raaijmakers21}.  

\subsection{Constraining power of \XMM}\label{sec:XMM constraining power}

The likelihood function given \NICER and \textit{XMM} data sets is dominated by the information from the former. However, longer \textit{XMM} exposure times naturally yield greater constraining power. A deep exposure exists for the rotation-powered millisecond PSR~J0030$+$0451 \citep{Bogdanov09}, and as suggested by \citet{riley19a}, the associated spectroscopic (and timing) event data can be jointly modeled with the \NICER event data set to address the open question regarding the contribution of the model hot regions to that set of events, which \citet{miller19} and \citet{riley19a} inferred to be (close to) minimal. The term {\it minimal} is used to mean that the hot regions contributed {\it only} to the pulsed component of the pulse profile and did not contribute to the unpulsed component. This present Letter offers a demonstration of how this can be executed, albeit with a contribution from \textit{XMM} that is less informative than the contribution from \NICER.

The \textit{XMM} data set for \joh is phase-averaged and sparse in terms of overall counts, which renders it less constraining than the \NICER data set. However, being an imaging telescope, \textit{XMM} facilitates better quantification of the contribution from the star (attributed to hot regions in our models) compared to the background, whereas the \NICER background is more difficult to constrain both \textit{a priori} and \textit{a posteriori}. The contribution of the hot regions to the unpulsed component of the \NICER pulse-profile is constrained by jointly modeling the \NICER and \textit{XMM} event data.

For \joh, the contribution from the hot regions is not minimal. Even when one considers only the \NICER data set, the hot regions are inferred to generate not only the pulsed component but also part of the unpulsed component of the pulse profile. The effect of including the \textit{XMM} data in the analysis is to increase the amplitude of the emission from the hot regions by reducing the contribution of the hot regions to the unpulsed component of the pulse profile. Smaller radius stars have larger gravitational fields and cause stronger gravitational lensing. The lensing makes it possible to see the hot regions for a larger fraction of the spin period, the resulting signal has a lower pulsed fraction than the signal from a larger star with less lensing. The samples where the \NICER-only unpulsed signal is brighter are those where \joh is more compact; by weighting away from these, the inclusion of \textit{XMM} data pushes the posterior towards less compact stars, where the radius is higher \citep[see also the discussion in][]{Miller16}. 

An interesting question is whether the analysis presented in this Letter tells us anything about the effect that a full joint inference analysis might have on the radius inferred for \jdbl. The emission from the hot regions for \jdbl conditional on only \NICER event data was minimal, meaning the unpulsed component of the combined signal from the hot regions was small. It follows that the inclusion of the \textit{XMM} constraints can only increase the contribution from the hot regions. However, the magnitude of the increase in brightness and the effect on the inferred radius is hard to predict because several parameters in the model are degenerate and changes in radius can be offset by, e.g., changes in hot region parameters.

{
    \begin{figure}[t!]
    \centering
    \includegraphics[clip, trim=0cm 0cm 0cm 0cm, width=\columnwidth]{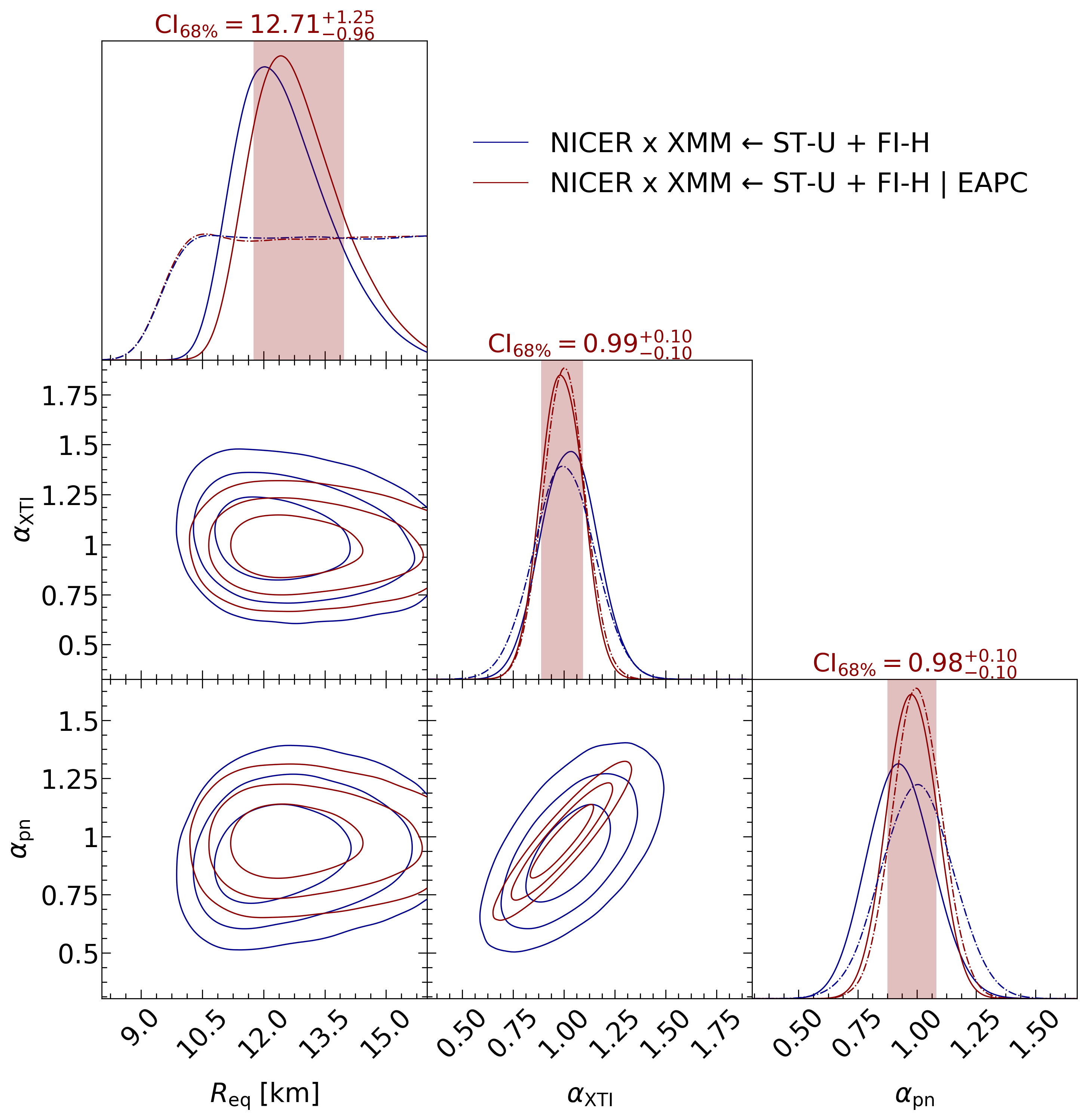}
    \caption{\small{One- and two-dimensional marginal posterior distributions of the \NICER and \textit{XMM} energy-independent effective area scaling factors $\alpha_{\textrm{XTI}}$ and $\alpha_{\textrm{pn}}$ respectively, and the equatorial radius. The \textit{XMM} scaling factor $\alpha_{\textrm{pn}}$ is shared by the three cameras and is denoted by $\alpha_{\textrm{XMM}}$ in Table~\ref{table: ST-U}. The \textit{blue} \NICER~$\times$~\textit{XMM} posterior is shown in Figures~\ref{fig:ST-U spacetime corner} as the headline posterior; the properties of this posterior are reported in Table~\ref{table: ST-U}. The \textit{red} \NICER~$\times$~\textit{XMM}~$|$~EAPC is conditional on a compressed effective area prior that aims to approximate the telescope calibration uncertainties discussed in Section~\ref{sec:response models}; the acronym EAPC simply means \textit{effective area prior compression}. The prior PDFs are displayed as the \textit{dash-dot} distributions in each on-diagonal panel.}}
    \label{fig:effective area prior compression}
    \end{figure}
}

Finally, we explore sensitivity to prior information about the \NICER and \textit{XMM} energy-independent effective area scaling factors. 
We importance-sample our joint \NICER and \textit{XMM} posterior to compress the joint prior on these scaling factors, as shown as Figure~\ref{fig:effective area prior compression}. The compressed joint prior approximates published telescope calibration uncertainties (see Section~\ref{sec:response models}) by using telescope-specific scaling factors, each with a Gaussian prior whose standard deviation is $3\%$. By compressing the joint prior, the marginal posterior distribution of the \textit{XMM} scaling factor median shifts from $\sim 0.93$ up to $\sim 0.98$. Consequently, to conserve the normalization of the high-likelihood count number spectra registered by each \textit{XMM} camera---which \textit{a posteriori} have larger typical effective areas after compressing the prior---the brightness of the signal incident on the telescope must decrease. It follows that subject to conserving the pulsed component of the combined signal from the hot regions as required by the \NICER event data, the brightness of the unpulsed component of the combined signal decreases as the high-likelihood regions of parameter space shift to slightly less compact stars---and thus to slightly higher radii given the informative mass prior. The overall shift in the posterior PDF of the radius due to the compression ($\sim +0.3$~km in the median, to $R = 12.71^{+1.25}_{-0.96}$~km) is much smaller than the measurement uncertainty\footnote{Such a small shift is not expected to have any remarkable effect on EOS inference, given typical EOS priors \citep{Greif19}. This is demonstrated in \citet{Raaijmakers21}, where EOS inference is carried out using both our \NICER-only inferred radius and our \NICER~$\times$~\textit{XMM} inferred radius. Despite an overall change in the median radius posterior inferred from the pulse profile modelling $\sim + 1.1$ km once the \textit{XMM} data set is included, the mass-radius band shifts by a much smaller amount than this, due to the strong influence of the priors on the EOS model.}. However it highlights that instrument cross-calibration is an important aspect of these analyses that warrants careful treatment.

Obtaining estimates of the absolute effective area of an X-ray instrument is a challenging task. Cross-calibration efforts by the International Astrophysical Consortium for High Energy Calibration (IACHEC) using observations from multiple concurrent X-ray telescopes have found offsets typically within $\pm 10\%$ but with occasional discrepancies reaching up to $\sim 20\%$   \citep[see, e.g.,][]{2011PASJ...63S.657I,2017A&A...597A..35P,2017AJ....153....2M}. The tails of the joint posterior PDF of the effective scaling parameters (see Figure~\ref{fig:effective area prior compression}) go beyond what these calibration measurements indicate, so the resulting radius credible intervals should be taken as conservative estimates.

\subsubsection{\NICER and \XMM backgrounds}\label{sec:NICERXMM_backgrounds}

The \NICER background is difficult to estimate directly, but there are two tools available. Figure~\ref{fig:background_spaceweather} shows the \NICER background estimated using the `space weather' model \citep{SpaceWeather} and the `3C50' model \citep{3C50}, see also \citet{bogdanov19a}. The former models background due to the space weather environment, which varies as \NICER moves through different geomagnetic latitudes, and depends on solar activity. The `3C50' model is empirical, taking into account two different types of particle-induced events, the cosmic X-ray background, and a soft X-ray noise component related to operation in sunlight. We also render a set of \NICER background estimates for comparison: using joint \NICER and \textit{XMM} posterior samples, we display the \NICER background spectrum that maximizes the conditional likelihood function, yielding a band that is a proxy for background variable posterior mass.

The background spectra displayed in Figure~\ref{fig:background_spaceweather} exceed the space weather model at low energies. This excess is not unreasonable given the presence of multiple other point sources in the \NICER field of view.\footnote{The \NICER background variables in principle also capture phase-invariant emission from the environment of \joh that does not originate from the surface hot regions in the \XPSI model. The \textit{XMM} background prior information is conservative (see Section~\ref{sec:XMM background}) and can also in principle capture such phase-invariant emission. The \textit{XMM} likelihood function is not purely marginalized with respect to an informative blank-sky background prior, which could attribute too much emission to the hot regions in the \XPSI model.} In higher channels, the background spectra appear to agree well with the space weather model. There is a possible small systematic under-prediction compared to the space weather model in channels 80$-$100, the level of agreement is satisfactory given the current uncertainties on the background modeling.  The level of agreement with the `3C50' model, which exceeds the space weather model at low energies and is consistent with it at higher energies, also appears good. Neither background model includes off-axis X-ray sources that might contaminate the target spectrum, and therefore inferred backgrounds that exceed the two estimates are not in principle a problem.  

Once the uncertainties on the two \NICER background models are understood more fully, we anticipate being able to use them to reduce systematic error in the pulse-profile modeling. The current radius measurement conditional on the \textit{XMM} data set---which yields an indirect \NICER background constraint---will therefore be superseded.  Efforts are also ongoing to quantify the level of background due to any off-axis sources in the field of view. 

\figsetstart
\figsetnum{15}
\figsettitle{\NICER background.}

\figsetgrpstart
\figsetgrpnum{15.1}
\figsetgrptitle{\textsl{NICER} background (\textsl{NICER}~$\times$~\textit{XMM} posterior).}
\figsetplot{f15_1.pdf}
\figsetgrpnote{The \textsl{NICER} background associated with the joint \textsl{NICER} and \textit{XMM} posterior. See the caption in the main text for details.}
\figsetgrpend

\figsetgrpstart
\figsetgrpnum{15.2}
\figsetgrptitle{\textsl{NICER} background (\textsl{NICER}~$\times$~\textit{XMM} high-likelihood).}
\figsetplot{f15_2.pdf}
\figsetgrpnote{The \textsl{NICER} background associated with the posterior samples with the highest background-marginalized likelihood values given the \textsl{NICER} and \textit{XMM} data sets. See the caption in the main text for details.}
\figsetgrpend

\figsetgrpstart
\figsetgrpnum{15.3}
\figsetgrptitle{\textsl{NICER} background (\textsl{NICER}~$\times$\textit{XMM} posterior given effective area compression).}
\figsetplot{f15_3.pdf}
\figsetgrpnote{The \textsl{NICER} background associated with the joint \textsl{NICER} and \textit{XMM} posterior given compression of the effective area prior. See the caption in the main text for details.}
\figsetgrpend

\figsetend

{
    \begin{figure}[t!]
    \centering
    \includegraphics[clip, trim=0cm 0cm 0cm 0cm, width=\columnwidth]{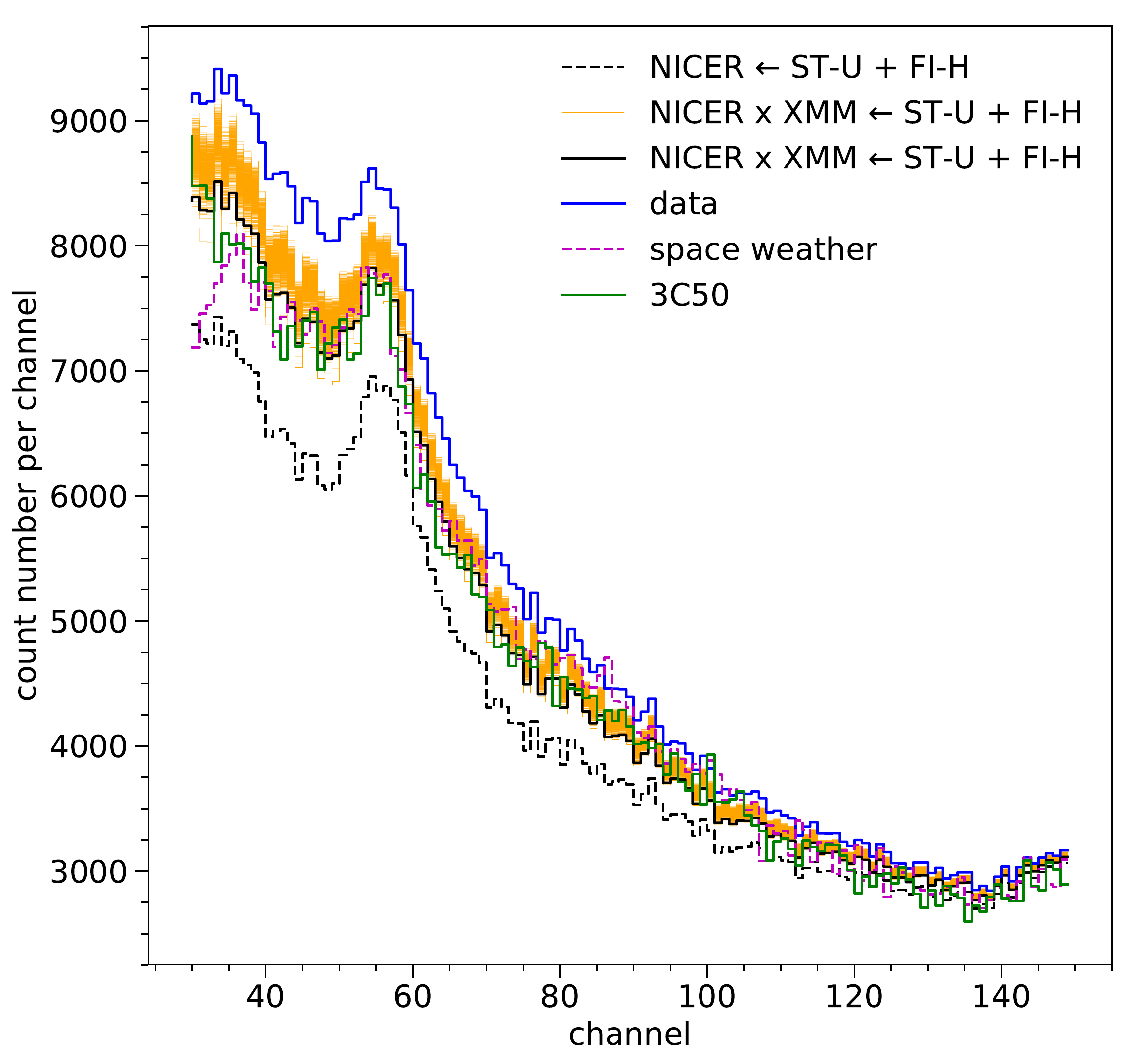}
    \caption{\small{A comparison of the \NICER background conditional on the \NICER likelihood function to the \NICER background conditional on the \NICER and \textit{XMM} likelihood function. The \textit{blue} step function is the total \NICER count spectrum, assumed to be generated by the surface hot regions and the phase-invariant background  in our modeling. The solid \textit{black} step function is the background that maximizes the conditional likelihood function given the parameter vector associated with the nested sample reporting the highest value of the background-marginalized likelihood function. The \textit{orange} step functions (of which there are $10^{3}$) that form a band are defined similarly, but each is conditional on a sample from the joint \NICER and \textit{XMM} posterior. To ensure tractability, we marginalize over background parameters in order to define a sampling space with $\mathcal{O}(10)$ dimensions; it follows that we cannot estimate marginal posterior PDFs from our posterior samples for each background variable due to information loss. Strictly, the \textit{orange} band should therefore \textit{not} be interpreted as a collection of posterior PDFs---one per background variable---but are indicative of where posterior mass will be concentrated. We compare the backgrounds to estimates of the \NICER background generated using the \NICER Space Weather background estimation tool \citep{SpaceWeather} and the `3C50' model \citep{3C50}. We provide a supplementary figure that shows the \NICER background for the $10^{3}$ highest-likelihood nested samples, given the \NICER and \textit{XMM} likelihood function; we also provide a figure that shows the \NICER background for a set of $10^{3}$ posterior samples after compression of the joint prior PDF of the telescope effective areas (see Figure~\ref{fig:effective area prior compression} and associated text). The complete figure set ($3$ images) is available in the online journal.}}
    \label{fig:background_spaceweather}
    \end{figure}
}

In Figure~\ref{fig:XMM background} we display the \textit{XMM} pn background spectra that maximize the conditional likelihood function given nested samples from the joint \NICER and \textit{XMM} posterior, together with supplementary information. Graphical checking of the spectra against the blank-sky estimate and the event data does not reveal any problems \textit{a posteriori}.

\figsetstart
\figsetnum{16}
\figsettitle{\XMM background.}

\figsetgrpstart
\figsetgrpnum{16.1}
\figsetgrptitle{\textit{XMM} pn.}
\figsetplot{f16_1.pdf}
\figsetgrpnote{The \textit{XMM} pn data and model spectra. See the caption in the main text for details.}
\figsetgrpend

\figsetgrpstart
\figsetgrpnum{16.2}
\figsetgrptitle{\textit{XMM} MOS1.}
\figsetplot{f16_2.pdf}
\figsetgrpnote{The \textit{XMM} MOS1 data and model spectra. See the caption in the main text for additional details.}
\figsetgrpend

\figsetgrpstart
\figsetgrpnum{16.3}
\figsetgrptitle{\textit{XMM} MOS2.}
\figsetplot{f16_3.pdf}
\figsetgrpnote{The \textit{XMM} MOS2 data and model spectra. See the caption in the main text for details.}
\figsetgrpend

\figsetend

{
    \begin{figure}[t!]
    \centering
    \includegraphics[clip, trim=0cm 0cm 0cm 0cm, width=\columnwidth]{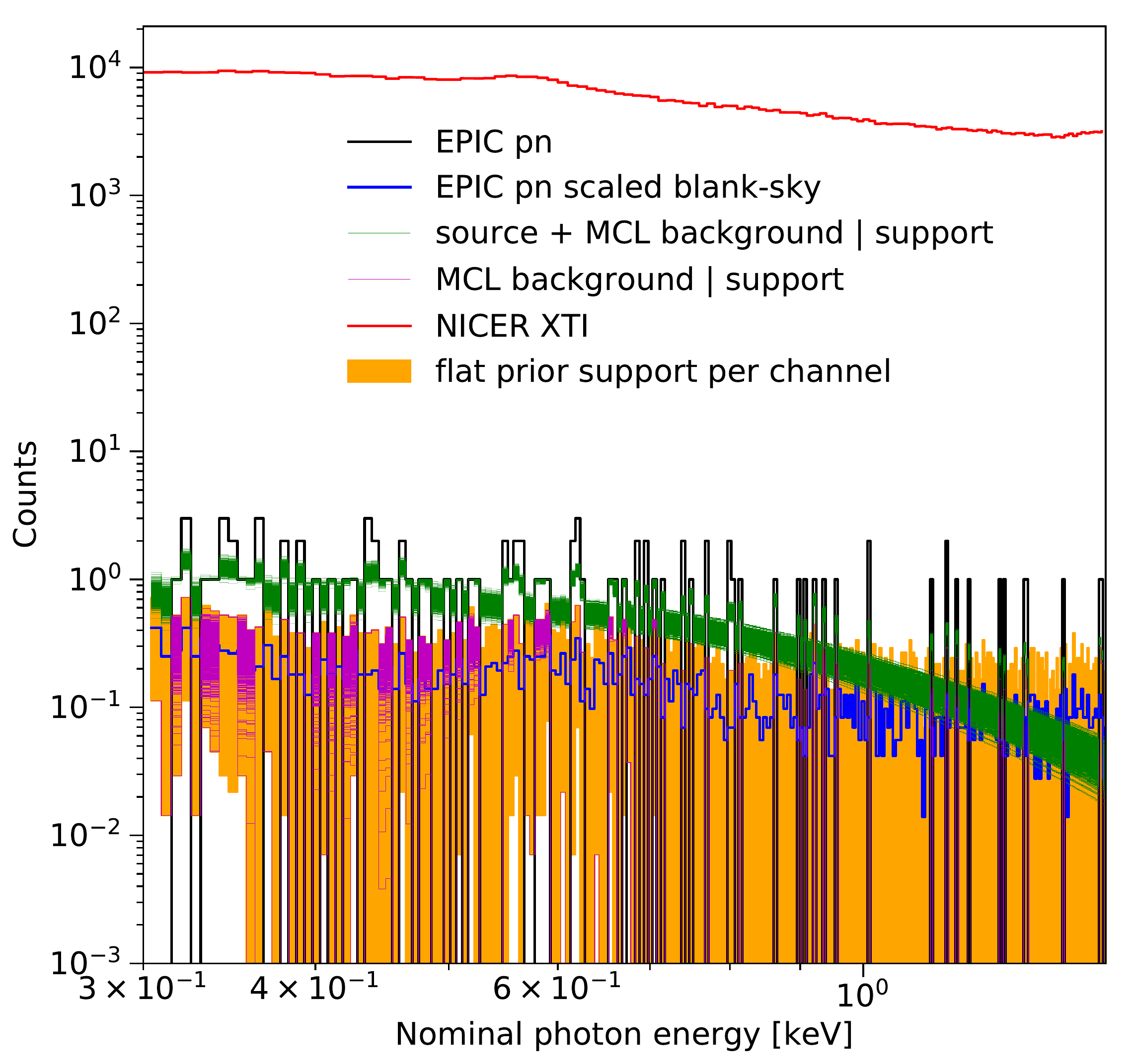}
    \caption{\small{We add to Figure~\ref{fig:pn count spectrum} the \textit{XMM} pn background spectra that maximize the conditional likelihood function (denoted by MCL in the legend) given nested samples from the joint \NICER and \textit{XMM} posterior, but subject to the prior support of the background variables. That is, if the maximum of the conditional likelihood function is not within the prior support, the nearest value of the background to the maximum that is within the prior support is used in the spectrum. We also show the total spectra as the sum of the \textit{XMM} pn source spectra (given the nested samples from the joint \NICER and \textit{XMM} posterior) and the \textit{XMM} pn background spectra that maximize the conditional likelihood function. The complete figure set ($3$ images) for the three \textit{XMM} cameras is available in the online journal.}}
    \label{fig:XMM background}
    \end{figure}
}

\subsection{Hot region configuration}

The hot region configuration is assumed to be related to the star's magnetic field structure. In our previous analysis of \jdbl, the superior model was \TT{ST+PST}: one of the hot regions was a small spherical cap and the other a long extended arc. A configuration in which the hot regions were antipodal was strongly disfavored. Although the hot regions were separated by approximately 180$^\circ$ in longitude, both were in the same hemisphere of the star.

An antipodal configuration is also disfavored for \joh, once again arguing against a simple dipolar model (although the configuration is closer to antipodal than it is for \jdbl). The location of the emitting regions is however rather different.  The expected number of counts contributing to the total expected \NICER signal for \joh is not (close to) minimal \textit{a posteriori}, despite the diffuse \textit{XMM} likelihood function (Section~\ref{sec:XMM constraining power}). Only one of the hot regions vanishes from sight during the rotational cycle; the other remains visible at all times. For PSR~J0030$+$0451, the expected number of counts contributing to the total expected \NICER signal is (close to) minimal \textit{a posteriori} for all models that passed graphical posterior predictive checking \citep{riley19a}; in all cases the hot regions were inferred to dance around the stellar limb, each being entirely non-visible for a substantial fraction of a rotational cycle.

We considered a range of shapes for the hot regions, from circles to rings and arcs. For \joh the \texttt{ST-U} model (in which both hot regions are circles) provides a reasonable description of the data in terms of e.g. residuals. A more complex model, \texttt{ST+PST} (the superior model for \jdbl) did not offer any improvement in model quality measures nor did it lead to changes in the inferred radius or hot region geometry. For \jdbl \texttt{ST-U} provided a reasonable description of the data, but we were sensitive \textit{a posteriori} to additional complexity in the structure of one of the hot regions; consequently, a large shift in the inferred radius was reported. No extended arc structure is inferred for the secondary hot region for \joh, although the hot region could well be a ring instead of a spherical cap.

The pulse profile modeling presented in this work constrains the location and shape of the hot regions on the neutron star surface. These regions arise either via heating by magnetospheric currents \citep{Kalapotharakos20}, or through complex magneto-thermal evolution in the stellar crust \citep{deGrandis20}. Thus, the information obtained can be used as input for modeling magnetic field structure both in the magnetosphere and inside the star, however, there are currently many unknowns in the picture.

Qualitatively, the pulsars which spin faster have more compact magnetospheres and larger (and more complex if the field has a substantial non-dipolar component) open field line regions. If heating happens at the open
field line footprints, then one would expect heated regions of \joh (with the ratio of light cylinder to neutron star radii, $R_\mathrm{LC}/R_\mathrm{NS}=11.1$) to be larger than those of \jdbl ($R_\mathrm{LC}/R_\mathrm{NS}=18.3$), provided that both pulsars have field configurations of similar complexity (i.e. similar relative magnitude of higher-order components), contrary to what is being inferred from the data.

Detailed modeling of pulsar magnetic fields similar that performed by \citet{Kalapotharakos20}, together with an analysis of crustal thermal evolution would be interesting from an evolutionary point of view. \joh has a white dwarf companion while \jdbl is solitary and the difference in recent accretion history may play a role in field configuration and residual heating pattern. Their masses are also substantially different, and according to the  population study by \citet{Antoniadis16}, such a large difference cannot be attributed to accretion alone and must stem partly from the difference in progenitor properties.

\subsection{Analysis cross-check}

An independent analysis carried out within the \NICER collaboration by \citet{IMJ0740} reports a \joh radius of $13.71^{+2.62}_{-1.50}$ km, derived from their combined \NICER and \textit{XMM} analysis.  The 68\% credible intervals overlap with those that we report in this Letter, but the differences deserve some explanation. Recall that our pulse-profile modeling involves several elements: the \NICER phase-resolved data set; the \textit{XMM} phase-averaged data set; a model for the generation of the count data (including priors on the model parameters); and statistical samplers.

Let us first focus on the analysis of the \NICER data. The two teams make different choices on what energy channels to include in the \NICER data set: we use channels [30,150) whereas \citet{IMJ0740} use channels [30,123] (although \citet{IMJ0740} report that including higher channels does not lead to notable changes in their results).

The two teams also make a number of different prior choices. \citet{IMJ0740} assume priors on the mass, distance, and inclination with larger spread to account for potential systematic error on top of the values reported by \citet{Fonseca20}. \citet{IMJ0740}, unlike us, do not impose a hard upper-limit on the prior support of the radius (see Section~\ref{sec:radius prior design}): they assume a flat prior on the reciprocal of the compactness $R_\mathrm{eq}/r_g (M) \sim U(3.2, 8.0)$.\footnote{The absence of prior support for high radii is effectively incorporated at a later stage, in the EOS analysis carried out by \citet{IMJ0740}.} We define the prior support so as to exclude hot-region exchange degeneracy---thus halving the number of posterior modes---whereas \citet{IMJ0740} do not exclude exchange degeneracy. We also condition on different prior PDFs of the hot region center colatitudes and effective temperatures: the prior PDFs of our hot region center colatitudes are isotropic;\footnote{Meaning uniform in the cosine of the hot region center colatitude.} and our prior PDFs of the logarithms of the effective temperatures are uniform. Finally, we use a marginal prior distribution of the energy-independent \NICER effective area scaling factor that has a larger spread and broader prior support than \citet{IMJ0740}; as we discuss below, however, this is not thought to be important. Our prior PDFs are defined in Table~\ref{table: ST-U}.

The two teams implement different statistical sampling protocols. We use nested sampling (MultiNest) with high-resolution settings, whilst \cite{IMJ0740} use a hybrid nested sampling (MultiNest) and parallel-tempering ensemble MCMC scheme; \citet{IMJ0740} use far more core hours during the ensemble sampling phase of their computations than during the nested sampling phase. Where both teams use MultiNest, our resolution settings are higher:\footnote{A caveat is that the performance of nested sampling with given resolution settings, on the same target distribution, is also dependent on the structure of the likelihood function in the native sampling space---the native sampling space is not however unique because different transformations can be defined to inverse-sample a particular joint prior PDF.} we use up to $8\times10^{4}$ live-points with a bounding hypervolume expansion factor of $10$, and we eliminate hot-region exchange degeneracy. For the same target distribution, using a higher number of live-points and eliminating hot-region exchange degeneracy (and thus the halving the number of modes) both yield lower likelihood-isosurface bounding approximation error; a higher number of live points also yields finer sampling of the distribution. Potential variation arising from sampler choice is nicely illustrated in the exploratory study by \citet{Bilby}, which investigates the effect on estimation of parameters of gravitational wave signals from compact binary coalescences.

These different modeling choices lead to some minor differences in the reported spreads of the radius posterior PDFs conditional on \NICER data. However, the degree of overlap is high: we find $R = 11.29_{-0.81}^{+1.20}$ km (see Table \ref{table: NICER ST-U}); \citet{IMJ0740} find $R = 11.51_{-1.13}^{+1.87}$ km.

The radius posterior PDF differences become more pronounced once the \textit{XMM} data set is included. Our posterior is shifted down in radius relative to the \citet{IMJ0740} posteriors which extend to higher radii. Once again, the two groups make a number of different choices that have more of an impact for a smaller data set. We use different formulations of the prior on the \textit{XMM} background:  \citet{IMJ0740} use a distribution based on the assumedly Poissonian observed numbers of blank-sky counts whereas we use a flat prior as described in Section~\ref{sec:XMM background}. Our posterior is also conditional on a broader prior for the cross-calibration uncertainty of the two instruments than \citet{IMJ0740} (who restrict the maximum relative calibration offset to $\pm 10$ \%); this permits lower inferred radii in our analysis (see Section \ref{sec:XMM constraining power}, where we study the effect of narrowing the cross-calibration uncertainty). And as already mentioned, \citet{IMJ0740} are more agnostic in terms of priors on the radius (allowing $R>16$ km): the \textit{XMM} likelihood function permits larger radii (see Figure \ref{fig:ST-U spacetime corner}) and hence their posterior PDF extends accordingly to higher radii.

Despite these differences, the inclusion of the \textit{XMM} data is still extremely valuable, because in both analyses there is a consistent increase in radius, with the lowest radii being ruled out. Moreover, there are good prospects for improving this: once the uncertainties on the \NICER background estimates mentioned in Section \ref{sec:NICERXMM_backgrounds} are clear, we anticipate being able to use those to supplement the indirect constraint on the \NICER background provided by the \textit{XMM} data set.

\section{Conclusions}\label{sec:conclusion}

We have derived a posterior distribution of the radius of the massive rotation-powered millisecond pulsar PSR~J0740$+$6620, conditional on \NICER XTI pulse-profile modeling, joint NANOGrav and CHIME/Pulsar wideband radio timing, and \textit{XMM} EPIC spectroscopy. The radius that we infer for \joh is $12.39_{-0.98}^{+1.30}$ km with an inferred mass (dominated by the radio-derived prior) of $2.072_{-0.066}^{+0.067}$ M$_{\odot}$. A measurement of radius for such a high-mass pulsar should provide a strong constraint on dense matter EOS models \citep[see][]{Raaijmakers21}, with the derived radius favoring models of intermediate stiffness. We anticipate being able to improve this measurement in the near future thanks to the ongoing development of detailed models of the \NICER background. This will be incorporated into future pulse-profile modeling, improving upon the current indirect constraint provided by the \textit{XMM} data set.   

Pulse-profile modeling also enables us to infer the properties of the X-ray emitting hot regions, which are assumed to be linked to the magnetic field structure. The two hot regions are not antipodal, arguing against a simple dipole magnetic field. There is however no evidence for extended crescents, as indicated by pulse profile modeling for \jdbl \citep{riley19a}; simple circular hot regions (spherical caps) suffice to describe the \joh data adequately. How this relates to the evolutionary history of the two sources remains to be determined. 

The analysis presented here also includes improvements to our pulse-profile modeling methodology and software, most notably the ability to include (in this case) phase-averaged X-ray data from \textit{XMM} EPIC. For \joh, the inclusion of this data set led to a remarkable shift in the inferred radius.  We have also investigated the sensitivity to uncertainties in instrumental cross-calibration, an area where we may be able improve our modeling in the future. \textit{XMM} EPIC data sets exist for other \NICER pulse-profile modeling targets, including the source analyzed in \citet{riley19a}, \jdbl. In that analysis the \textit{XMM} EPIC data was used retrospectively, as a check on the consistency of the inferred model {\it a posteriori}; more formally this information should be used to form a likelihood function that is a product of likelihood function factors over telescopes, as in this present work. In future work, we will use the improved pipeline presented in this Letter to perform joint analysis of the \NICER and \textit{XMM} data sets for \jdbl and other sources.

\begin{acknowledgments}
This work was supported in part by NASA through the \NICER mission and the Astrophysics Explorers Program. T.E.R. and A.L.W. acknowledge support from ERC Consolidator Grant No.~865768 AEONS (PI: Watts). T.E.R. also acknowledges support from the Nederlandse Organisatie voor Wetenschappelijk Onderzoek (NWO) through the VIDI and Projectruimte grants (PI: Nissanke). This work was sponsored by NWO Exact and Natural Sciences for the use of supercomputer facilities, and was carried out on the Dutch national e-infrastructure with the support of SURF Cooperative. This work was granted access to the HPC resources of CALMIP supercomputing center under the allocation 2016- P19056. S.B.~was funded in part by NASA grants NNX17AC28G and 80NSSC20K0275. S.M.M. thanks NSERC for support.  W.C.G.H. appreciates use of computer facilities at the Kavli Institute for Particle Astrophysics and Cosmology and acknowledges support through grant 80NSSC20K0278 from NASA.  R.M.L. acknowledges the support of NASA through Hubble Fellowship Program grant HST-HF2-51440.001.  Support for H.T.C. was provided by NASA through the NASA Hubble Fellowship Program grant \#HST-HF2-51453.001 awarded by the Space Telescope Science Institute, which is operated by the Association of Universities for Research in Astronomy, Inc., for NASA, under contract NAS5-26555.  T.T.P. is a NANOGrav Physics Frontiers Center Postdoctoral Fellow funded by the National Science Foundation award number 1430284. The National Radio Astronomy Observatory is a facility of the National Science Foundation operated under cooperative agreement by Associated Universities, Inc. S.M.R. is a CIFAR Fellow and is supported by the NSF Physics Frontiers Center award 1430284. Pulsar research at UBC is supported by an NSERC Discovery Grant and by the Canadian Institute for Advanced Research. This research has made extensive use of NASA's Astrophysics Data System Bibliographic Services (ADS) and the arXiv. We would also like to acknowledge the administrative and facilities staff whose labor supports our work.
\end{acknowledgments}

\facility{NICER XTI~\citep{Gendreau16}, NANOGrav, Green Bank Telescope, CHIME/Pulsar, \XMM EPIC.}

\software{Python/C~language~\citep{python2007}, GNU~Scientific~Library~\citep[GSL;][]{Gough:2009}, NumPy~\citep{Numpy2011}, Cython~\citep{cython2011}, SciPy~\citep{Scipy}, OpenMP~\citep{openmp}, MPI~\citep{MPI}, \project{MPI for Python}~\citep{mpi4py}, Matplotlib~\citep{Hunter:2007,matplotlibv2}, IPython~\citep{IPython2007}, Jupyter~\citep{Kluyver:2016aa}, \TEMPO~\citep[\TT{photons};][]{Hobbs06}, PINT~(\TT{photonphase}; \url{https://github.com/nanograv/PINT}), \MultiNest~\citep{MultiNest_2009}, \textsc{PyMultiNest}~\citep{PyMultiNest}, \project{GetDist}~\citep[][\url{https://github.com/cmbant/getdist}]{Lewis19}, \project{nestcheck}~\citep{higson2018nestcheck,higson2018sampling,higson2019diagnostic}, \project{fgivenx}~\citep{fgivenx}, NICERsoft~(\url{ https://github.com/paulray/NICERsoft}), \XPSI~\texttt{v0.7} (\url{https://github.com/ThomasEdwardRiley/xpsi}; \citealt{xpsi}).}

\bibliographystyle{aasjournal}
\bibliography{nicer_pulse_profile_modeling}

\appendix

\begin{longtable}{l|l|l|lll}
\caption{Summary table for \TT{ST-U} \TT{NSX} fully-ionized hydrogen hot regions plugged into the \NICER~$\times$~\textit{XMM} likelihood function, conditional on the NANOGrav~$\times$~CHIME/Pulsar prior PDF. The description in this caption is largely adopted from \citet{riley19a} for consistency. We provide: (i) the parameters that constitute the sampling space, with symbols, units, and short descriptions; (ii) any notable derived or fixed parameters; (iii) the joint prior distribution, including hard truncation bounds and constraint equations that define the hyperboundary of the support; (iv) one-dimensional (marginal) $68.3\%$ credible interval estimates symmetric in posterior mass about the \textit{median} ($\widehat{\textrm{CI}}_{68\%}$); (v) KL-divergence estimates in \textit{bits} ($\widehat{D}_{\textrm{KL}}$) representing prior-to-posterior information gain (see the appendix of \citet{riley19a} for high-level description of the divergence); and (vi) the parameter vector ($\widehat{\textrm{ML}}$) estimated to maximize the background-marginalized likelihood function, corresponding to a nested sample. Note that strictly, the target of nested sampling is not to generate a maximum likelihood estimator---it is a by-product of the sampling process for evidence estimation. Moreover, there is degeneracy in the likelihood function and high-likelihood solutions with remarkably different parameter values---such as a radius near or above the posterior median---may be retrieved from the public sample information. Constraint equations in terms of two or more parameters result in \textit{marginal} distributions that are not equivalent to those inverse-sampled.}\label{table: ST-U}\\
\hline\hline
Parameter & Description & Prior PDF (density and support) & $\widehat{\textrm{CI}}_{68\%}$ & $\widehat{D}_{\textrm{KL}}$ &
$\widehat{\textrm{ML}}$\\
\hline
\endfirsthead
\multicolumn{6}{c}%
{\tablename\ \thetable\---\textit{Continued from previous page}} \\
\hline
\endhead
\hline \multicolumn{6}{r}{\textit{Continued on next page}} \\
\endfoot
\hline\hline
\endlastfoot
$P$ $[$ms$]$ &
coordinate spin period &
$P=2.8857$,\footnote{\citet{Cromartie19, Wolff20}.} fixed &
$-$ &
$-$ &
$-$\\
\hline
$M$ $[\textrm{M}_{\odot}]$ &
gravitational mass &
$M, \cos(i)\sim N(\boldsymbol{\mu}^{\star},\boldsymbol{\Sigma}^{\star}) $ &
$2.072_{-0.066}^{+0.067}$ &
$0.01$ &
$2.070$ \\
\hline
&joint prior PDF $N(\boldsymbol{\mu}^{\star},\boldsymbol{\Sigma}^{\star})$  & $\boldsymbol{\mu}^{\star}=[2.082,0.0427]^{\top}$\\
&&$\boldsymbol{\Sigma}^{\star}=
\begin{bmatrix}
0.0703^{2} & 0.0131^{2} \\
0.0131^{2} & 0.00304^{2}
\end{bmatrix}$\\
\hline
$R_{\textrm{eq}}$ $[$km$]$ &
coordinate equatorial radius &
$R_{\textrm{eq}}\sim U(3r_{\rm g}(1),16)$\footnote{The function $r_{\rm g}(M)\coloneqq GM/c^{2}$ denotes the gravitational radius with dimensions of length.} &
$12.39_{-0.98}^{+1.30}$ &
$0.58$ &
$11.02$ \\
\hline
&compactness condition\footnote{The coordinate polar radius of the source $2$-surface, $R_{\textrm{polar}}(M,R_{\textrm{eq}},\Omega)$, is a quasi-universal function adopted from \citet[][]{AlGendy2014}, where $\Omega\coloneqq2\pi/P$ is the coordinate angular rotation frequency.} & $R_{\textrm{polar}}/r_{\rm g}(M)>3$\\
&effective gravity condition\footnote{The range of effective gravity from the equator (minimum gravity) to the pole (maximum gravity) must lie within \texttt{NSX limits}. A quasi-universal function is adopted from \citet[][]{AlGendy2014} for effective gravity $g_{\textrm{eff}}(\theta;M,R_{\textrm{eq}},\Omega)$ in units of cm~s$^{-2}$ in the table, where $\Omega\coloneqq2\pi/P$ is the coordinate angular rotation frequency.} & $13.7\leq \log_{10}g_{\textrm{eff}}(\theta)\leq15.0$,~$\forall\theta$\\
\hline
$\Theta_{p}$ $[$radians$]$ &
$p$ region center colatitude &
$\cos(\Theta_{p})\sim U(-1,1)$ &
$1.35_{-0.39}^{+0.46}$ &
$0.25$ &
$1.622$ \\
$\Theta_{s}$ $[$radians$]$ &
$s$ region center colatitude &
$\cos(\Theta_{s})\sim U(-1,1)$ &
$1.89_{-0.46}^{+0.40}$ &
$0.24$ &
$2.303$ \\
$\phi_{p}$ $[$cycles$]$ &
$p$ region initial phase\footnote{With respect to the meridian on which Earth lies.} &
$\phi_{p}\sim U(-0.5,0.5)$, wrapped\footnote{The periodic boundary is admitted and handled by \MultiNest. However, this is an unnecessary measure because we straightforwardly define the mapping from the native sampling space to the space of a phase parameter $\phi$ such that the likelihood function maxima are not in the vicinity of this boundary.} &
bimodal &
$3.52$ &
$0.185$\\
$\phi_{s}$ $[$cycles$]$ &
$s$ region initial phase\footnote{With respect to the meridian on which the Earth antipode lies.} &
$\phi_{s}\sim U(-0.5,0.5)$, wrapped &
bimodal &
$3.51$ &
$0.243$ \\
$\zeta_{p}$ $[$radians$]$ &
$p$ region angular radius &
$\zeta_{p}\sim U(0,\pi/2)$ &
$0.147_{-0.041}^{+0.070}$ &
$2.18$ &
$0.093$ \\
$\zeta_{s}$ $[$radians$]$ &
$s$ region angular radius &
$\zeta_{s}\sim U(0,\pi/2)$ &
$0.146_{-0.042}^{+0.071}$ &
$2.15$ &
$0.127$ \\
\hline
&no region-exchange degeneracy & $\Theta_{s}\geq\Theta_{p}$\\
&non-overlapping hot regions & function of $(\Theta_{p}, \Theta_{s}, \phi_{p}, \phi_{s}, \zeta_{p}, \zeta_{s})$\\
\pagebreak
$\log_{10}\left(\mathcal{T}_{p}\;[\textrm{K}]\right)$ &
$p$ region \TT{NSX} effective temperature &
$\log_{10}\left(\mathcal{T}_{p}\right)\sim U(5.1,6.8)$, \TT{NSX} limits &
$5.988_{-0.059}^{+0.048}$ &
$2.95$ &
$6.080$ \\
$\log_{10}\left(\mathcal{T}_{s}\;[\textrm{K}]\right)$ &
$s$ region \TT{NSX} effective temperature &
$\log_{10}\left(\mathcal{T}_{s}\right)\sim U(5.1,6.8)$, \TT{NSX} limits &
$5.992_{-0.058}^{+0.047}$ &
$2.98$ &
$6.058$ \\
$\cos(i)$ &
cosine Earth inclination to spin axis &
$M,\cos(i)\sim N(\boldsymbol{\mu}^{\star},\boldsymbol{\Sigma}^{\star}) $ &
$0.0424_{-0.0029}^{+0.0029}$ &
$0.01$ &
$0.044$ \\
$D$ $[$kpc$]$ &
Earth distance &
$D\sim \texttt{skewnorm(1.7, 1.0, 0.23)}$\footnote{Specifically, the PDF defined as \texttt{scipy.stats.skewnorm.pdf(D, 1.7, loc=1.002, scale=0.227)}, truncated to the interval $D\in[0,1.7]$~kpc.} &
$1.21_{-0.15}^{+0.15}$ &
$0.10$ &
$0.995$ \\
$N_{\textrm{H}}$ $[10^{20}$cm$^{-2}]$ &
interstellar neutral H column density &
$N_{\textrm{H}}\sim U(0,10)$ &
$1.587_{-1.092}^{+1.953}$ &
$0.91$ &
$0.216$ \\
$\alpha_{\rm{NICER}}$ &
NICER effective-area scaling &
$\alpha_{\rm{NICER}},\alpha_{\rm{XMM}}\sim N(\boldsymbol{\mu},\boldsymbol{\Sigma})$ &
$1.026_{-0.137}^{+0.136}$ &
$0.03$ &
$1.111$\\
$\alpha_{\rm{XMM}}$ &
\textit{XMM} effective-area scaling &
$\alpha_{\rm{NICER}},\alpha_{\rm{XMM}}\sim N(\boldsymbol{\mu},\boldsymbol{\Sigma})$ &
$0.93_{-0.13}^{+0.14}$ &
$0.17$ &
$0.638$ \\
\hline
&joint prior PDF $N(\boldsymbol{\mu},\boldsymbol{\Sigma})$  & $\boldsymbol{\mu}=[1.0,1.0]^{\top}$\\
&&$\boldsymbol{\Sigma}=
\begin{bmatrix}
0.150^{2} & 0.106^{2} \\
0.106^{2} & 0.150^{2}
\end{bmatrix}$\\
\hline
\hline
&Sampling process information&&& \\
\hline
&number of free parameters:\footnote{In the sampling space; the number of background count rate variables is equal to the number of channels defined by the \NICER and \textit{XMM} data sets.} $15$ &&& \\
&number of processes:\footnote{The mode-separation \MultiNest variant was deactivated, meaning that isolated modes are not evolved independently and nested sampling threads contact multiple modes. In principle this also allows us to combine the processes in a post-processing phase using \project{nestcheck}~\citep{higson2018nestcheck}, if more than one process is available for a given posterior; the posteriors derived in the production analysis are high-resolution, so we neglect combining repeat processes.} $1$ &&& \\
&number of live points: $4\times10^{4}$ &&& \\
&hypervolume expansion factor: $0.1^{-1}$ &&& \\
&termination condition: $10^{-1}$ &&& \\
&evidence:\footnote{Defined as the prior predictive probability $p(d_{\rm N}, d_{\rm X},\{\mathscr{B}_{\rm X}\}\,|\,\TT{ST-U})$. Note, however, that in order to complete the reported evidence for comparison to models other than those defined in this work, upper-bounds for the \NICER background parameters need to be specified.} $\widehat{\ln\mathcal{Z}}= -20714.61\pm0.02$ &&&\\
&number of core\footnote{Intel$\textsuperscript{\textregistered}$ Xeon$\textsuperscript{\textregistered}$ E5-2697A v4.} hours: $28320$ &&& \\
&likelihood evaluations: $23710136$ &&& \\
&nested replacements: $1250386$ &&& \\
&effective sample size:\footnote{The effective sample size estimator invoked, following \textbf{DNest4}~\citep[][\url{https://github.com/eggplantbren/DNest4}]{Brewer2018}, is the \textit{perplexity} measure $$\widehat{\textrm{ESS}}\coloneqq\exp\left(-\mathop{\sum}_{i}^{I}w_{i}\log w_{i}\right),$$ where the $\{w_{i}\}_{i=1,\ldots,I}$ are the sample weights \citep[e.g.,][]{Martino2017}.} $420281$ &&& \\
\end{longtable}

\newpage
\begin{longtable}{l|l|l|lll}
\caption{Summary table for \TT{ST-U} \TT{NSX} fully-ionized hydrogen hot regions plugged into the \NICER likelihood function, conditional on the NANOGrav~$\times$~CHIME/Pulsar prior PDF. See the caption of Table~\ref{table: ST-U} and the associated footnotes for details.}\label{table: NICER ST-U}\\
\hline\hline
Parameter & Description & Prior PDF (density and support) & $\widehat{\textrm{CI}}_{68\%}$ & $\widehat{D}_{\textrm{KL}}$ & $\widehat{\textrm{ML}}$\\
\hline
\endfirsthead
\multicolumn{6}{c}%
{\tablename\ \thetable\---\textit{Continued from previous page}} \\
\hline
\endhead
\hline \multicolumn{6}{r}{\textit{Continued on next page}} \\
\endfoot
\hline\hline
\endlastfoot
$P$ $[$ms$]$ &
coordinate spin period &
$P=2.8857$, fixed &
$-$ &
$-$ &
$-$ \\
\hline
$M$ $[\textrm{M}_{\odot}]$ &
gravitational mass &
$M, \cos(i)\sim N(\boldsymbol{\mu},\boldsymbol{\Sigma}) $ &
$2.078_{-0.063}^{+0.066}$ &
$0.01$ &
$2.125$ \\
\hline
&joint prior PDF $N(\boldsymbol{\mu}^{\star},\boldsymbol{\Sigma}^{\star})$  & $\boldsymbol{\mu}^{\star}=[2.082,0.0427]^{\top}$\\
&&$\boldsymbol{\Sigma}^{\star}=
\begin{bmatrix}
0.0703^{2} & 0.0131^{2} \\
0.0131^{2} & 0.00304^{2}
\end{bmatrix}$\\
\hline
$R_{\textrm{eq}}$ $[$km$]$ &
coordinate equatorial radius &
$R_{\textrm{eq}}\sim U(3r_{\rm g}(1),16)$ &
$11.29_{-0.81}^{+1.20}$ &
$0.72$ &
$10.90$ \\
\hline
&compactness condition & $13.7\leq \log_{10}g_{\textrm{eff}}(\theta)\leq15.0$,~$\forall\theta$\\
\hline
$\Theta_{p}$ $[$radians$]$ &
$p$ region center colatitude &
$\cos(\Theta_{p})\sim U(-1,1)$ &
$1.13_{-0.47}^{+0.63}$ &
$0.13$ &
$0.425$ \\
$\Theta_{s}$ $[$radians$]$ &
$s$ region center colatitude &
$\cos(\Theta_{s})\sim U(-1,1)$ &
$1.98_{-0.62}^{+0.49}$ &
$0.13$ &
$1.610$ \\
$\phi_{p}$ $[$cycles$]$ &
$p$ region initial phase\footnote{With respect to the meridian on which Earth lies.} &
$\phi_{p}\sim U(-0.5,0.5)$, wrapped &
bimodal &
$3.46$ &
$-0.267$ \\
$\phi_{s}$ $[$cycles$]$ &
$s$ region initial phase\footnote{With respect to the meridian on which the Earth antipode lies.} &
$\phi_{s}\sim U(-0.5,0.5)$, wrapped &
bimodal &
$3.47$ &
$-0.309$ \\
$\zeta_{p}$ $[$radians$]$ &
$p$ region angular radius &
$\zeta_{p}\sim U(0,\pi/2)$ &
$0.191_{-0.057}^{+0.102}$ &
$1.69$ &
$0.233$ \\
$\zeta_{s}$ $[$radians$]$ &
$s$ region angular radius &
$\zeta_{s}\sim U(0,\pi/2)$ &
$0.189_{-0.058}^{+0.104}$ &
$1.68$ &
$0.132$ \\
\hline
&no region-exchange degeneracy & $\Theta_{s}\geq\Theta_{p}$\\
&non-overlapping hot regions & function of $(\Theta_{p}, \Theta_{s}, \phi_{p}, \phi_{s}, \zeta_{p}, \zeta_{s})$\\
\hline
$\log_{10}\left(\mathcal{T}_{p}\;[\textrm{K}]\right)$ &
$p$ region \TT{NSX} effective temperature &
$\log_{10}\left(\mathcal{T}_{p}\right)\sim U(5.1,6.8)$, \TT{NSX} limits &
$6.014_{-0.063}^{+0.048}$ &
$2.90$ &
$6.068$ \\
$\log_{10}\left(\mathcal{T}_{s}\;[\textrm{K}]\right)$ &
$s$ region \TT{NSX} effective temperature &
$\log_{10}\left(\mathcal{T}_{s}\right)\sim U(5.1,6.8)$, \TT{NSX} limits &
$6.017_{-0.062}^{+0.048}$ &
$2.89$ &
$6.086$ \\
$\cos(i)$ &
cosine Earth inclination to spin axis &
$M,\cos(i)\sim N(\boldsymbol{\mu}^{\star},\boldsymbol{\Sigma}^{\star}) $ &
$0.0426_{-0.0028}^{+0.0029}$ &
$0.01$ &
$0.045$ \\
$D$ $[$kpc$]$ &
Earth distance &
$D\sim \texttt{skewnorm(1.7, 1.0, 0.23)}$\footnote{The PDF defined as \texttt{scipy.stats.skewnorm.pdf(D, 1.7, loc=1.002, scale=0.227)}, truncated to the interval $D\in[0,1.7]$~kpc.} &
$1.19_{-0.14}^{+0.14}$ &
$0.08$ &
$1.145$ \\
$N_{\textrm{H}}$ $[10^{20}$cm$^{-2}]$ &
interstellar neutral H column density &
$N_{\textrm{H}}\sim U(0,10)$ &
$1.34_{-0.93}^{+1.87}$ &
$1.07$ &
$0.029$ \\
$\alpha_{\rm{NICER}}$ &
NICER effective-area scaling &
$\alpha_{\rm{NICER}}\sim N(1,0.15^{2})$ &
$0.97_{-0.13}^{+0.15}$ &
$0.04$ &
$1.083$ \\
\hline\hline
&Sampling process information&&& \\
\hline
&number of free parameters:\footnote{In the sampling space; the number of background count rate variables is equal to the number of channels defined by the \NICER data set.} $14$ &&& \\
&number of processes:\footnote{The mode-separation \MultiNest variant was deactivated, meaning that isolated modes are not evolved independently and nested sampling threads contact multiple modes.} $1$ &&& \\
&number of live points: $4\times10^{4}$ &&&& \\
&hypervolume expansion factor: $0.1^{-1}$ &&&& \\
&termination condition: $10^{-1}$ &&&& \\
&number of core\footnote{Intel$\textsuperscript{\textregistered}$ Xeon$\textsuperscript{\textregistered}$ E5-2697A v4.} hours: $24000$ &&&& \\
&likelihood evaluations: $26858453$ &&&& \\
&nested replacements: $1208371$ &&&& \\
&effective sample size: $303905$ &&&& \\
\end{longtable}

\listofchanges
\end{document}